\documentclass[review]{elsarticle}

\usepackage{lineno,hyperref}
\modulolinenumbers[5]

\journal{Journal of Quantitative Spectroscopy and Radiative Transfer}

\usepackage{graphicx}
\usepackage{longtable}
\usepackage{multirow}
\usepackage{lscape}

\begin{document}

\begin{frontmatter}

\title{Atlas and wavenumber tables for visible part of the multiline 
electronic-vibro-rotational emission spectrum of the $D_2$ molecule 
measured with moderate resolution in deuterium plasma with minor hydrogen impurity.}

\author{B.P. Lavrov, I.S. Umrikhin}
\address{Faculty of Physics, St.-Petersburg State University, St.-Petersburg, 198504, Russia}

\begin{abstract}
The visible part ($\approx 419-696$ nm) of the multiline 
electronic-vibro-rotational emission spectrum of the $D_2$ molecule 
was recorded with moderate resolution mainly limited by Doppler 
broadening of spectral lines (line widths FWHM less than $0.013$ nm). 
After numerical deconvolution of the recorded intensity distributions 
and proper calibration of the spectrometer the new set of wavenumber 
values was obtained. It is shown that new data are significantly more 
precise than those available for visible part of the $D_2$ spectrum.
Final results are reported as a supplement in the form of an atlas divided into 158 sections 
(each section covers about 1.5 nm) containing pictures of images in the 
focal plane of the spectrometer, intensity distributions in linear and 
logarithmic scales and the table containing wavenumber and relative 
intensity values for 11941 spectral lines together with available and 
new line assignments for the $D_2$ and $HD$ molecules. 
\end{abstract}


\end{frontmatter}


\section{Introduction}

Any activity in practical spectroscopy starts from recording certain spectrum 
and recognizing lines, branches and bands interesting for an experimentalist. 
Most straightforward, dependable and easy way for the recognition is direct 
comparison of an observed spectrum with certain reference atlases of spectra 
for various atoms and molecules. Currently for molecular hydrogen such an 
atlas is available only for the $H_2$ isotopologue and 
only for very small part of the vacuum ultraviolet (VUV) emission spectrum 
$78.60 - 171.35$ nm \cite{RL1994}. Present work reports an atlas of the multiline 
electronic-vibro-rotational (rovibronic) spectrum of the $D_2$ molecule for visible 
part of the emission spectrum ($419 - 696$ nm), which is known to be most suitable 
and therefore most used for practical applications in studies of deuterium containing 
ionized gases and plasmas (See e.g. \cite{LKOR1997, LMKR1999, RDKL2001, PBS2005, LPR2006, NTPCP2013}).

Experimental studies of the $D_2$ emission spectrum have been started soon 
after discovery of the heavy isotope of atomic hydrogen \cite{Urey1932_1, Urey1932_2}. 
Wavenumber values of rovibronic radiative transitions obtained by emission 
spectroscopy in visible 
\cite{DiekeBlue1935, Dieke1935_1, Dieke1935_2, Dieke1935_3, Dieke1936, DiekeLewis1937}
and infrared (IR) \cite{DiekePorto, GloersenDieke, DiekeCunningham} parts of the $D_2$ 
spectrum together with those obtained by VUV 
\cite{Wilkinson1968, Monfils1968, BredohlHerzberg1972, DabrowskiHerzberg1974, TT1975, LLR1980}
and anticrossing 
\cite{JLDRFMZ1976, MFZ1976, MZF1978, MF1977, FMZ1976} spectroscopic experiments were 
collected and analyzed in the review paper \cite{FSC1985}. Later only few fragmentary 
measurements were made 
in the middle infrared part of the spectrum (about $4.5$ $\mu m$) by FTIR (Fourier transform infrared) 
\cite{DabrHerz} and tunable laser \cite{Davies} spectroscopy. Measurements of the wavenumber 
values for separate rovibronic lines in VUV and empirical determination of singlet rovibronic 
term values are in progress up to now 
\cite{RLTBjcp2006, RLTBjcp2007, RIVLTUmol2008, LSJUMchp2010, GJRT2011, DIUROJNTGSKEmol2011}. 

\begin{figure}[!ht]
\begin{center}
\includegraphics[width=\textwidth]{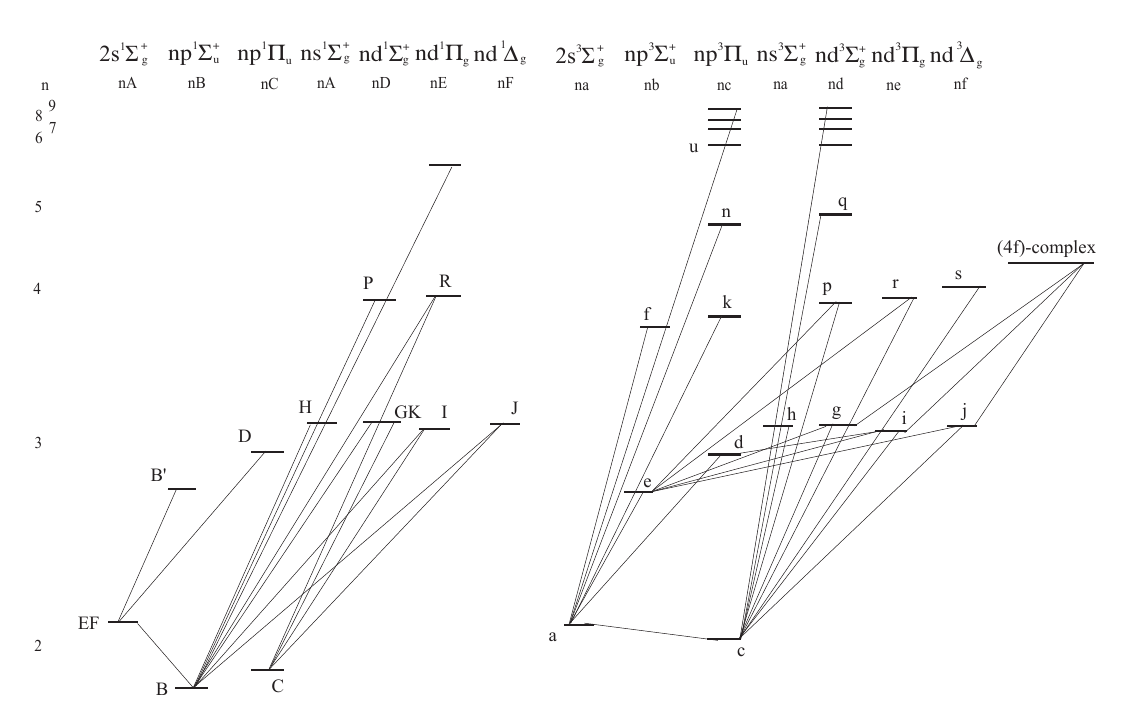}
\caption{\label{allgrotr}
Grotrian diagram representing currently known bound electronic states and
band systems of the $D_2$ molecule.}
\end{center}
\end{figure}

The spectrum of the $D_2$ molecule is caused by both singlet-to-singlet and triplet-to-triplet 
rovibronic transitions. The intercombination lines were not observed yet. The most interesting 
resonance singlet-to-singlet band systems connected with the ground electronic state are 
located in vacuum ultraviolet. Singlet-to-singlet and triplet-to-triplet transitions 
between excited electronic states are responsible for light emission of ionized gases and 
plasmas in near infrared, visible and near ultraviolet. Grotrian diagram for currently known 
bound electronic states and studied band systems of the $D_2$ molecule is shown in figure~\ref{allgrotr}. 
One may see that a lot of electronic states and band systems are not observed, recognized and 
assigned yet. It should be noted that atomic and molecular emission lines in the
visible part of the spectrum are most often used for spectroscopic diagnostics 
of non-equilibrium plasmas (see e.g. \cite{RDKL2001, NTPCP2013}).

{\def\baselinestretch{1.0} 
\begin{table}
\caption{The list of notations used for a designation of electronic states of the $D_2$ 
molecule corresponding to various electron configurations.}\label{notations}
\begin{tabular}{@{}ccrlcrl}

\hline
Electron & 
\multicolumn{3}{c}{Singlet electronic states} & 
\multicolumn{3}{c}{Triplet electronic states} \\
\cline{2-7}
configuration  & 
   \multicolumn{2}{c}{Traditional} & Dieke &
   \multicolumn{2}{c}{Traditional} & Dieke \\
\cline{2-7}
 & \cite{Huber} & \multicolumn{2}{c}{\cite{FSC1985}} &
   \cite{Huber} & \multicolumn{2}{c}{\cite{FSC1985}} \\
\hline

$1s\sigma^2$         & $X^1\Sigma_g^+ 1s\sigma$   & $X(1s)^1\Sigma_g^+$    & $1A$     & \multicolumn{3}{c}{$-$} \\
                                                                       
$1s\sigma 2s\sigma$  & $E^1\Sigma_g^+ 2s\sigma$   & 
                                    \multirow{2}{*}{$EF^1\Sigma_g^+$}      & 
                                                             \multirow{2}{*}{$EF$}    & $a^3\Sigma_g^+ 2s\sigma$   & $a(2s)^3\Sigma_g^+$   & $2a$     \\
$2p\sigma^2$         & $F^1\Sigma_g^+ 2p\sigma^2$ &                        &          &                            &                       &          \\
$1s\sigma 2p\sigma$  & $B^1\Sigma_u^+ 2p\sigma$   & $B(2p)^1\Sigma_u^+$    & $2B$     & $b^3\Sigma_u^+ 2p\sigma$   & $b(2p)^3\Sigma_u^+$   & $2b$     \\
$1s\sigma 2p\pi$     & $C^1\Pi_u 2p\pi$           & $C(2p)^1\Pi_u^\pm$     & $2C^\pm$ & $c^3\Pi_u 2p\pi$           & $c(2p)^3\Pi_u^\pm$    & $2c^\pm$ \\
                                                                       
$1s\sigma 3s\sigma$  &                            & $H(3s)^1\Sigma_g^+$    & $3A$     &                            & $h(3s)^3\Sigma_g^+$   & $3a$     \\
$1s\sigma 3p\sigma$  & $B'^1\Sigma_u^+ 3p\sigma$  & $B'(3p)^1\Sigma_u^+$   & $3B$     & $e^3\Sigma_u^+ 3p\sigma$   & $e(3p)^3\Sigma_u^+$   & $3b$     \\
$1s\sigma 3p\pi$     & $D^1\Pi_u 3p\pi$           & $D(3p)^1\Pi_u^\pm$     & $3C^\pm$ & $d^3\Pi_u 3p\pi$           & $d(3p)^3\Pi_u^\pm$    & $3c^\pm$ \\
$1s\sigma 3d\sigma$  & $G^1\Sigma_g^+ 3d\sigma$   & 
                                    \multirow{2}{*}{$GK^1\Sigma_g^+$}      & 
                                                             \multirow{2}{*}{$GK$}    & $g^3\Sigma_g^+ 3d\sigma$   & $g(3d)^3\Sigma_g^+$   & $3d$     \\
                     & $K(^1\Sigma_g^+)$          &                        &          &                            &                       &          \\
$1s\sigma 3d\pi$     & $I^1\Pi_g^ 3d\pi$          & $I(3d)^1\Pi_g^\pm$     & $3E^\pm$ & $i^3\Pi_g 3d\pi$           & $i(3d)^3\Pi_g^\pm$    & $3e^\pm$ \\
$1s\sigma 3d\delta$  &                            & $J(3d)^1\Delta_g^\pm$  & $3F^\pm$ & $j^3\Delta_g 3d\delta$     & $j(3d)^3\Delta_g^\pm$ & $3f^\pm$ \\
                     
$1s\sigma 4p\sigma$  & $B''^1\Sigma_u^+ 4p\sigma$ & $B''(4p)^1\Sigma_u^+$  & $4B$     & $f^3\Sigma_u^+ 4p\sigma$   & $f(4p)^3\Sigma_u^+$   & $4b$     \\
$1s\sigma 4p\pi$     & $D'^1\Pi_u 4p\pi$          & $D'(4p)^1\Pi_u^\pm$    & $4C^\pm$ & $k^3\Pi_u 4p\pi$           & $k(4p)^3\Pi_u^\pm$    & $4c^\pm$ \\
$1s\sigma 4d\sigma$  &                            & $P(4d)^1\Sigma_g^+$    & $4D$     & $p^3\Sigma_g^+ 4d\sigma$   & $p(4d)^3\Sigma_g^+$   & $4d$     \\
$1s\sigma 4d\pi$     &                            & $R(4d)^1\Pi_g^\pm$     & $4E^\pm$ & $r^3\Pi_g 4d\pi$           & $r(4d)^3\Pi_g^\pm$    & $4e^\pm$ \\
$1s\sigma 4d\delta$  &                            &                        &          &                            & $s(4d)^3\Delta_g^\pm$ & $4f^\pm$ \\
                                                                       
$1s\sigma 5p\sigma$  &                            & $B'''(5p)^1\Sigma_u^+$ & $5B$     &                            &                       &          \\
$1s\sigma 5p\pi$     & $D''^1\Pi_u 5p\pi$         & $D''(5p)^1\Pi_u^\pm$   & $5C^\pm$ & $n^3\Pi_u 5p\pi$           & $n(5p)^3\Pi_u^\pm$    & $5c^\pm$ \\
$1s\sigma 5d\sigma$  &                            &                        &          & $q(^3\Sigma_g^+) 5d\sigma$ & $q(5d)^3\Sigma_g^+$   & $5d$     \\
$1s\sigma 5d\pi$     &                            &                        &          & $w(^3\Pi_g) 5d\sigma$      &                       &          \\
                                                                       
$1s\sigma 6p\sigma$  &                            & $(6p)^1\Sigma_u^+$     & $6B$     &                            &                       &          \\
$1s\sigma 6p\pi$     &                            & $(6p)^1\Pi_u^\pm$      & $6C^\pm$ & $u^3\Pi_u 6p\pi$           & $u(6p)^3\Pi_u^\pm$    & $6c^\pm$ \\
$1s\sigma 6d\sigma$  &                            &                        &          &                            & $(6d)^3\Sigma_g^+$    & $6d$     \\
                                         
$1s\sigma 7p\sigma$  &                            & $(7p)^1\Sigma_u^+$     & $7B$     &                            &                       &          \\
$1s\sigma 7p\pi$     &                            & $(7p)^1\Pi_u^\pm$      & $7C^\pm$ &                            & $(7p)^3\Pi_u^\pm$     & $7c^\pm$ \\
$1s\sigma 7d\sigma$  &                            &                        &          &                            & $(7d)^3\Sigma_g^+$    & $7d$     \\
                                         
$1s\sigma 8p\sigma$  &                            & $(8p)^1\Sigma_u^+$     & $8B$     &                            &                       &          \\
$1s\sigma 8p\pi$     &                            & $(8p)^1\Pi_u^\pm$      & $8C^\pm$ &                            & $(8p)^3\Pi_u^\pm$     & $8c^\pm$ \\
$1s\sigma 8d\sigma$  &                            &                        &          &                            & $(8d)^3\Sigma_g^+$    & $8d$     \\
                                         
$1s\sigma 9p\sigma$  &                            & $(9p)^1\Sigma_u^+$     & $9B$     &                            &                       &          \\
$1s\sigma 9p\pi$     &                            & $(9p)^1\Pi_u^\pm$      & $9C^\pm$ &                            & $(9p)^3\Pi_u^\pm$     & $9c^\pm$ \\
$1s\sigma 9d\sigma$  &                            &                        &          &                            & $(9d)^3\Sigma_g^+$    & $9d$     \\
\hline

\end{tabular}
\end{table}
}

For labeling the electronic states of the $D_2$ molecule two different types 
of a notation are used at present time. Traditional notation \cite{Huber} does 
not need any additional explanations. It is based on the assumption for 
applicability of an adiabatic approximation and Hund's case 'b' for angular 
momentum coupling scheme. The notation earlier introduced by G.H.Dieke 
\cite{Dieke1972} (and later made more exact in \cite{FSC1985}) is based on the 
same assumptions but it is much more compact what is very important for long 
tables of spectral lines. For example, the $(s\sigma)\Sigma_g^+$ states are 
denoted by uppercase letter "A" for singlets and by lowercase letter "a" 
for triplets, the $(p\sigma)\Sigma_u^+$ states are marked by "B" and "b", 
the $(p\pi)\Pi_u$ states by "C" and "c", the $(d\sigma)\Sigma_g^+$ states 
by "D" and "d" etc. with the principal quantum number $n$ of the excited 
electron for united atom limit case included as a prefix.

The relation between the notations from \cite{Huber} and \cite{FSC1985}
is presented in table~\ref{notations}. One may see that Dieke's notation is
much more compact than the traditional one. Only main part of all investigated 
electronic states of the $D_2$ molecule is listed in table~\ref{notations}. 
Singlet electronic states $(np)^1\Sigma_u^+$ up to $n=46$ were studied by 
absorption spectroscopy in the VUV region of the $D_2$ spectrum \cite{TT1975},
the study of these states is outside the scope of our paper and they are not
listed in the table~\ref{notations}.

For the rovibronic transitions in the tables of the present paper, we used modified 
Dieke's notation from \cite{FSC1985}. It consists of: the multiplicity (all lines 
are assigned to either singlet-to-singlet or a triplet-to-triplet transitions, 
indicated by capital letter S or T respectively), the reflection (Kr\"onig) 
symmetry of the upper electronic state (is given as $+$ or $-$), an electronic 
transition in Dieke's notation with the upper electronic state coming first, 
a designation of the band (the vibrational quantum numbers of upper $v'$ and 
lower $v''$ vibrational levels in the parentheses), and finally a designation 
of rovibronic spectral line including a branch ($P$, $Q$, or $R$) followed 
by the rotational quantum number $N''$ (of total angular momentum excluding 
spins of electrons and nuclei) for lower rovibronic level. For example the 
notation "T+  4b-2a  (2-3) P1" represents the transition between 
$f(4p)^3\Sigma_u^+$, $v'=2$, $N'=0$ and $a(2s)^3\Sigma_g^+$, $v''=3$, $N''=1$ 
rovibronic levels in the traditional notation.

\section{Experimental}

Our experimental setup and techniques of data processing were briefly described
in our previous paper \cite{LUZ2012}. This short communication reported measurements
of high-resolution spectra of the $D_2$ molecule emitted by plasma of extremely 
low current density glow discharge with cold cathode and water-cooled walls to provide low
Doppler broadening of spectral lines. Such statement of the experiment made it
possible to observe and analyze the partly resolved fine structure of rovibronic lines
connected with vibro-rotational levels of the $c^3\Pi_u^-$ electronic state. 
At the same time, low current density and low translational (gas) temperature of plasma
leads to the considerable decrease of the total number of observable rovibronic lines, 
because only rotational levels with small quantum numbers are sufficiently populated.

The goal of the present paper was to study as much rovibronic lines as possible. 
That was reached by increasing the current density in plasma to obtain higher 
gas temperature, which leads to an increase of population densities of high 
rotational levels in the ground and excited electronic-vibrational states. 
That greatly increased Doppler broadening of the lines and mainly determined 
the moderate resolution of observed spectra. Therefore, we have to describe 
in detail the setup, experimental conditions and data processing technique 
used in the present work for measurements and analysis of the spectra.

Emission of plasma located inside additional cylindrical molybdenum 
electrode (with inner diameter \O{2} mm) introduced between the anode and the hot
cathode of the low-pressure arc discharge was used as a light source. The flux of 
radiation through a hole in the anode was focused by achromatic lens on the 
entrance slit of the spectrometer. The hot-cathode arc discharge
lamp LD-2D described in \cite{GLT1982} was filled with $\approx 6$ Torr of 
chemically pure deuterium with small impurity of hydrogen ($\leq 6\%$). 
Main measurements were made with the discharge current 
300 mA. Gas temperature $T = 1900 \pm 170 $ K
was obtained from the intensity distribution in the Q-branch of
the $(2-2)$ band of the $d^3\Pi_u^- - a^3\Sigma_g^+$ electronic transition (see e.g. 
\cite{L1980, AKKK1996}). Thus, in the wavelength range $\lambda = 419 - 696$ nm the
Doppler linewidths (FWHM) in wavenumbers were within the range $\Delta \nu_D = 0.22 - 0.37$ cm$^{-1}$.

Detailed description of the self-made high 
resolution automatic spectrometer and corresponding software was described in detail in
\cite{LMU2011}. The Ebert-type spectrograph equipped with additional camera lens
(the effective focal length $F \approx 6.8 m$) has the reverse linear dispersion 
$0.076 - 0.065$ nm/mm in the wavelength range under the study
and highest possible resolving power up to $1.8 \times 10^5$. However, 
actual resolving power in the experimental conditions reported in the present 
paper was mainly limited by Doppler broadening of the $D_2$ spectral lines due 
to small reduced mass of nuclei and high translational temperature of molecules 
(see above).

The dynamic range of measurable intensities for every single recording (one shot) 
depends on the range of linear response of photodetectors and the noise amplitude. 
But linear dependence of the photoelectric signal on the exposure duration makes 
it possible to increase signal-to-noise ratio by means of summing the results of 
multiple shots for the same "window" in the spectrum. For example, by summing 150 
shots it was possible to reach the dynamic range value of $10^4$ \cite{LMU2011}. 
Taking into account the goal of the present work, we restricted ourselves to 
obtaining the dynamic range of only $\approx 100$ to be able to cover completely 
the visible part of the spectrum in reasonable time scale. On the other hand, 
that value of the dynamic range is typical for applied researches in plasma spectroscopy
(see e.g. \cite{NTPCP2013, OLR2002}).

For determination of wavenumbers of rovibronic lines, we used the optimizational 
technique based on linear response of a multichannel photo detector and digital 
recording of intensities (see \cite{LUZ2012, LMU2011, ALMU2008, LU2008, LU2009}). 
It has certain advantage over traditional photographic recording previously used for 
obtaining wavenumber values in visible part of the $D_2$ spectrum. The technique gives 
the opportunity to study line profiles. In the case of an overlap between neighboring line profiles  
(the so-called effect of the line blending) it is possible to make numerical deconvolution (inverse to the convolution)
operation and thus to obtain correct wavenumber values of rovibronic transitions even for blended lines. 
Such additional digital resolution is sufficiently higher than the optical resolution of the spectrometer.

We are treating the inverse spectroscopic problem of determining the 
wavenumber values of rovibronic transitions from measured intensity 
distributions as that of the so-called conditional optimization. This 
technique consists of the following components.

i.  The first one is a parameterization of a model describing measured illuminance 
distribution in the focal plane of the spectrometer --- a dependence of the photoelectric 
signal $I_k^{expt}$ of the k-th column of CMOS matrix photodetectors on their position $x_k$. 
In the present work, we assumed that inside every small enough part of 
the spectrum ($\approx 0.5$ nm wide) the distribution may be represented 
as a sum of a constant level of the background intensity $I_{bg}$ and a 
finite number $M$ of spectral lines having Gaussian shape of the line profiles:

\begin{equation}
I^{calc}(x) = I_{bg} + \sum\limits_{i=1}^{M} A_i f_i(x, x^0_i, \Delta x_i), \label{eq:icalc}
\end{equation}

where $A_i$ is the intensity of the i-th spectral line corresponding to its centre $x^0_i$, 
and $\Delta x_i$ - the line width (FWHM).

ii.  The second step is the formulation of reasonable restricting conditions 
for the optimization process. Experimental studies of the shape of profiles 
for bright and unblended spectral lines of the $D_2$ molecule were carried out. 
They shown that in our conditions the line profiles may be described by 
Gaussian shape, corresponding to the convolution of Doppler and instrumental 
broadening. Moreover, within small parts of recorded spectra ($\approx 0.5-1$ nm) 
the measured values of the line widths $\Delta x_i$ of all such lines of the $D_2$ 
molecule are in coincidence within error bars. Thus, optimal approximations 
of every small part of the spectrum were obtained under the condition of 
Gaussian shape of the profiles and common value of the $\Delta x_i = \Delta x$ 
for all (blended and unblended) lines emitted by the $D_2$ molecule.

iii.  Then we obtained an optimal set of the parameters by searching a 
global minimum of the following function
\begin{equation}
\Phi(\{A_i, x^0_i\}_{i=1...M}, \Delta x, I_{bg}) = 
  \sum\limits_{k=1}^{K} (I_k^{expt} - I^{calc}(x_k))^2, \label{eq:r2}
\end{equation}

which represents the sum of squares of deviations between observed and model (\ref{eq:icalc})
illuminance distribution. Here $K$ indicates the order number of experimental intensity values $I_k^{expt}$ 
in the spectral region choosed for the optimizational analysis. 
The least-squares criterion (\ref{eq:r2}) corresponds to the 
maximum likelihood principle in the case of random errors of measured illuminance distributions.

The problem of finding the number of distinguishable spectral lines $M$ giving 
random distribution of the differences $\Delta I_k = I_k^{expt} - I^{calc}(x_k)$  
for each $k = 1,2,3,...,K$ may be solved only in the framework of interactive 
analysis of $\Delta I_k$ values. On the first iteration of such an analysis
we tried to find minimal number of Gaussian profiles $M_0$, approximating experimental 
illuminance distribution. Every local maximum in the recorded distribution, 
which could be distinguished by the naked eye, was approximated by one profile. 
In the case of an asymmetric shape or too thick line profile, we assumed that it is 
due to overlap of two spectral lines having different intensities. In the cases of 
more complex overlaps, we tried to find minimum number of the line profiles providing 
a random distribution of the differences $\Delta I_k$.

By minimizing the function (\ref{eq:r2}) we obtained an optimal set of
parameters $\{A_i, x^0_i\}_{i = 1...M_0}$, $\Delta x$, and $I_{bg}$ for certain 
value of $M_0$. Then the differences $\Delta I_k$ and their distributions were 
analyzed. In the case of systematic deviations in a dependence of the $\Delta I_k$ 
values on $x_k$ we added more line profiles for compensating those deviations by 
the second iteration. Then by minimizing (\ref{eq:r2}), we obtained new set of 
parameters $\{A_i, x^0_i\}_{i = 1...M_1}$, $\Delta x$, and $I_{bg}$, and analyzed 
$\Delta I_k$ again. This iterative procedure lasts until we achieved random distribution 
of $\Delta I_k$ within every spectral region under the study. Thus, the optimal 
sets of parameters $\{A_i, x^0_i\}_{i = 1...M_m}$, $\Delta x$ for all spectral 
lines and certain background intensity value $I_{bg}$ were obtained.

Due to small hydrogen impurity in plasma under the study, we may observe not only easy 
distinguishable lines of hydrogen atoms. Some strong lines of the $HD$ molecule could 
be recorded as well. Doppler widths of the $HD$ lines are $15.4\%$ larger than those 
of the $D_2$ lines (vast majority in each spectral "window"). Thus, our optimization 
condition of the identical width for all lines within every small spectral range is 
not good for the $HD$ lines. Unfortunately, wavenumber values and assignments of 
the $HD$ lines are currently available only for rather limited number of bands studied 
in \cite{DiekeBlue1935, Dieke1935_2, DiekeLewis1937, KB1987}. 
Therefore, when we suspected that some line in our 
spectrum belongs to the $HD$ molecule (because it was earlier assigned as belonging), 
we allowed to our program to consider this line as an exclusion and to determine her 
width regardless of the common width determined for the $D_2$ lines located in the 
same window. In the cases of secluded and bright $HD$ lines the values of their widths 
became $10-30\%$ larger than before the exclusion. For blended and weak $HD$ lines, our 
optimizational determination of their characteristics was ill-conditioned problem. Thus, 
some lines which existence seems questionable to authors of the present work are marked 
by asterisk in final tables. 

Wavelength dependence along the dispersion direction for a long-focus spectrometer 
is close to linear in the centre of the focal plane (see e.g. the formula (2) in \cite{LMU2011}).
It may be approximated by polynomial of the ratio $x/F$, which in our case is less than $0.002$ 
(here $x$ is the small displacement from the centre of the matrix detector, and $F$ is effective focal 
length of the spectrometer). On the other hand, the wavelength dependence of the 
refractive index of air  $n(\lambda)$ is also close to linear inside a small enough part of the 
spectrum. Inside narrow spectral intervals the product 
$\lambda_{vac}(x) = \lambda(x) n(\lambda(x))$ may be also approximated by low degree power series expansion.
Therefore, we calibrated the spectrometer directly in vacuum wavelengths $\lambda_{vac} = 1 / \nu$.

Another peculiarity of our calibration technique is using the experimental 
vacuum wavelength values $1 / \nu$ from \cite{FSC1985} as the reference marks. 
For bright unblended spectral lines these wavelength values show small 
random spread around smooth curve approximating the dependence of the 
wavelengths of the lines against their positions in the focal plane of the 
spectrometer. Moreover, those random errors are in good accordance with normal 
(Gaussian) distribution function. Thus, it is possible to obtain a precision for 
new wavenumber values even a little bit better than those of the reference data due to smoothing.

To be sure that the data from \cite{FSC1985} are free from systematic errors 
we have had to perform special experiments with capillary-arc lamp analogous 
to that described in \cite{LSh1979} (capillary diameter $d = 1.5$ mm and current 
density $j = 30$ A/cm$^2$) but filled with the $H_2+D_2+Ne$ mixture (1:1:2) under 
total pressure $P \approx 8$ Torr \cite{LU2011}.

\begin{figure}[!ht]
\begin{center}
\includegraphics[width=0.5\textwidth]{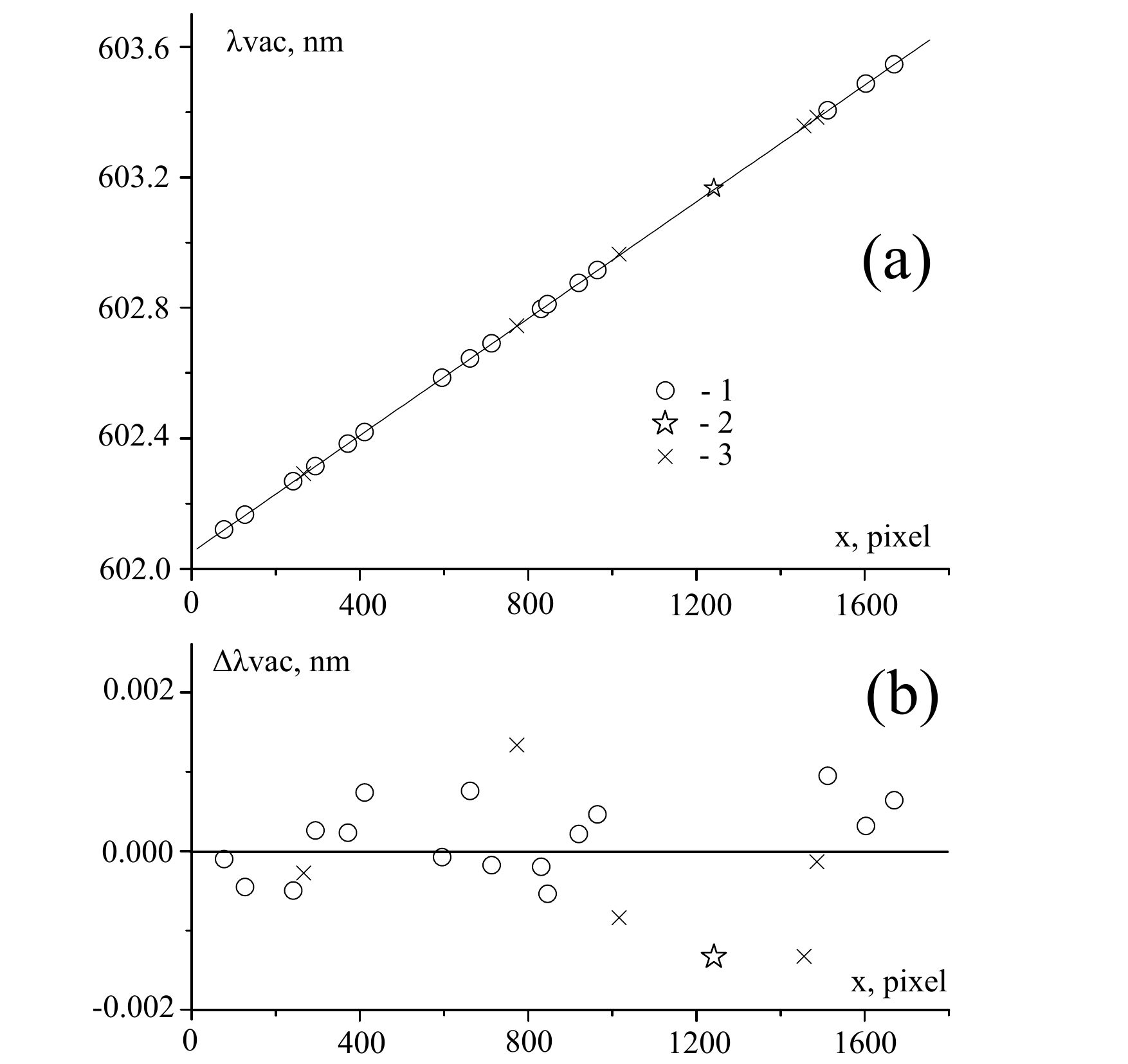}
\caption{Dependences of the vacuum wavelengths $\lambda_{vac}$ of the brightest 
$D_2$, $H_2$ and $Ne$ spectral lines on the coordinate (in pixels) in the focal plane 
of the spectrometer (a) and their deviations $\Delta \lambda_{vac}$ from 
the calibration curve (b); 
$1$ are the experimental values for the $D_2$ molecule from \cite{FSC1985},
$2$ --- for the $Ne$ atom from \cite{SS2004},
$3$ --- for the $H_2$ molecule from \cite{Dieke1972};
solid line represents the approximation of experimental data \cite{LU2011}.}\label{d2h2ne}
\end{center}
\end{figure}

For vacuum wavelength calibration we used bright and free of blending lines of 
the $D_2$ and $H_2$ molecules as well as $Ne$ spectral lines with reference 
data from \cite{FSC1985, Dieke1972, SS2004} respectively. As an example, the 
distribution of vacuum wavelengths of the lines against their positions on the 
CMOS matrix in pixels obtained for one narrow spectral interval is shown in 
figure~\ref{d2h2ne}(a). One may see that the dependence of the 
wavelength for most of the lines on the coordinate $x$ is monotonic and close 
to linear. The calibration curve of the spectrometer was obtained by the 
polynomial least-squares fitting the data. Our measurements showed that, 
using the linear hypothesis is inadequate, third--degree polynomial is 
excessive, while an approximation by the second-degree polynomial provides 
calibration accuracy better than $2 \times 10^{-3}$ nm. Such a way of the 
wavelength calibration allows us to get new experimental values for the 
rovibronic line wavenumbers. The differences $\Delta \lambda_{vac}$ between 
new values and used reference data are shown in figure~\ref{d2h2ne}(b).
One may see that these differences have certain spread around calibration curve, 
that does not exceed $0.002$ nm. Thus, our measurements show that experimental 
wavenumber values from \cite{FSC1985, Dieke1972, SS2004} are in good agreement 
with each other. Therefore in our studies of the $D_2$ spectrum, the wavenumber 
values from \cite{FSC1985} were used as the reference data set.
Such "internal reference source" gave us an opportunity to eliminate 
experimental errors caused by the shift between a spectrum under the study 
and the reference spectrum from another light source, due to a different 
illumination of the grating by different lamps (see e.g.\cite{RLTBjcp2006}).

Each experimental wavenumber value measured in the framework of the procedure 
described above is obtained with the uncertainty (one standard deviation) 
determined by quality of an approximation of a recorded intensity distribution 
and quality and quantity of standard reference data within every small fragment 
of the spectrum ($\approx 0.5$ nm) selected for data processing.

\section{Results and discussion}

The visible part of the emission spectrum of deuterium plasma with minor 
hydrogen impurity was recorded and analyzed by means of the optimizational 
technique. The results thus obtained are presented in the 
Appendix to the present paper in the form of the atlas and accompanying table 
with certain data for the wavelength range $\approx 419 - 696$ nm.
One may see that the spectrum contains three Balmer series lines of atomic 
deuterium $D_\alpha$, $D_\beta$, and $D_\gamma$, corresponding lines of atomic 
hydrogen (impurity) and 11935 molecular rovibronic lines.
The overwhelming majority of molecular lines belong to the $D_2$ molecule but
some $HD$ lines were distinguished and marked in the tables with 
corresponding assignments according to
\cite{DiekeBlue1935, Dieke1935_2, DiekeLewis1937, KB1987} together 
with some new assignments made in the present work. The atlas is divided into 
158 sections. Each section covers about 1.5 nm. It contains pictures of 
images in the focal plane of the spectrometer, and intensity distributions 
in linear and logarithmic scales. Positions of the line centres obtained by 
the deconvolution are presented as "stick diagrams" indicating their 
wavenumbers and the intensity amplitudes. The numbering of the lines 
(for every fifth line) is shown below the linear scale intensity distributions. 
The vacuum wavelength and wavenumber scales are valid for both the images and 
the graphs. The accompanying table contains measured wavenumber and relative 
intensity values for recognized spectral lines together with available and 
new line assignments. Wavenumber values from \cite{FSC1985} are included into 
the table only for spectral lines used as the reference points in the process 
of calibrating our spectrometer.

\begin{landscape}

\begin{figure}[!ht]
\includegraphics[angle=0, totalheight=0.8\textheight]{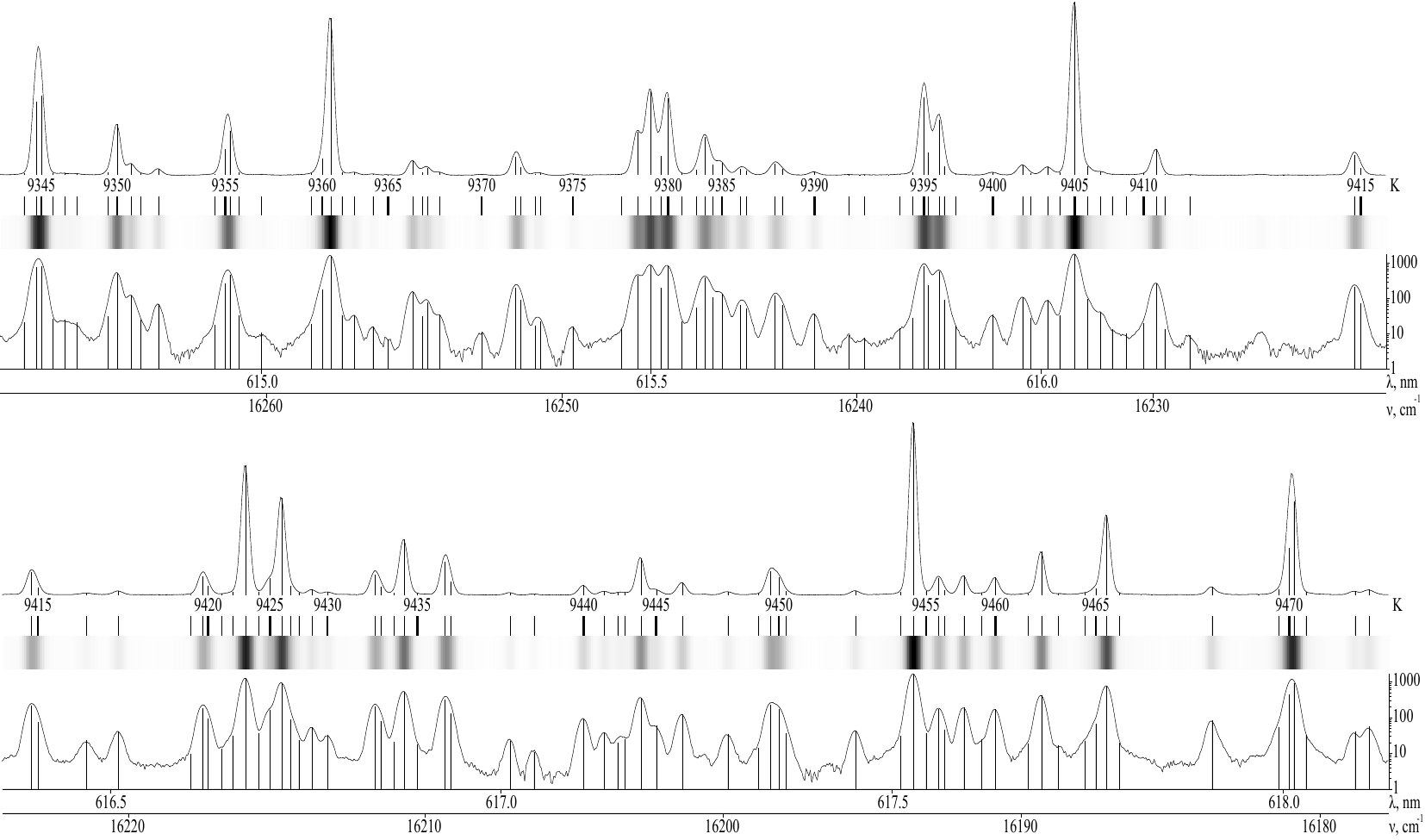}
\caption{
Typical fragment of the $D_2$ spectrum in the range $\approx 614.3 - 618.3$ nm containing 
first 6 lines of the Q-branch for $(2-2)$ band of the 
$d^3\Pi_u^- \to a^3\Sigma_g^+$ electronic transition.
Spectral line centres obtained by the deconvolution are presented as the "stick diagrams" 
indicating their wavenumber positions and relative intensities.}\label{atlex}
\end{figure}
\end{landscape}

As an example, one typical fragment of the spectrum having the length equal 
to one section of the atlas is shown in the figure~\ref{atlex}. This fragment 
was chosen as an example because it contains first 6 spectral lines of the 
(2-2) Q-branch of the Fulcher-$\alpha$ band system 
(the $d(3p)^3\Pi_u^- - a(2s)^3\Sigma_g^+$ electronic transition) of the $D_2$ 
molecule. The Q-branches of this band system are most often used for 
spectroscopic determination of gas temperature in non-equilibrium plasmas 
containing deuterium \cite{LKOR1997, LMKR1999, RDKL2001, NTPCP2013, AKKK1996}.

One may see from the figure~\ref{atlex} that "the digital resolution" obtained by 
the optimization solution of the inverse problem is much higher than the moderate 
"optical resolution" of the recorded spectrum, which is determined by Doppler 
broadening and the instrumental function of our spectrometer. That digital 
deconvolution of recorded intensity distributions made it possible to obtain 
wavenumber values of close-lying and considerably blended lines. It should be emphasized 
that quite often this technique allows distinguishing lines separated closer 
than the linewidth (one may find about dozen of such cases in the 
figure~\ref{atlex}).

Obtained wavenumber and intensity values for all lines in the presented 
spectral region are shown in the table~\ref{wavenumberdata}. Column 1 of the table 
(K) gives spectral line index which was used for numbering of recognized lines 
throughout the atlas. Column 2 ($\nu$) gives measured spectral line wavenumber 
values with the uncertainty (one standard deviation) shown in brackets in units 
of the last significant digit. Some spectral lines included into the table may 
appear as false solutions of the inverse problem or may belong to $HD$ molecule (see above).
Their wavenumber values are marked by asterisk in the table. Such spectral lines 
are beyond the scope of the present paper. They need additional studies in high 
resolution experiments similar to those reported in \cite{LUZ2012, LU2011}.
Wavenumber values for spectral lines assigned as pseudo-doublet components of 
partly resolved fine structure of triplet-to-triplet lines (see \cite{LUZ2012}) 
are marked by letters "s" and "w" for strong and weak components respectively. 
Column 3 (I) gives measured relative line intensities. The uncertainty (one standard 
deviation) is shown in brackets in units of the last 
significant digit. Column 4 ($\nu_R$) gives wavenumber values from the appendix C of \cite{FSC1985} 
close to our values. Wavenumber values used as standard reference 
data for the reverse wavenumber calibration of our spectrometer are presented in the bold face.
Column 5 (Assignment) gives assignments for the $D_2$ lines from \cite{FSC1985} and for the $HD$ 
lines from \cite{DiekeBlue1935, Dieke1935_2, DiekeLewis1937, KB1987}. 
Assignments for the triplet-to-triplet lines confirmed by the
statistical analysis analogous to that reported in \cite{LU2008} are shown in the bold face, 
while new assignments made in the present work are italicized.

It should be stressed that our technique makes it possible to find out the "personal"
error estimate for every wavenumber value (see the table~\ref{wavenumberdata}).
This is most sharp distinction of the data reported in the present work from those
earlier obtained by traditional photographic technique. In original papers
reported the data for visible part of the $D_2$ spectrum
experimental errors for wavenumber values are or rough estimates or not mentioned at all.
In the compilation \cite{FSC1985} the uncertainty of experimental wavenumber values was 
reported as single value $\sigma = 0.05$ cm$^{-1}$ common for all spectral lines under 
the study (from UV up to IR). That is obvious underestimation of real error bars because 
about 28000 lines are distributed in the wavelength range $\approx 300 - 2800$ nm 
and were studied by spectrometers having different resolving power. 

\begin{figure}[!ht]
\begin{center}
\includegraphics[width=0.4\textwidth]{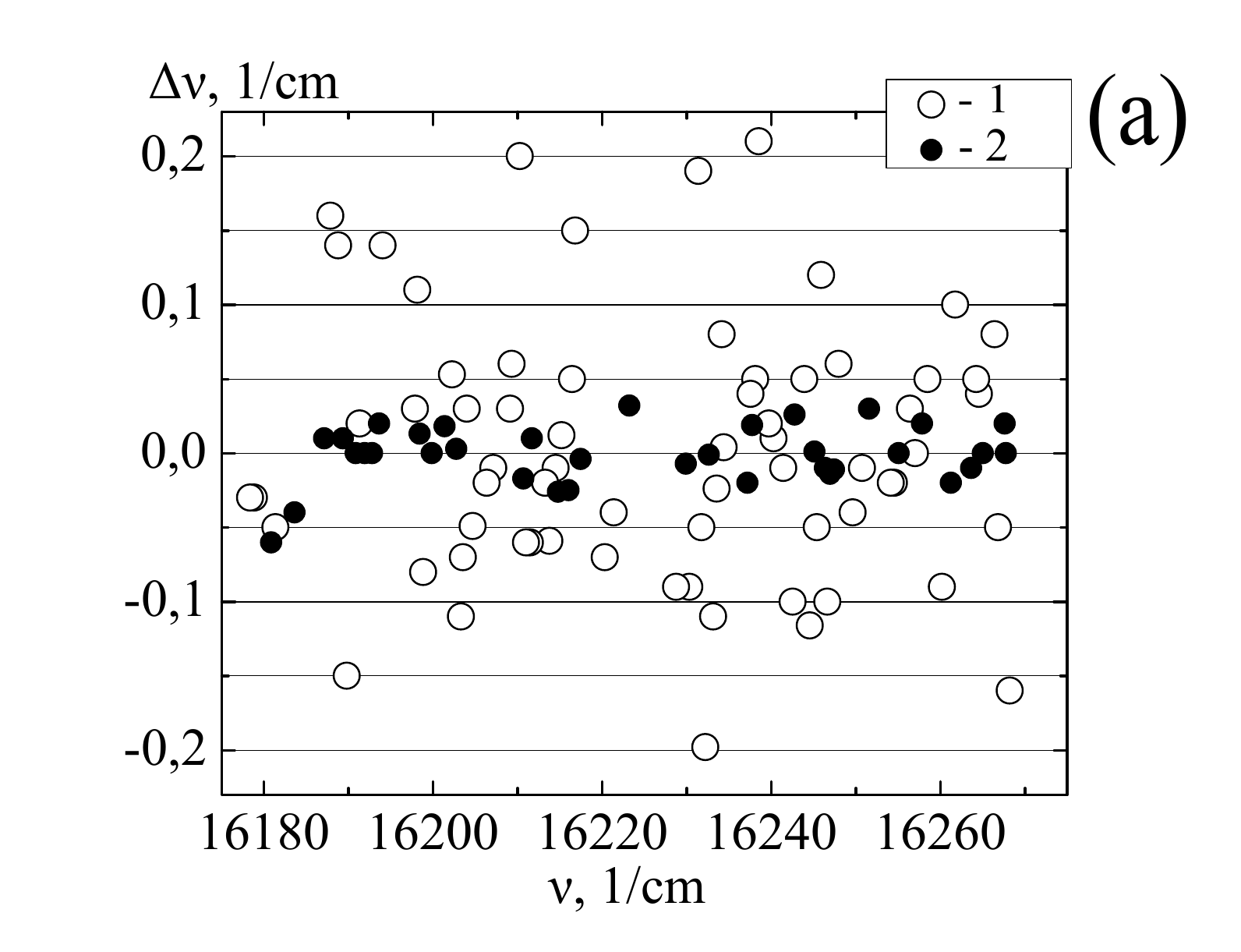}
\includegraphics[width=0.4\textwidth]{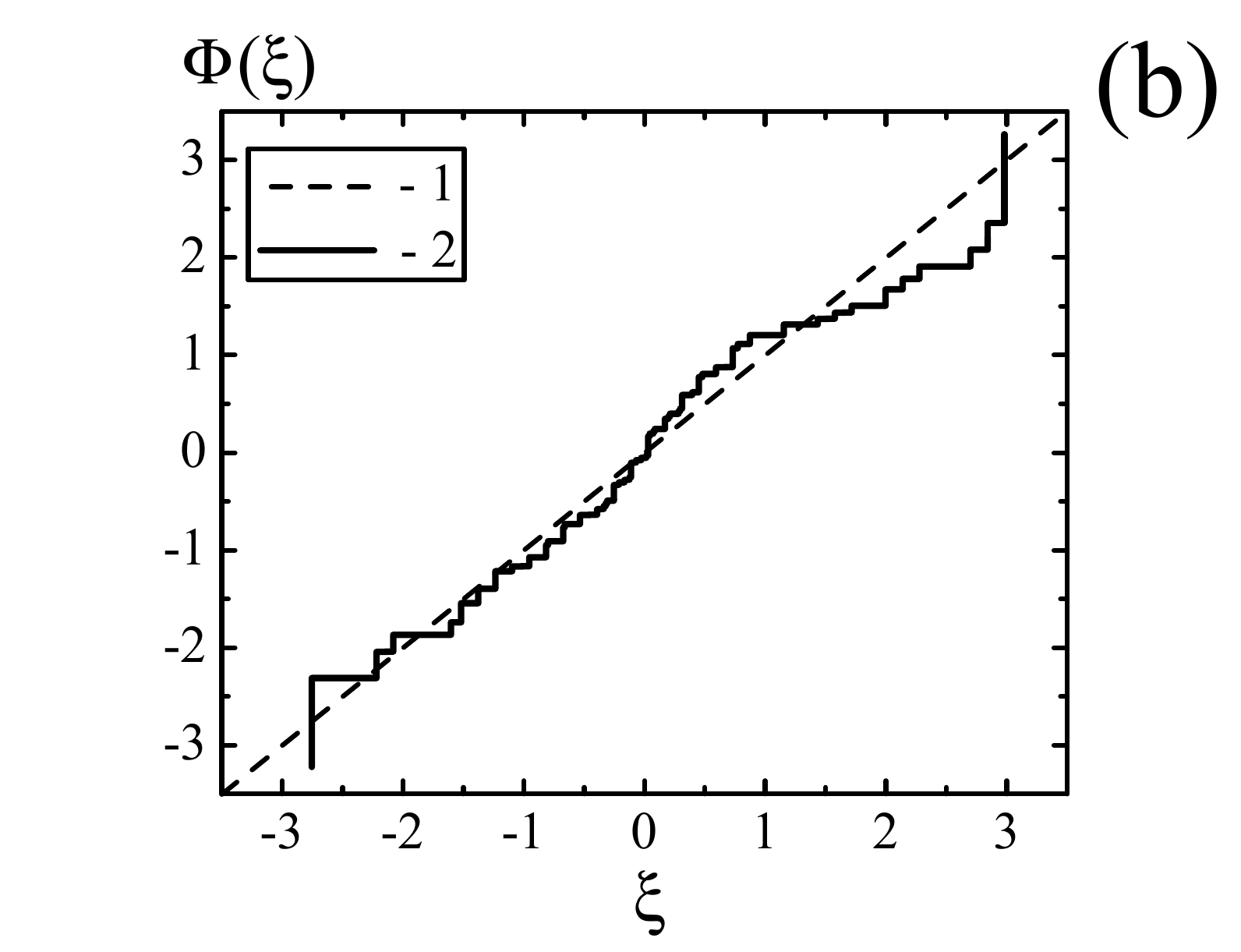}
\caption{(a) -- The differences between wavenumber values obtained in the present work 
and those from \cite{FSC1985} for the wavelength region presented in figure~\ref{atlex}. 
The points 2 correspond 
to the lines used as reference data, and points 1 to all other lines.
(b) -- The cumulative distribution of the differences in the form of the
function $\Phi(\xi)$ from \cite{LU2008}, where the dot line 1 represents Gaussian 
and the bold line 2 --- empirical distributions.}\label{diff}
\end{center}
\end{figure}

The differences $\Delta \nu$ between wavenumber values obtained in the present 
work and those from \cite{FSC1985} for the fragment shown in figure~\ref{atlex} 
are presented in figure~\ref{diff}(a). One may see that the deviations of lines 
used in the present work as a reference wavenumber values are smaller than 
$\pm 0.05$ cm$^{-1}$. But for all other lines the deviations from our calibration 
curve (representing the dispersion curve of our spectrometer, which is obviously 
a monotonic function of the coordinate) may reach $0.2$ cm$^{-1}$.
The cumulative distribution function of the differences $\Delta \nu$ is shown in 
figure~\ref{diff}(b) together with the straight line representing Gaussian (so called 
normal) distribution (see e.g. \cite{LRJetf2005}). One may see that the empirical distribution of the deviations 
is close to normal. Therefore, the deviations are random and obviously represent 
reading errors of the data collected in \cite{FSC1985}. Thus, our data are more 
precise due to the deconvolution and the smoothing procedure in determination of 
the spectrometer dispersion curve.

All measured in the present work wavenumber values for the assigned triplet spectral 
lines together with all available wavenumber values from 
\cite{DiekeBlue1935, Dieke1935_2, DiekePorto, GloersenDieke, FMZ1976, FSC1985, DabrHerz, Davies}
were used as input data for obtaining the set of optimal rovibronic energy 
levels by means of the method of statistical analysis proposed in \cite{LRJetf2005}. 
Detailed description of the analysis will be reported elsewhere. It was
similar to our previous work \cite{LU2008}, but the observation of pseudo 
doublets for lines coming to vibration-rotational levels of the $c^3\Pi_u$ 
electronic state \cite{LUZ2012} forced us to carry out the optimization in 
two stages. At the first stage, the 2718 wavenumber values of rovibronic lines 
belonging to band systems having the $a^3\Sigma_g^+$ as a common low electronic state  
were analyzed for all known $n^3\Lambda_g - a^3\Sigma_g^+$ electronic transitions 
with the quantum numbers of electronic angular momentum projection about the internuclear 
axis $\Lambda = 0, 1$ and the principal quantum numbers (in the united atom model) 
$n = 3 - 9$. Then optimal values of the rovibronic 
energy levels obtained on the first stage were fixed and included into new input 
data set together with the 1263 wavenumber values of all other spectral lines. 
Finally the optimal set of the wavenumber values for all triplet rovibronic 
levels of the $D_2$ molecule were obtained by the optimization procedure.
Such a two-stage procedure gave us an opportunity to obtain with high 
precision 595 energy values for vibro-rotational levels of the 
$a^3\Sigma_g^+$ and $n^3\Lambda_u$ (with $\Lambda = 0, 1$ and $n = 3 - 9$)
electronic states having negligibly small values of triplet splitting. Other 450 values of 
energy for vibro-rotational levels   the $c^3\Pi_u$ and $n^3\Lambda_g$ 
(with $\Lambda = 0, 1, 2$ and $n = 3 - 9$) electronic states are less precise 
because in the input data we used wavenumber values for the strong components 
of the pseudo doublets. Those pseudo doublets represent partly resolved triplet 
structure of the lines mainly caused by triplet splitting of the $c^3\Pi_u, v, N$ 
rovibronic levels \cite{LUZ2012}.

Our statistical analysis shows good agreement (in the framework of the maximum 
likelihood and Rydberg-Ritz principles) between 1045 values of obtained rovibronic 
energy levels and 3981 experimental wavenumber 
values of spectral lines, which are spread over very wide wavenumber range 
0.896--28166.84 cm$^{-1}$ (from radio frequencies up to the near ultraviolet), 
obtained for various band systems, by various methods and authors, and in various 
works.

Originally, we intended and designed the atlas and the wavenumber tables mainly 
for experimentalists working in the great variety of spectroscopic applications 
with ionized gases and plasmas containing deuterium. We hope that they appreciate 
an appearance of the source of information providing an opportunity of the 
timesaving and comfortable work with the $D_2$ spectra.

However, the wavenumber values obtained in the present work could be interesting 
and useful for theoreticians as well. From the data reported in the present 
paper one may see that most of the $D_2$ spectral lines are not assigned so far. 
Therefore, a nature of these unassigned lines is still uncertain. We consider this situation 
as abnormal for an isotopologue of simplest neutral molecule. We hope that our results may 
stimulate interest of theoreticians dealing with high accuracy non-adiabatic 
\textit{ab initio} calculations.

\section{Acknowledgements}

Present work was supported, in part, by 
the Russian Foundation for Basic Research, Grant No. 13-03-00786-a.

{\def\baselinestretch{1.0}

\begin{landscape}
\begin{table}
\footnotesize		
\setlength{\tabcolsep}{3pt}
\caption{Experimental wavenumber, intensity values and assignments for spectral lines of 
$D_2$ and $HD$ molecules (see text).}\label{wavenumberdata}
\begin{tabular}{@{}llcclllccl}  
\hline                              

K & $\nu$, cm$^{-1}$ & $I$, counts & $\nu_R$, cm$^{-1}$ & Assignment &
K & $\nu$, cm$^{-1}$ & $I$, counts & $\nu_R$, cm$^{-1}$ & Assignment \\
\hline

9343	& 16268.21(3)*	& 19.0(10) 
& 16268.37     	&
& 
9373	& 16250.94(5)*	& 15(3)
&      	&
\\

9344	& 16267.78(2)	& 780(9)
& \textbf{16267.78}	& \textbf{T+  3c-2a  (1-1) P5   } 
& 
9374	& 16250.75(4)*	& 19(3)
& 16250.76     	&  
\\

9345	& 16267.62(2)	& 837(9)
& \textbf{16267.60}	& \textbf{T-  3c-2a  (1-1) Q10   }
& 
9375	& 16249.64(3)	& 13.4(7)
& 16249.68     	&  
\\

9346	& 16267.22(3)*	& 24.2(10)
&      	&
& 
9376	& 16247.96(3)*	& 11.8(8)
& 16247.90     	&  
\\

9347	& 16266.82(3)*	& 22.9(9)
& 16266.87     	&  
& 
9377	& 16247.409(16)	& 453.5(9)
& \textbf{16247.42}	& \textbf{T+  3c-2a  (0-0) P9   } 
\\

9348	& 16266.42(3)*	& 18.1(9)
& 16266.34     	& S-  3E-2B  (2-10) Q7    
& 
9378	& 16246.976(16)	& 889(2)
& \textbf{16246.99}	& \textbf{T-  3c-2a  (2-2) Q3   } 
\\

9349	& 16265.37(3)*	& 29.1(12)
&      	& HD T+  3c-2a  (2-2) R2
& 
& & 
& & S-  3F-2C  (3-1) R1    
\\

9350	& 16265.05(2)	& 539.2(13)
& \textbf{16265.05}	& \textbf{T-  3c-2a  (2-2) Q1   } 
&
9379	& 16246.63(2)*	& 200(4)
& 16246.73     	&  
\\

9351	& 16264.58(3)	& 122.0(11)
& 16264.54     	&  
& 
9380	& 16246.380(17)	& 815(4)
& \textbf{16246.39}	& \textbf{T+  3b-2a  (5-1) P6   } 
\\

9352	& 16264.25(3)*	& 22.2(11)
& 16264.20     	& HD T+  3d-2c  (1-1) Q6
& 
9381	& 16245.91(3)*	& 20.4(9)
& 16245.79     	&
\\

9353	& 16263.66(3)	& 68.6(9)
& \textbf{16263.67}	& \textbf{T+  3b-2a  (8-3) P2   } 
& 
9382	& 16245.41(3)*	& 53(3)
& 16245.46     	&  
\\

9354	& 16261.76(4)*	& 15.1(13)
& 16261.66     	& S+  GK-2B  (2-7) P2    
& 
9383	& 16245.131(19)	& 400(3)
& \textbf{16245.13}	& \textbf{T+  3d-2c  (3-3) Q2   } 
\\

9355	& 16261.39(3)	& 271(17)
&      	&
& 
9384	& 16244.87(3)*	& 108(4)
&      	&
\\

9356	& 16261.24(3)	& 469(16)
& \textbf{16261.26}	& \textbf{T+  3d-2c  (3-3) R4   } 
& 
9385	& 16244.554(19)	& 134(2)
& 16244.67     	&
\\

9357	& 16260.93(3)*	& 30(2)
&      	&
& 
9386	& 16243.94(2)*	& 62(4)
& 16243.89     	&  
\\

9358	& 16260.18(4)*	& 8.2(9)
& 16260.27     	& \textbf{T+  3d-2c  (1-1) P7   } 
& 
9387	& 16243.73(3)*	& 49(4)
&      	&
\\

9359	& 16258.49(3)*	& 16.6(7)
& 16258.44     	& \textbf{T+  3b-2a  (8-3) R5   } 
& 
9388	& 16242.776(19)	& 120(2)
& \textbf{16242.75}	& \textbf{T-  3c-2a  (0-0) Q15  } 
\\

9360	& 16258.12(3)	& 176(2)
&      	&
& 
9389	& 16242.52(2)	& 64(2)
& 16242.62     	&
\\

9361	& 16257.83(2)	& 1658(3)
& \textbf{16257.81}	& \textbf{T-  3c-2a  (2-2) Q2   } 
&
9390	& 16241.43(2)	& 35.8(8)
& 16241.44     	& \textbf{T+  3d-2c  (2-2) P5   } 
\\

9362	& 16257.45(3)*	& 31.4(9)
&      	&
& 
9391	& 16240.27(4)	& 7.0(8)
& 16240.26     	&  
\\

9363	& 16257.02(3)	& 31.7(6)
& 16257.02     	& \textbf{T+  3b-2a (10-4) P2   } 
& 
9392	& 16239.73(5)	& 5.6(8)
& 16239.71     	&  
\\

9364	& 16256.41(3)	& 13.0(5)
& 16256.38     	&  
& 
9393	& 16238.55(3)*	& 12.0(8)
& 16238.34     	&
\\

9365	& 16255.89(5)*	& 4.2(5)
&      	&
& 
9394	& 16238.12(3)*	& 26.6(13)
& 16238.07     	&  
\\

9366	& 16255.06(2)	& 152.2(11)
& \textbf{16255.06}	& \textbf{T+  3d-2c  (3-3) P1   } 
& 
9395	& 16237.729(19)	& 820(26)
& \textbf{16237.71}	& \textbf{T+  3d-2c  (3-3) R5   } 
\\

9367	& 16254.74(5)*	& 30(4)
&      	& 
& 
& & 
&	& \textbf{T+  3d-2c  (2-2) R11  } 
\\

9368	& 16254.55(3)	& 76(5)
& 16254.57     	&  
&
9396	& 16237.57(3)*	& 240(24)
& 16237.53     	&  HD T+  3c-2a  (1-1) P3
\\

9369	& 16254.15(3)	& 32.0(6)
& 16254.17     	&  
& 
9397	& 16237.21(2)	& 587(15)
& \textbf{16237.23}	&  
\\

9370	& 16252.73(4)*	& 8.5(5)
&      	&
& 
9398	& 16237.04(4)*	& 90(18)
&      	&
\\

9371	& 16251.59(3)	& 196(5)
& \textbf{16251.56}	& \textbf{T-  3e-2c  (4-4) P5 s  } 
& 
9399	& 16236.65(3)*	& 14.9(12)
&      	&
\\

9372	& 16251.43(3)	& 88(5)
&      	& \textbf{\textit{T-  3e-2c  (4-4) P5 w}}
&
9400	& 16235.40(2)	& 33.0(8)
&      	& S+  EF-2B (17-1) P5    
\\
\hline                              
\end{tabular}
\end{table}

\begin{table}
\footnotesize		
\setlength{\tabcolsep}{3pt}
\begin{tabular}{@{}llcclllccl}  
\multicolumn{10}{l}{\bf Table \ref{wavenumberdata}. \rm (Continued.)} \\
\hline                              

K & $\nu$, cm$^{-1}$ & $I$, counts & $\nu_R$, cm$^{-1}$ & Assignment &
K & $\nu$, cm$^{-1}$ & $I$, counts & $\nu_R$, cm$^{-1}$ & Assignment \\
\hline

9401	& 16234.404(19)	& 105.0(16)
& 16234.40     	&  
& 
9431	& 16211.680(19)	& 195(4)
& \textbf{16211.67}	& \textbf{T+  3d-2c  (3-3) Q3 s   } 
\\

9402	& 16234.13(3)*	& 25.5(16)
& 16234.05     	&  
& 
9432	& 16211.48(2)	& 77(4)
& 16211.54     	& \textbf{\textit{T+  3d-2c  (3-3) Q3 w}}
\\

9403	& 16233.556(18)	& 89.4(8)
& \textbf{16233.58}    	&  
& 
9433	& 16211.03(3)*	& 16.3(12)
& 16211.09      & \textbf{\textit{T+ 3d-2c (3-3) R6}}      	
\\

9404	& 16233.15(2)*	& 30.7(8)
& 16233.26     	&
& 
9434	& 16210.693(17)	& 532.3(15)
& \textbf{16210.71}	& T+  3d-2c  (3-3) R6 (?)   
\\

9405	& 16232.639(16)	& 1829.0(13)
& \textbf{16232.64}	& \textbf{T-  3c-2a  (2-2) Q4   } 
& 
9435	& 16210.25(3)*	& 13.0(8)
& 16210.05     	&
\\

9406	& 16232.202(19)*	& 93.4(9)
& 16232.40     	&
& 
9436	& 16209.320(18)	& 318(5)
& 16209.26     	&
\\

9407	& 16231.77(2)*	& 40.3(9)
& 16231.82     	&  
& 
9437	& 16209.14(2)	& 124(5)
& 16209.11     	& \textbf{T+  3b-2a  (8-3) P3   } 
\\

9408	& 16231.38(4)*	& 11.3(9)
& 16231.19     	&
& 
9438	& 16207.14(2)	& 17.8(7)
& 16207.15     	&  
\\

9409	& 16230.90(4)*	& 8.0(8)
&      	&
& 
9439	& 16206.34(4)	& 5.7(7)
& 16206.36     	& S+  EF-2B (19-2) P7    
\\

9410	& 16230.32(3)*	& 18.0(8)
& 16230.41     	&
& 
9440	& 16204.671(18)	& 90.0(7)
& 16204.72     	&  
\\

9411	& 16229.893(17)	& 273.4(15)
& \textbf{16229.90}	& \textbf{T-  3c-2a  (1-1) Q11  } 
& 
9441	& 16203.98(2)	& 33.8(7)
& 16203.95     	&  
\\

9412	& 16229.61(4)*	& 11.4(16)
&      	&
& 
9442	& 16203.54(4)*	& 15(2)
& 16203.61     	&  
\\

9413	& 16228.77(4)	& 7.0(8)
& 16228.86     	&  
& 
9443	& 16203.29(3)*	& 20(2)
& 16203.40     	&
\\

9414	& 16223.242(18)	& 216(4)
& \textbf{16223.21}	& \textbf{T+  3d-2c  (2-2) Q7 s   } 
& 
9444	& 16202.753(17)	& 353.7(8)
& \textbf{16202.75}	& \textbf{T+  3c-2a  (2-2) P2   } 
\\

9415	& 16223.02(2)	& 72(4)
&      	& \textbf{\textit{T+  3d-2c  (2-2) Q7 w}}
& 
9445	& 16202.243(19)*	& 54.1(7)
& 16202.19     	&  HD T+  3d-2c  (2-2) Q4
\\

9416	& 16221.39(3)*	& 19.2(8)
& 16221.43     	& HD T-  3d-2c  (3-3) P5
& 
9446	& 16201.368(17)	& 117.6(7)
& \textbf{16201.35}	&  
\\

9417	& 16220.33(2)	& 37.7(8)
& 16220.40     	&
& 
9447	& 16199.85(2)*	& 30.3(7)
& \textbf{16199.85}    	&  HD T+  3c-2a  (2-2) R0
\\

9418	& 16217.86(5)*	& 4.8(7)
&      	&
& 
9448	& 16198.83(4)*	& 9.1(8)
& 16198.91     	&  
\\

9419	& 16217.476(19)	& 179(3)
& \textbf{16217.48}	& 
& 
9449	& 16198.413(18)	& 225.5(19)
& \textbf{16198.40}	& \textbf{T+  3d-2c  (2-2) R12  } 
\\

9420	& 16217.28(2)	& 89(3)
&      	& 
& 
9450	& 16198.14(2)	& 168.7(17)
& 16198.03     	&
\\

9421	& 16216.83(3)*	& 8.6(6)
& 16216.68     	&
& 
9451	& 16197.89(3)*	& 32(2)
& 16197.86     	&  
\\

9422	& 16216.46(2)*	& 26.5(6)
& 16216.41     	&  
& 
9452	& 16195.56(3)	& 38.4(8)
&      	&
\\

9423	& 16216.015(16)	& 1246.8(9)
& \textbf{16216.04}	& \textbf{T+  3c-2a  (1-1) P6   } 
& 
9453	& 16194.06(4)*	& 24.7(9)
& 16193.92     	&
\\

9424	& 16215.58(2)*	& 30.6(6)
&      	&
& 
9454	& 16193.61(3)	& 1655.0(14)
& \textbf{16193.59}	& \textbf{T-  3c-2a  (2-2) Q6   } 
\\

9425	& 16215.202(17)	& 156.0(6)
& 16215.19     	&  
& 
9455	& 16193.20(4)*	& 31.8(10)
&      	&
\\

9426	& 16214.804(18)	& 931(8)
& \textbf{16214.83}	& \textbf{T-  3c-2a  (2-2) Q5   } 
& 
9456	& 16192.78(4)	& 163(4)
& \textbf{16192.78}	&  
\\

9427	& 16214.50(2)*	& 81.5(11)
& 16214.51     	& \textbf{T+  3d-2c  (3-3) P2   } 
& 
9457	& 16192.58(4)	& 41(4)
&      	&
\\

9428	& 16214.23(3)*	& 18.8(14)
&      	&
& 
9458	& 16191.93(3)	& 185.7(8)
& \textbf{16191.93}	&  
\\

9429	& 16213.781(19)	& 49.7(6)
& 16213.84     	&
& 
9459	& 16191.33(4)*	& 18.8(8)
& 16191.31     	&  
\\

9430	& 16213.26(2)	& 28.3(7)
& 16213.28     	&  
&
9460	& 16190.87(3)	& 167.4(8)
& \textbf{16190.87}	& \textbf{T-  3c-2a  (0-0) Q16  } 
\\
\hline                              
\end{tabular}
\end{table}

\begin{table}
1\footnotesize		
\setlength{\tabcolsep}{3pt}
\begin{tabular}{@{}llcclllccl}  
\multicolumn{10}{l}{\bf Table \ref{wavenumberdata}. \rm (Continued.)} \\
\hline                              

K & $\nu$, cm$^{-1}$ & $I$, counts & $\nu_R$, cm$^{-1}$ & Assignment &
K & $\nu$, cm$^{-1}$ & $I$, counts & $\nu_R$, cm$^{-1}$ & Assignment \\
\hline

9461	& 16189.78(4)*	& 13.4(8)
& 16189.93     	&
& 
9468	& 16183.61(3)	& 77.6(12)
& \textbf{16183.65}	&  
\\

9462	& 16189.33(3)	& 413.6(9)
& \textbf{16189.32}	& \textbf{ \textit{T- 3c-2a (0-0) Q12}} 
& 
9469	& 16181.38(4)*	& 49.6(17)
& 16181.43     	&  
\\

9463	& 16188.77(5)*	& 11.8(8)
& 16188.63      &
& 
9470	& 16181.03(4)	& 449(15)
&      	
& \textbf{\textit{T+ 3c-2a (3-3) R5}} 
\\

9464	& 16187.87(4)*	& 17.3(9)
& 16187.71     	&
& 
9471	& 16180.87(3)	& 900(16)
& \textbf{16180.93}	& \textbf{T+  3d-2c  (3-3) R7   }    
\\

9465	& 16187.51(3)*	& 63.5(10)
&      	&
& 
& & &	& T+  3c-2a  (3-3) R5 (?) 
\\

9466	& 16187.14(3)	& 768.3(11)
& \textbf{16187.13}	& \textbf{T+  3c-2a  (0-0) P10  } 
& 
9472	& 
16180.47(4)*	& 27.4(14)
&      	&
\\

& & &	& S-  3F-2C  (3-1) Q2    
& 
9473	& 16178.83(4)*	& 34.6(12)
& 16178.86     	&  
\\

9467	& 16186.71(4)*	& 14.7(9)
&      	&
&
9474	& 16178.35(4)*	& 50.3(12)
& 16178.38     	&  
\\

\hline                              
\end{tabular}
\end{table}
\end{landscape}
}

\thispagestyle{empty}

\newpage

\bibliography{2015LU_JQSRT_f5}

\newpage

\appendix
\section*{Appendix}

Atlas and wavenumber table of the $D_2$ molecule emission spectrum for wavelength range $419 - 696$ nm.

The table contains: first column --- the order numbers $K$ of distinguished spectral lines; 
second and third column --- measured wavenumber $\nu$ and the relative
intensity $I$ values respectively, the standard deviations being shown in brackets 
in units of last significant digit; fourth column --- wavenumber values 
of lines from \cite{FSC1985} used as 
the reference data for the wavenumber calibrations; and the fifth column --- 
the line assignments in the Dieke's notation.
Confirmed by statistical analysis assignments for triplet lines are 
set in bold face and the new assignments proposed in the present work are italicized. 
Lines of the $HD$ molecule are marked by the $HD$ symbol.

\begin{landscape}
{\def\baselinestretch{1.0}
\footnotesize		
\setlength{\tabcolsep}{1pt}

\begin{longtable}[]{r|lr|l|r|r|lr|l|r}

\hline                              
K & $\nu$, cm$^{-1}$ & $I$, counts & $\nu_R$, cm$^{-1}$ & Assignment &
K & $\nu$, cm$^{-1}$ & $I$, counts & $\nu_R$, cm$^{-1}$ & Assignment \\
\hline
\endfirsthead

\hline
K & $\nu$, cm$^{-1}$ & $I$, counts & $\nu_R$, cm$^{-1}$ & Assignment &
K & $\nu$, cm$^{-1}$ & $I$, counts & $\nu_R$, cm$^{-1}$ & Assignment \\
\hline
\endhead

\input{NewLineTab01.tex}

\hline
\end{longtable}
}

\end{landscape}


\begin{figure}[!ht]
\includegraphics[angle=90, totalheight=0.9\textheight]{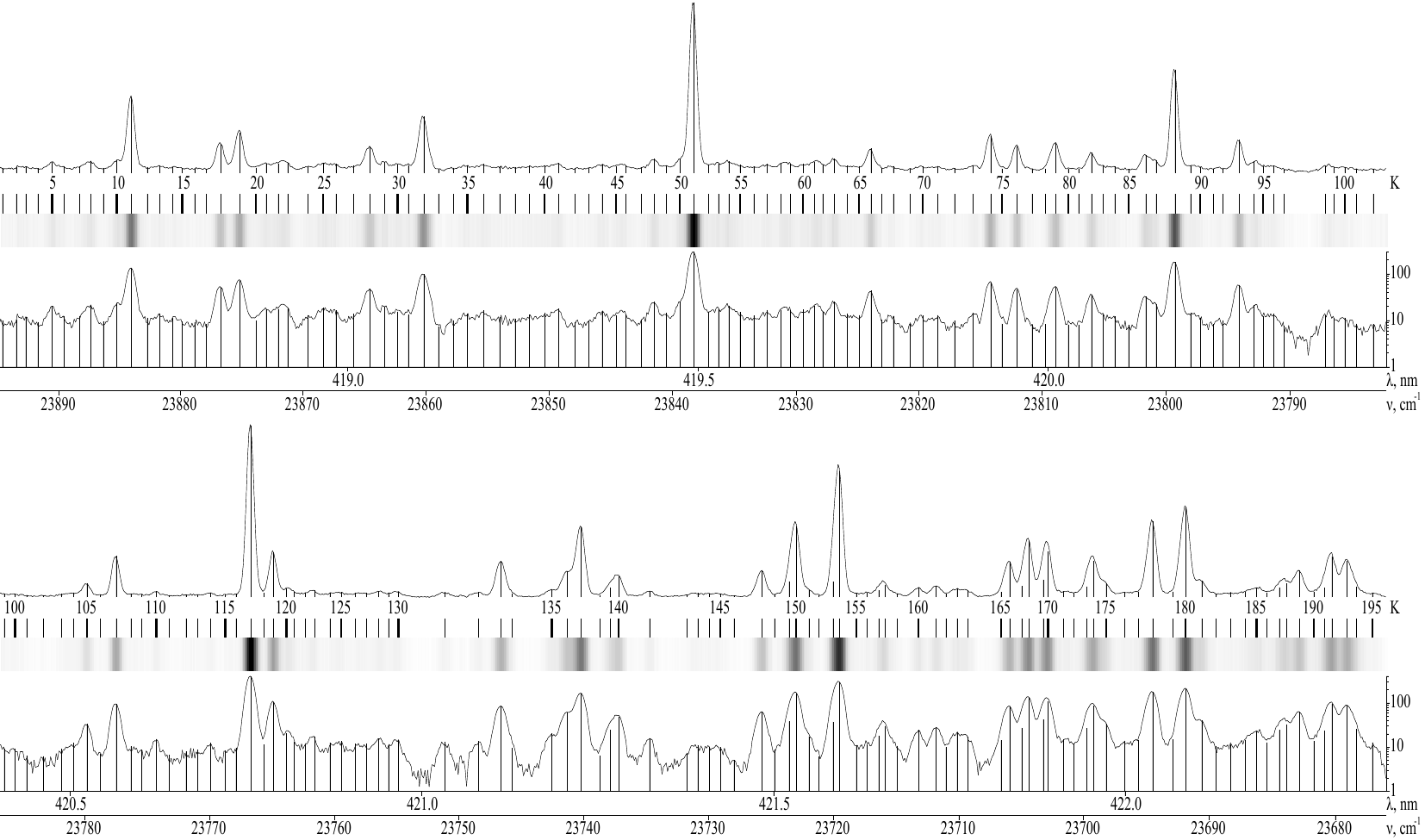}
\end{figure}

\newpage
\begin{figure}[!ht]
\includegraphics[angle=90, totalheight=0.9\textheight]{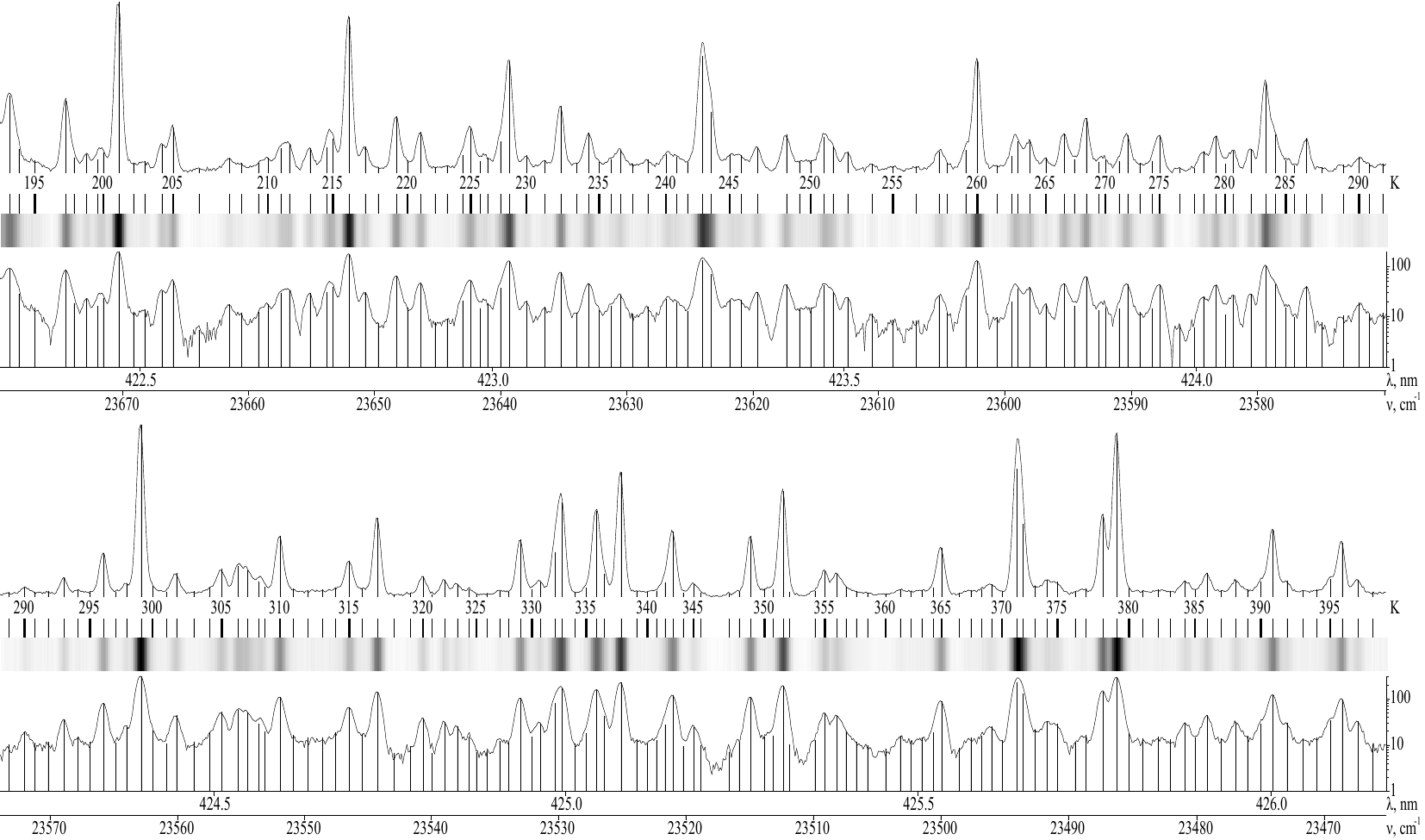}
\end{figure}

\newpage
\begin{figure}[!ht]
\includegraphics[angle=90, totalheight=0.9\textheight]{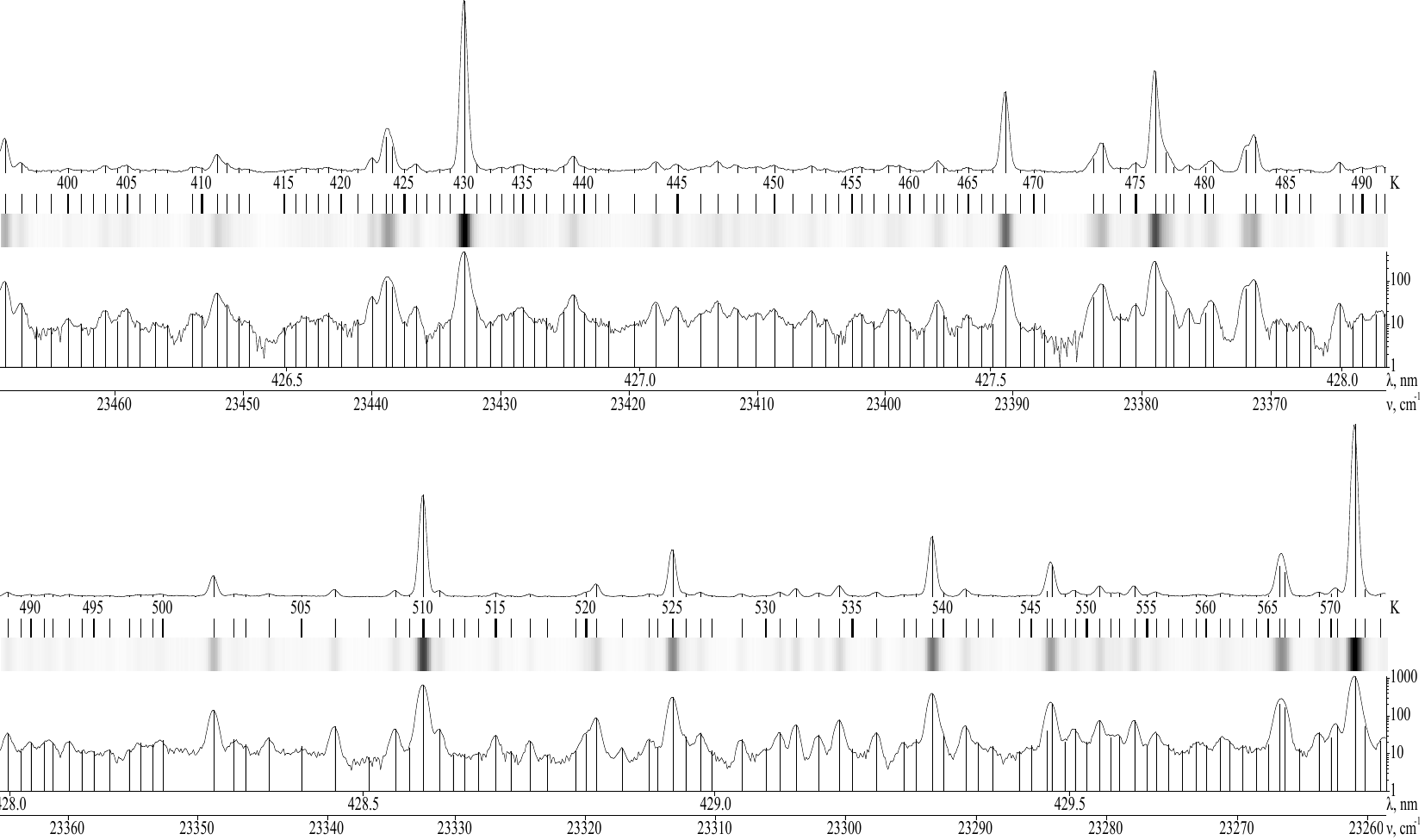}
\end{figure}

\newpage
\clearpage
\begin{figure}[!ht]
\includegraphics[angle=90, totalheight=0.9\textheight]{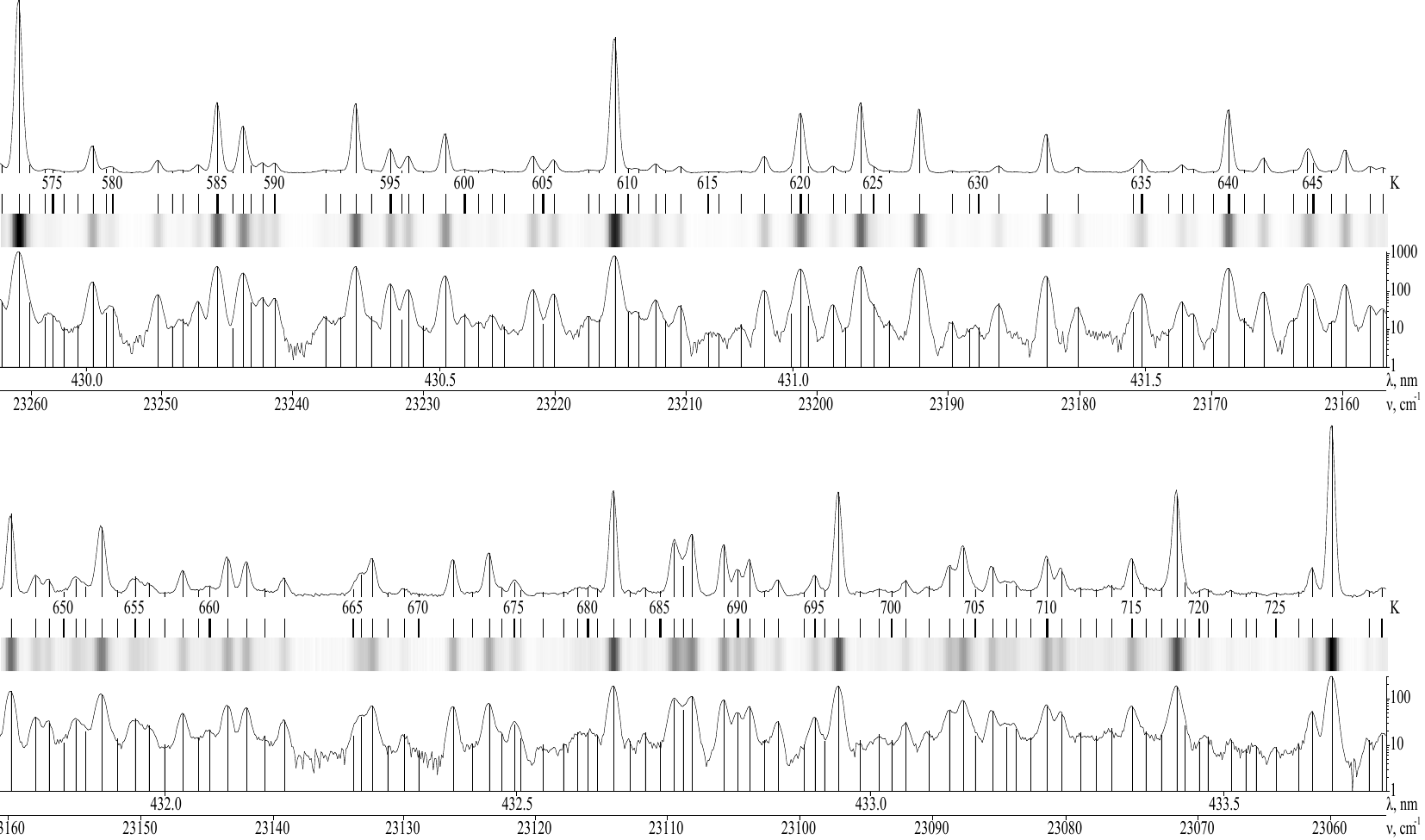}
\end{figure}

\newpage
\begin{figure}[!ht]
\includegraphics[angle=90, totalheight=0.9\textheight]{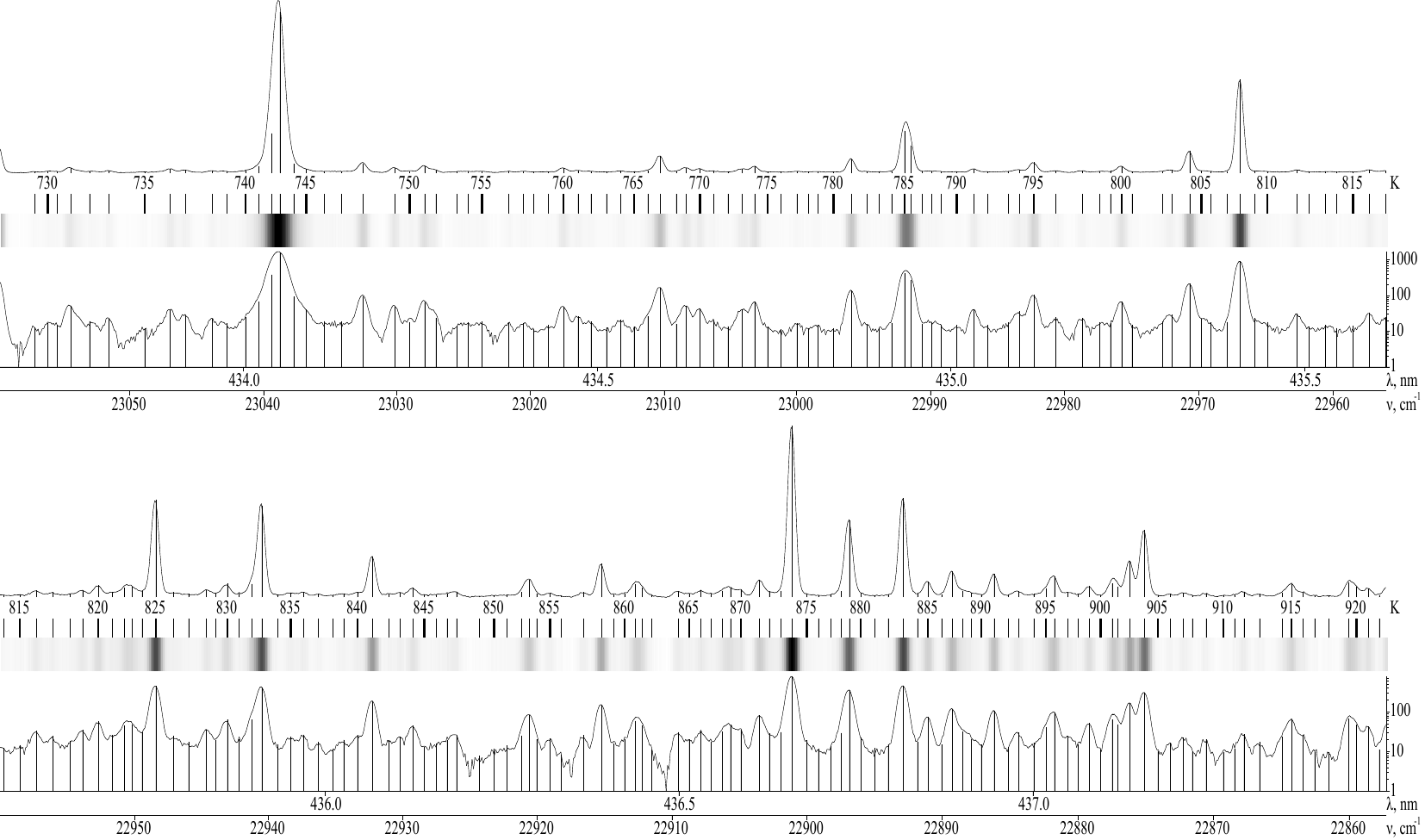}
\end{figure}

\newpage
\begin{figure}[!ht]
\includegraphics[angle=90, totalheight=0.9\textheight]{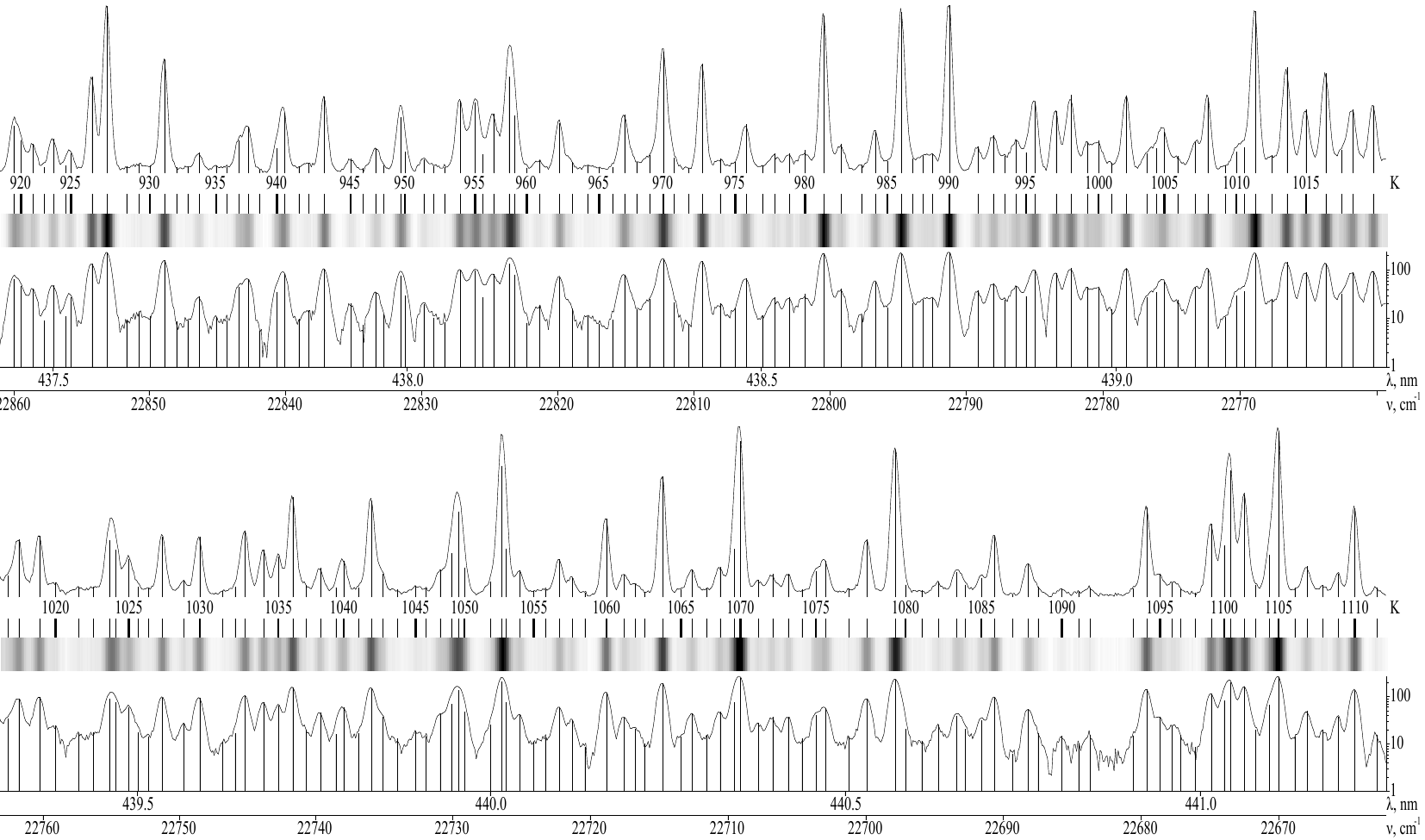}
\end{figure}

\newpage
\begin{figure}[!ht]
\includegraphics[angle=90, totalheight=0.9\textheight]{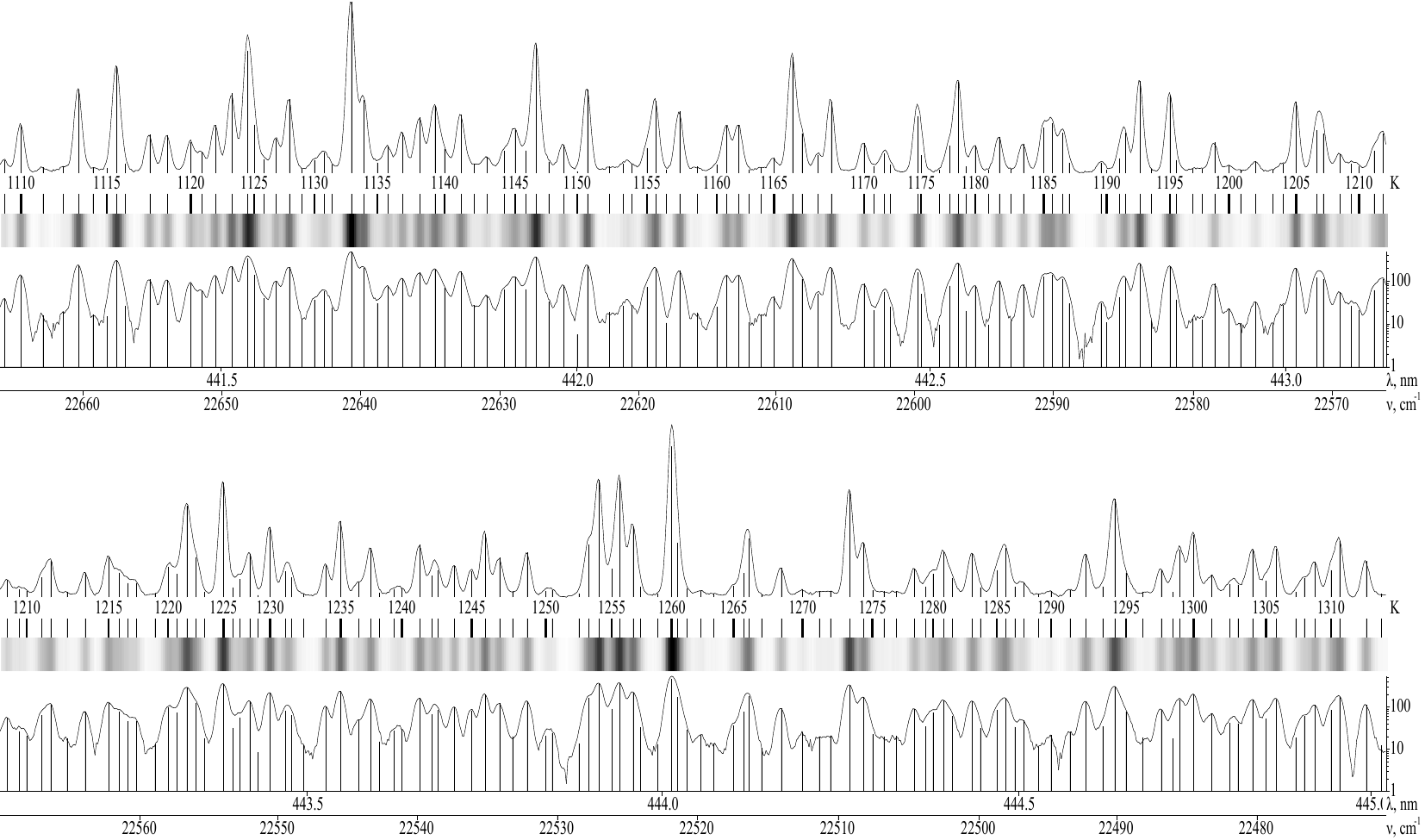}
\end{figure}

\newpage
\begin{figure}[!ht]
\includegraphics[angle=90, totalheight=0.9\textheight]{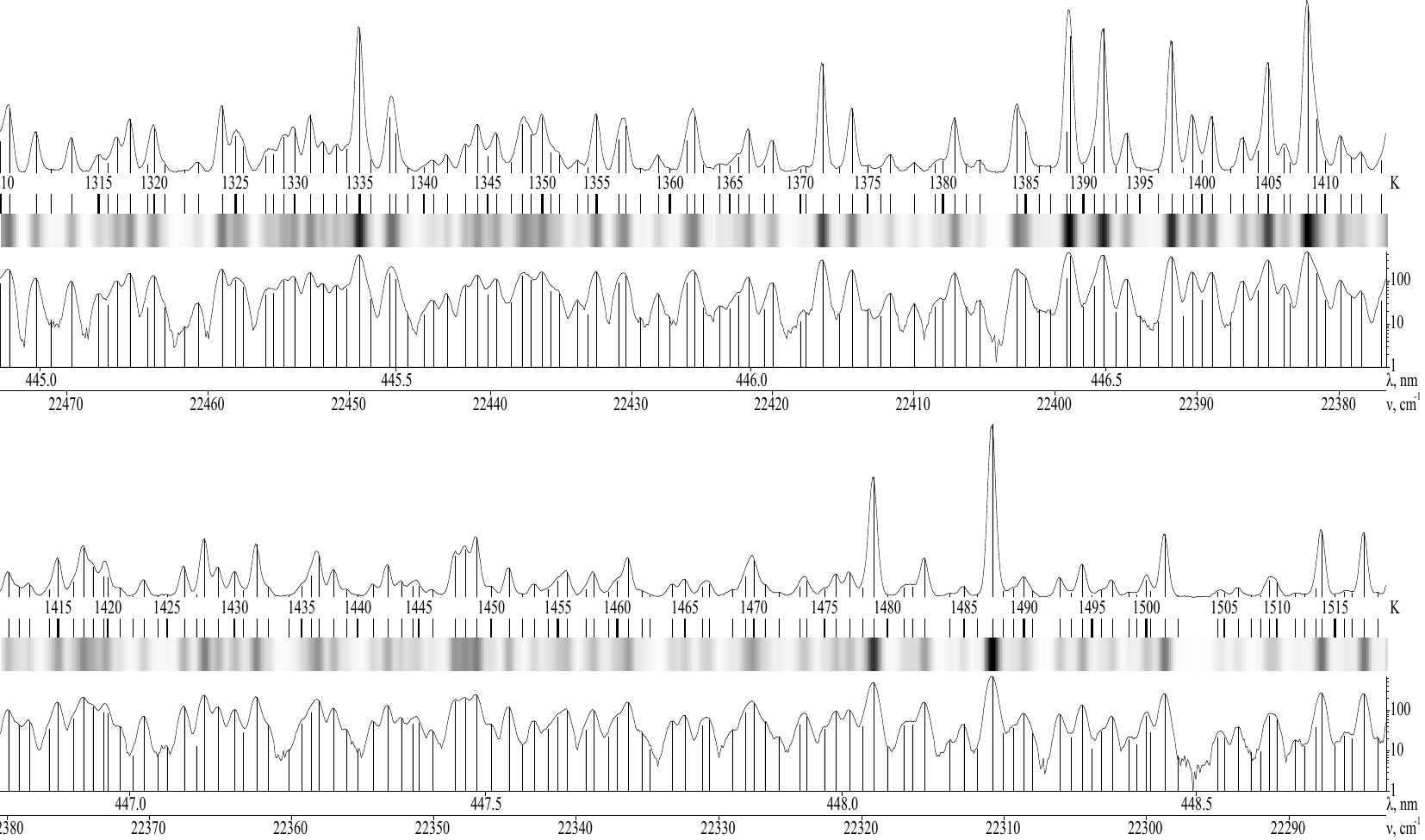}
\end{figure}

\newpage
\begin{figure}[!ht]
\includegraphics[angle=90, totalheight=0.9\textheight]{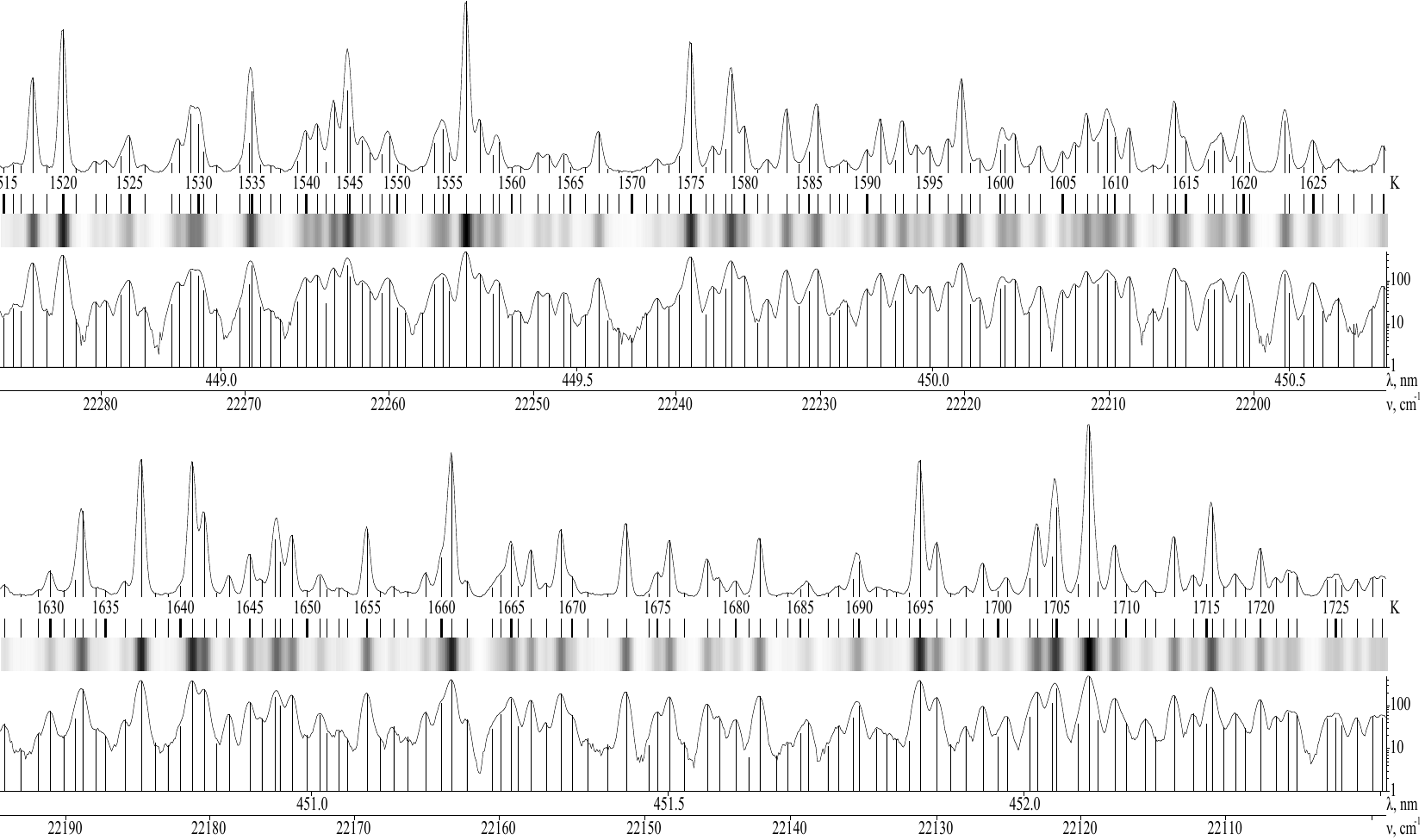}
\end{figure}

\newpage
\begin{figure}[!ht]
\includegraphics[angle=90, totalheight=0.9\textheight]{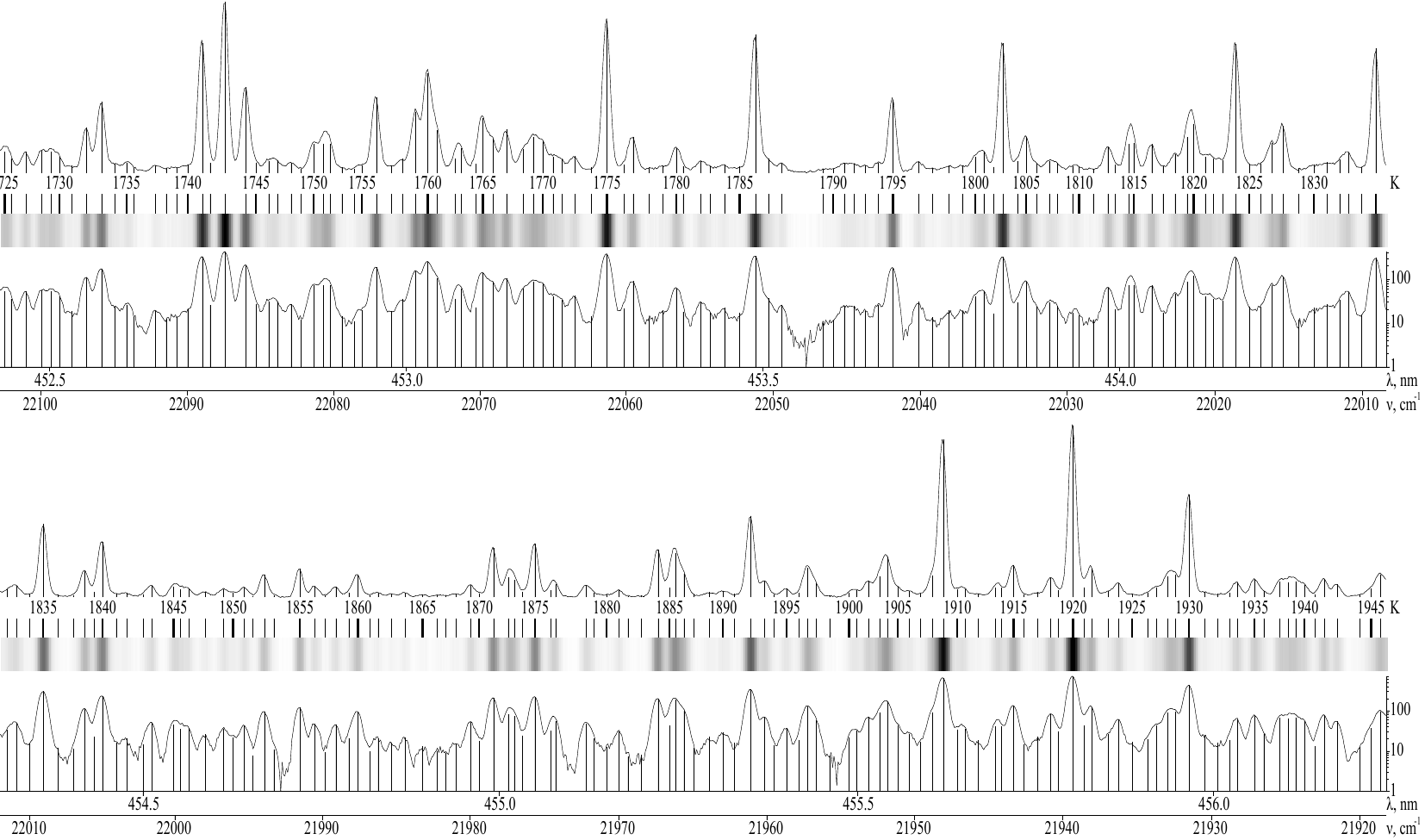}
\end{figure}

\newpage
\begin{figure}[!ht]
\includegraphics[angle=90, totalheight=0.9\textheight]{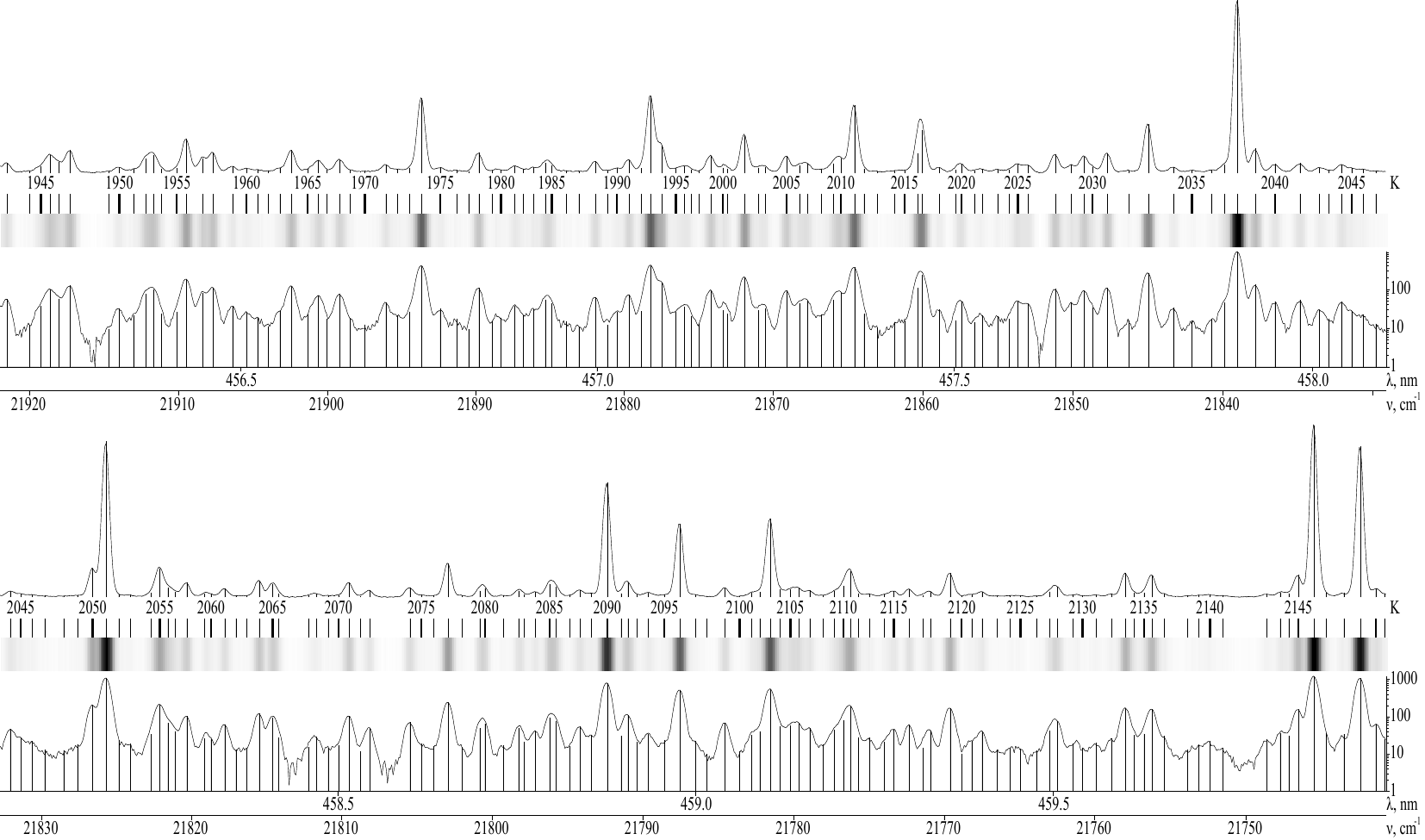}
\end{figure}

\newpage
\begin{figure}[!ht]
\includegraphics[angle=90, totalheight=0.9\textheight]{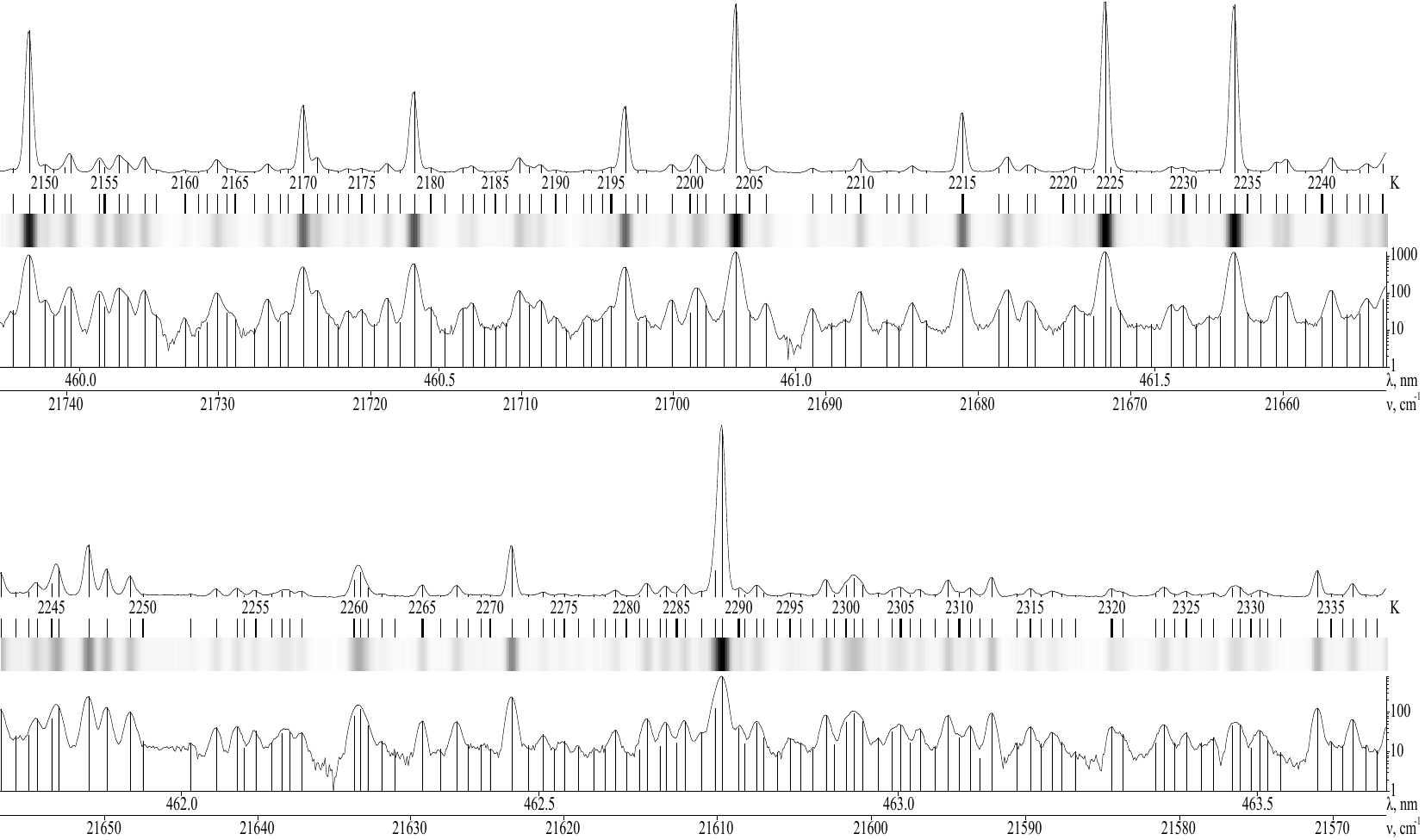}
\end{figure}

\newpage
\begin{figure}[!ht]
\includegraphics[angle=90, totalheight=0.9\textheight]{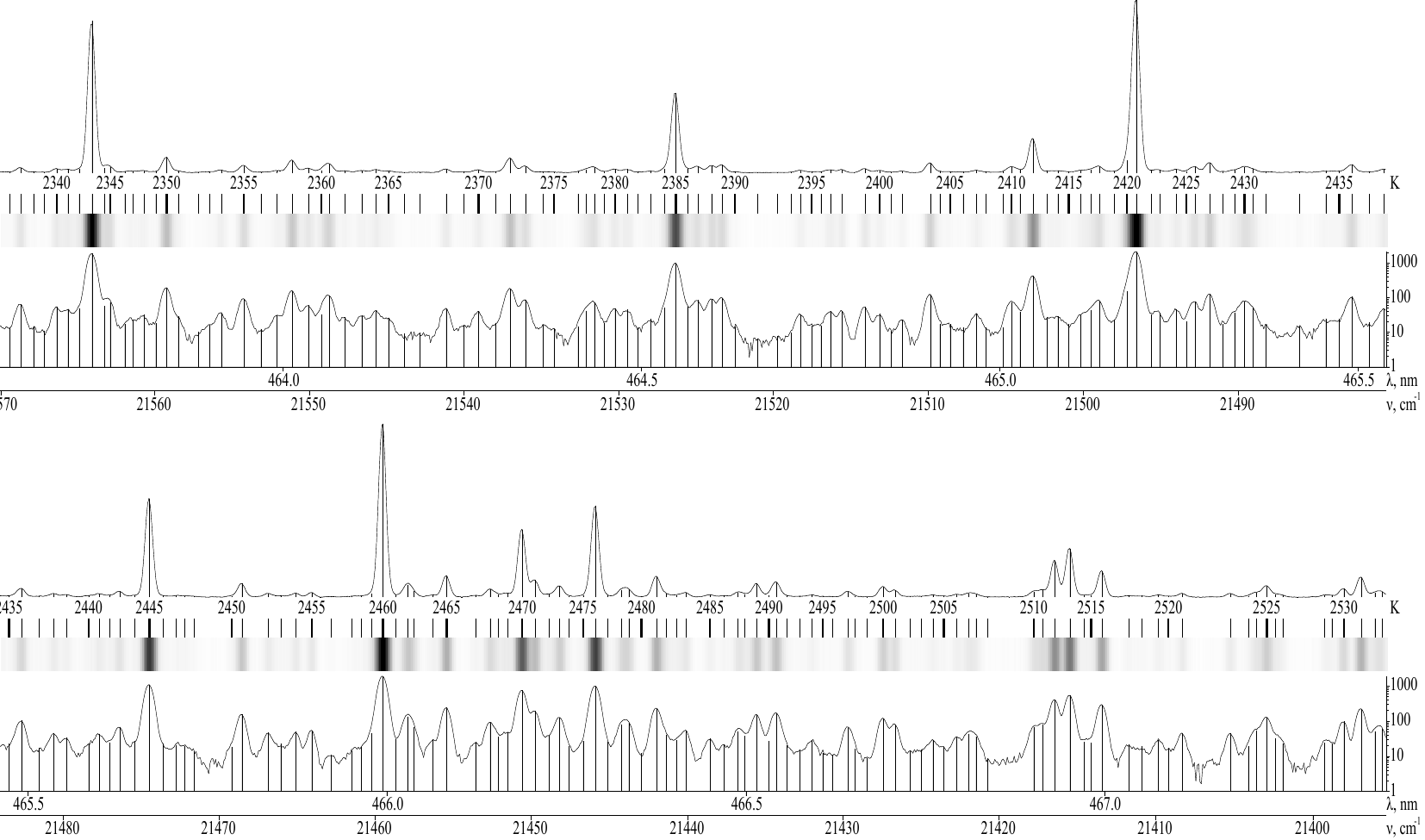}
\end{figure}

\newpage
\begin{figure}[!ht]
\includegraphics[angle=90, totalheight=0.9\textheight]{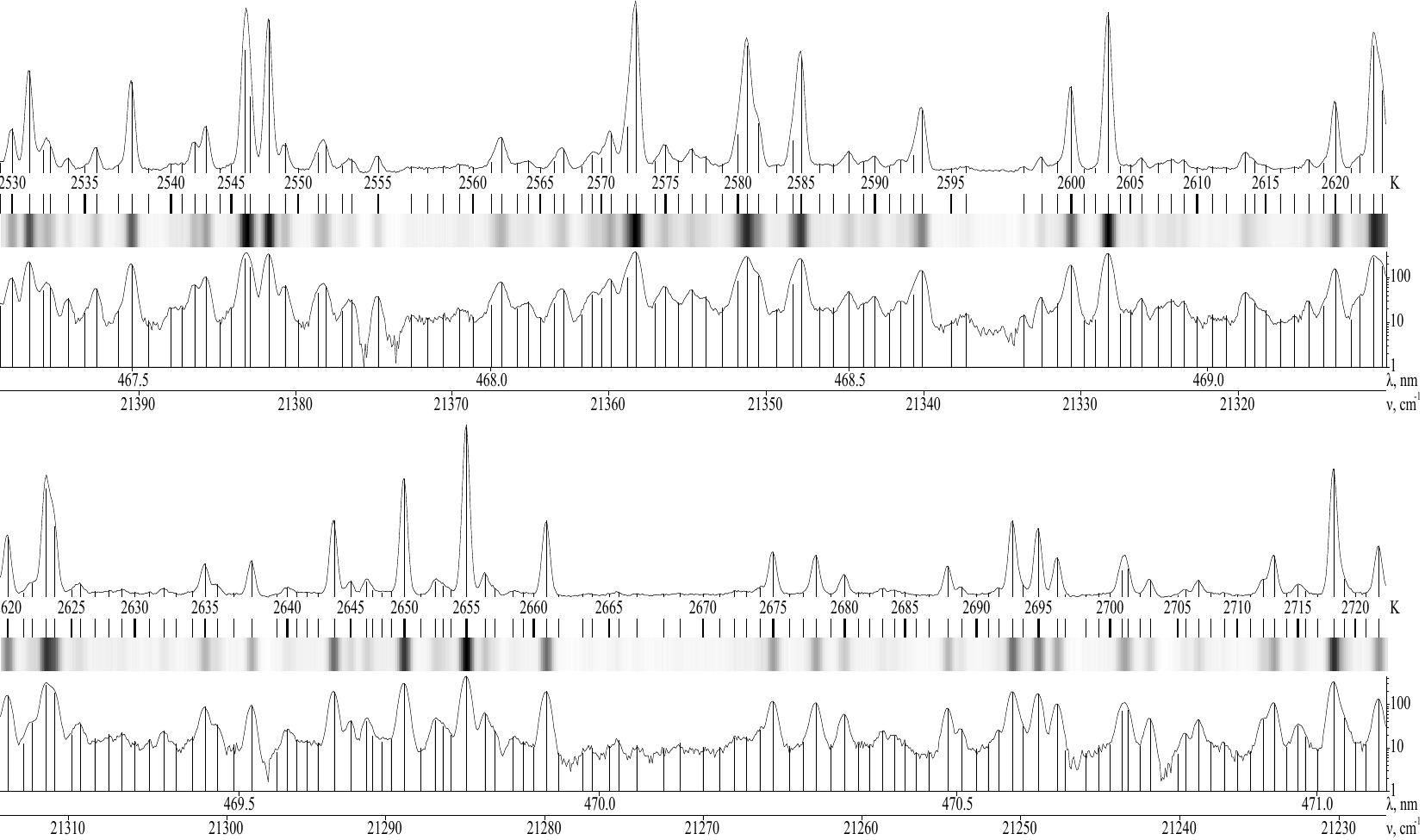}
\end{figure}

\newpage
\begin{figure}[!ht]
\includegraphics[angle=90, totalheight=0.9\textheight]{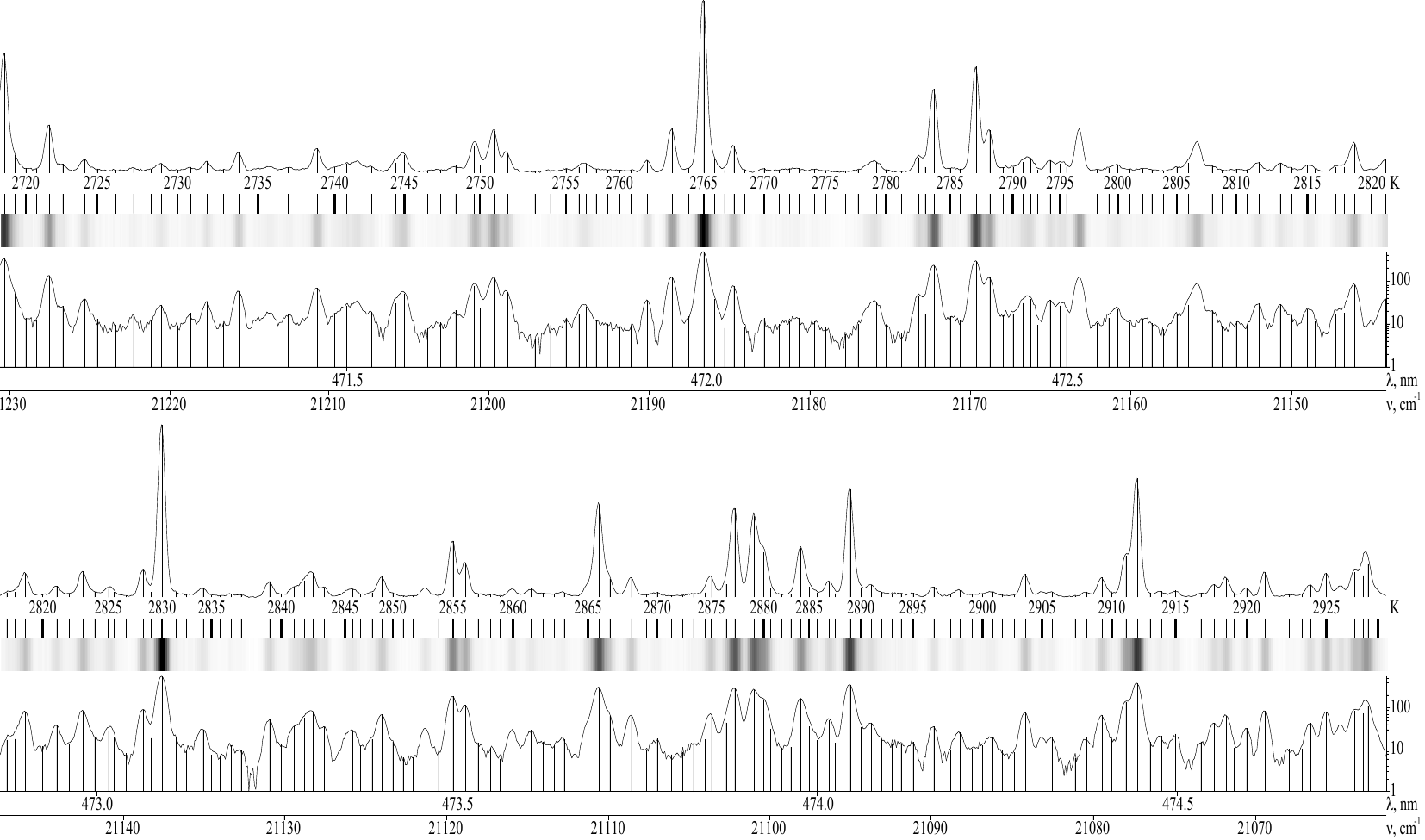}
\end{figure}

\newpage
\begin{figure}[!ht]
\includegraphics[angle=90, totalheight=0.9\textheight]{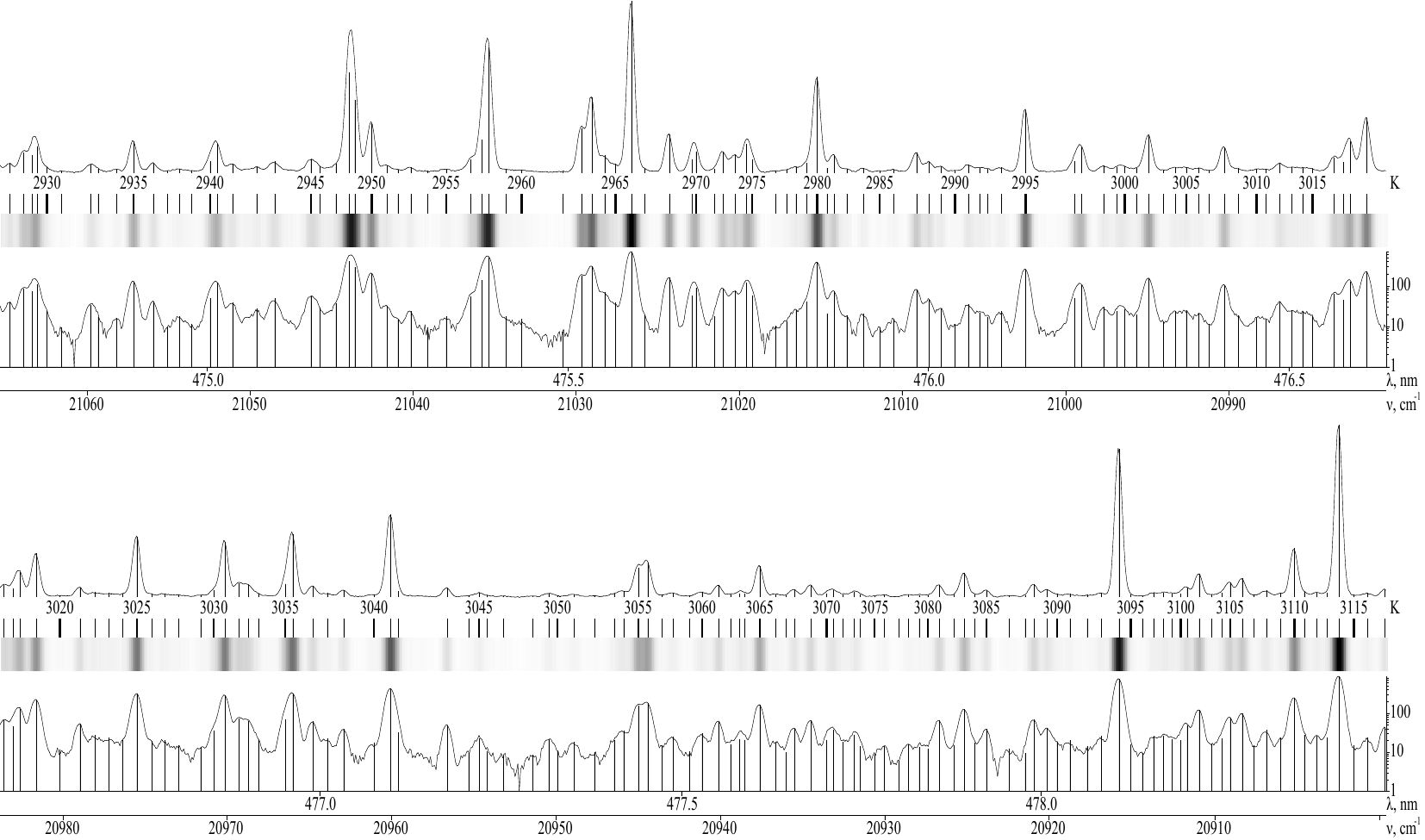}
\end{figure}

\newpage
\begin{figure}[!ht]
\includegraphics[angle=90, totalheight=0.9\textheight]{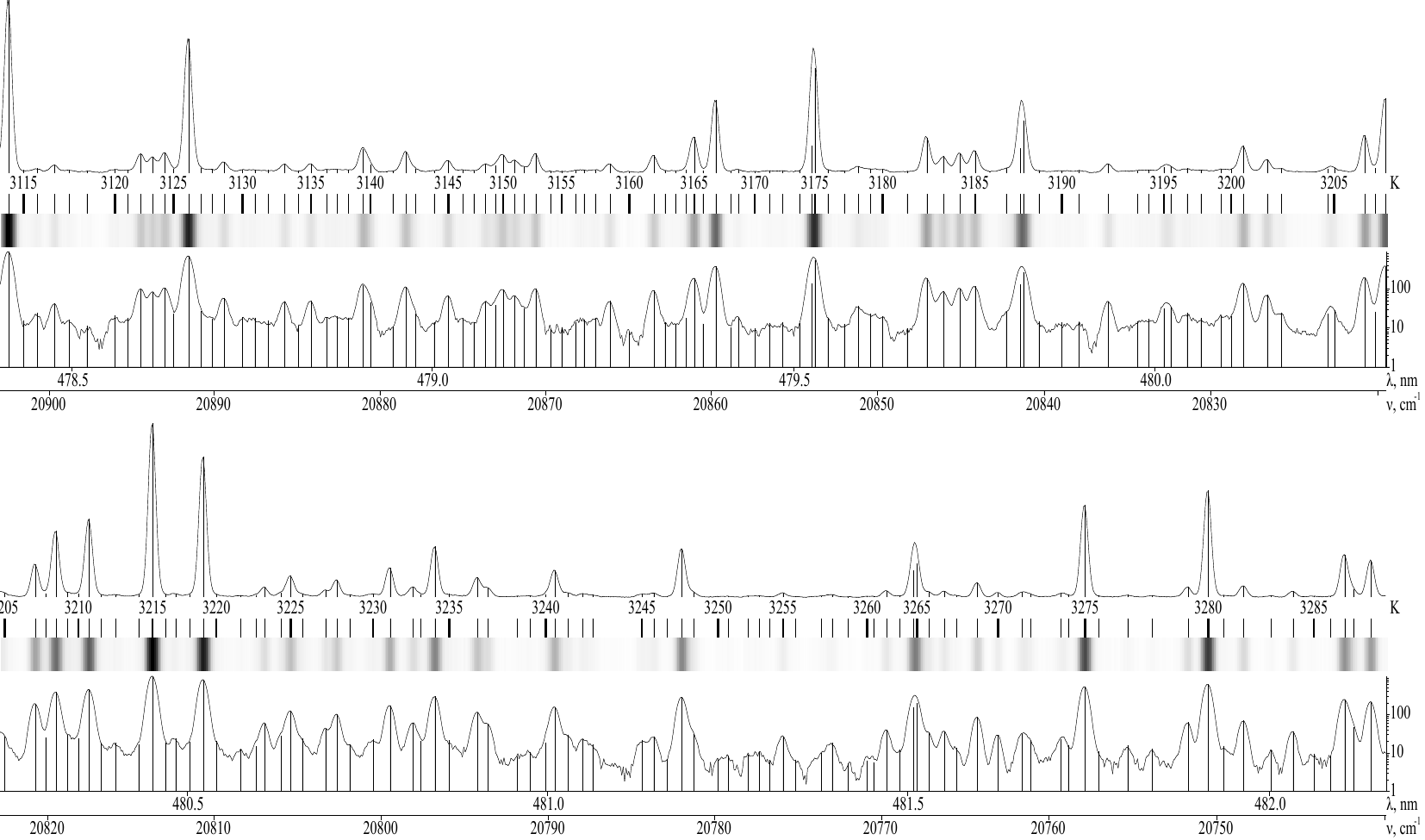}
\end{figure}

\newpage
\begin{figure}[!ht]
\includegraphics[angle=90, totalheight=0.9\textheight]{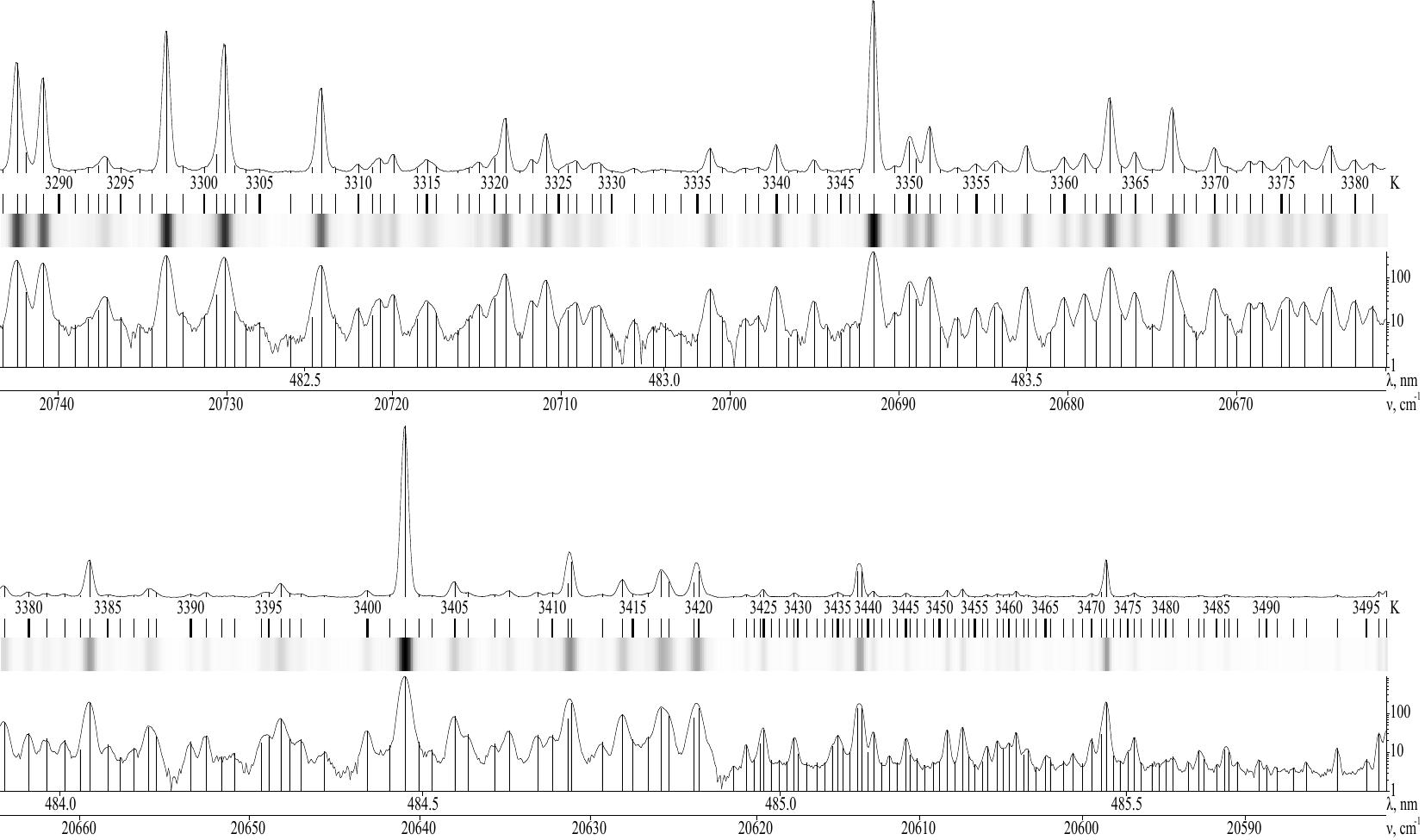}
\end{figure}

\newpage
\begin{figure}[!ht]
\includegraphics[angle=90, totalheight=0.9\textheight]{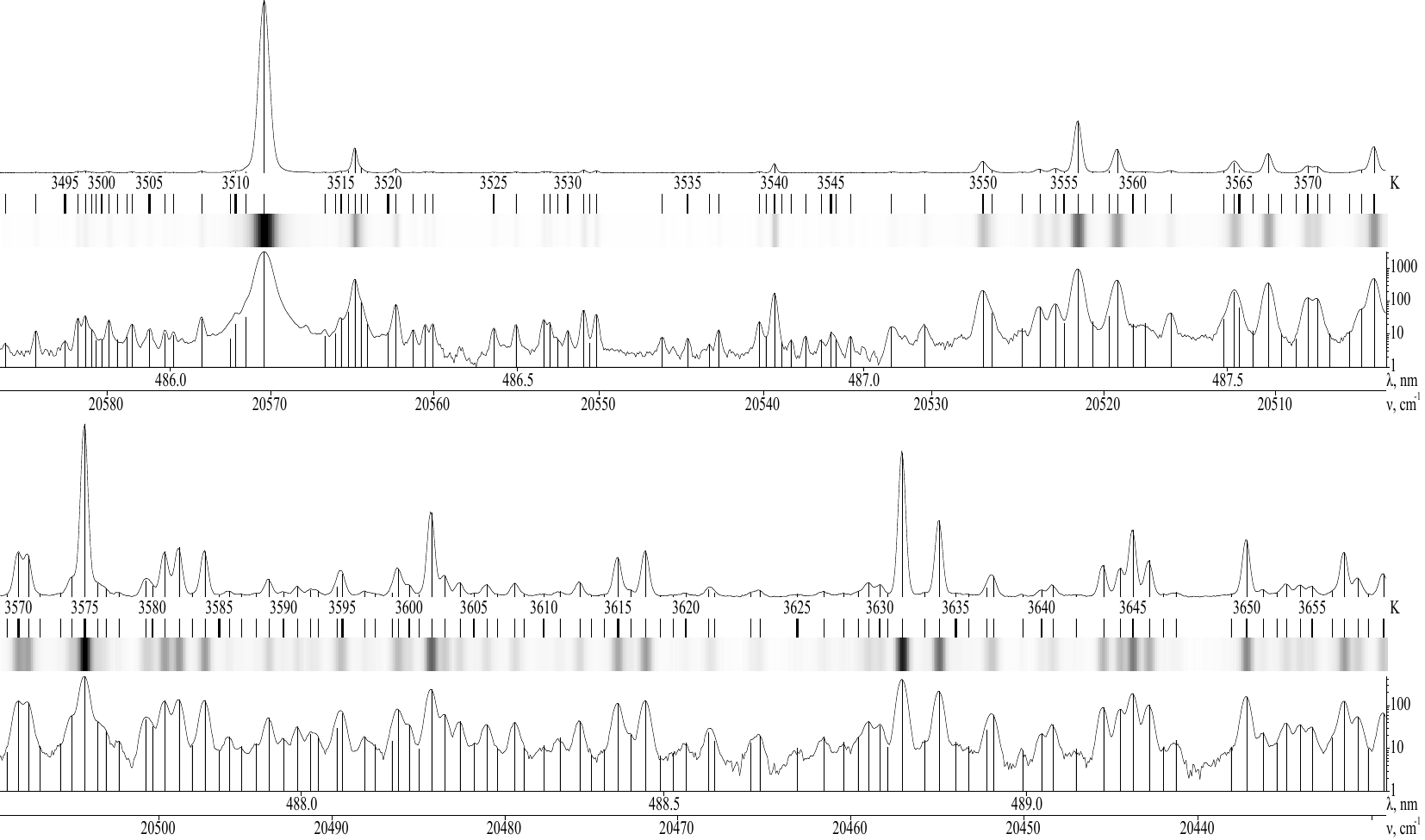}
\end{figure}

\newpage
\begin{figure}[!ht]
\includegraphics[angle=90, totalheight=0.9\textheight]{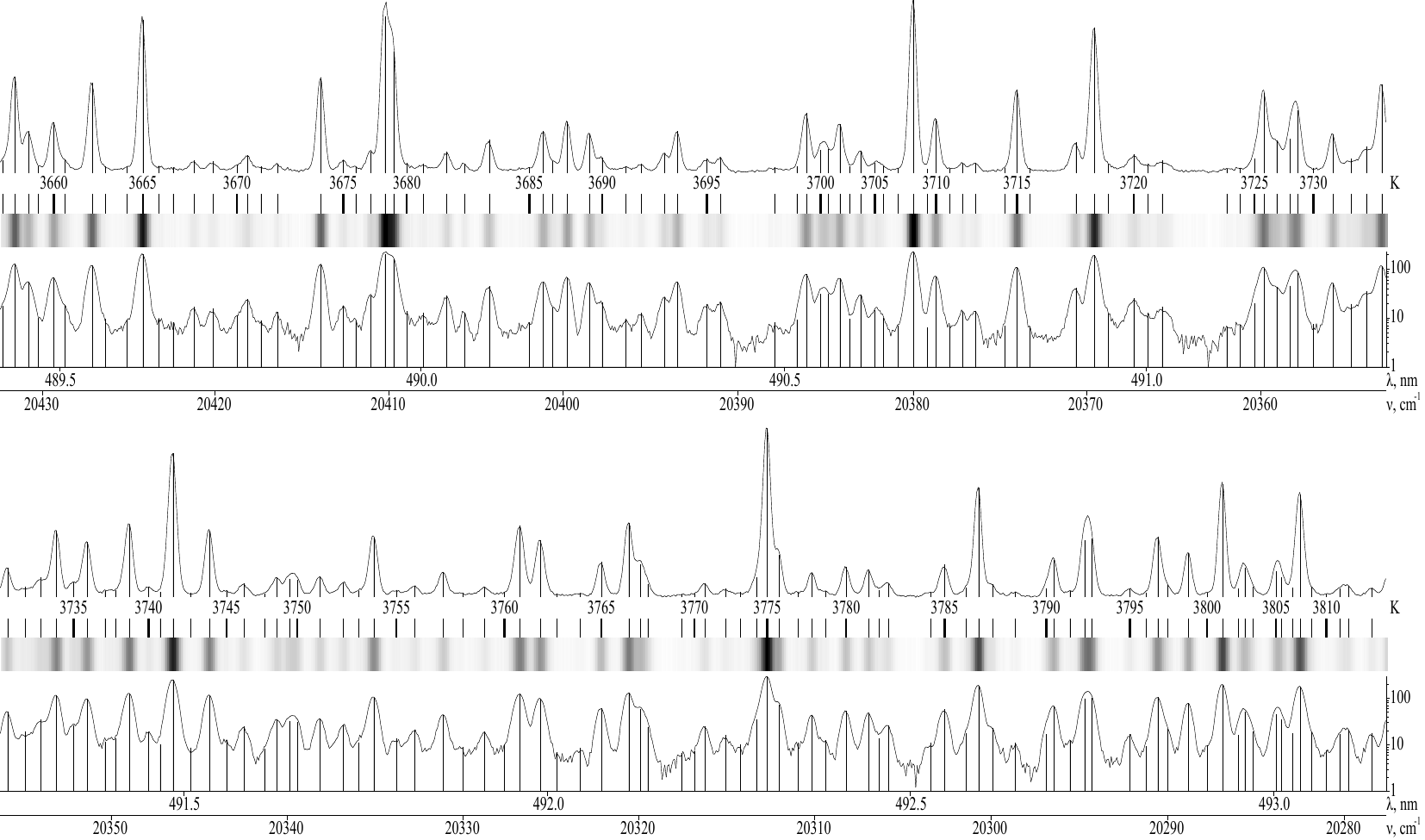}
\end{figure}

\newpage
\begin{figure}[!ht]
\includegraphics[angle=90, totalheight=0.9\textheight]{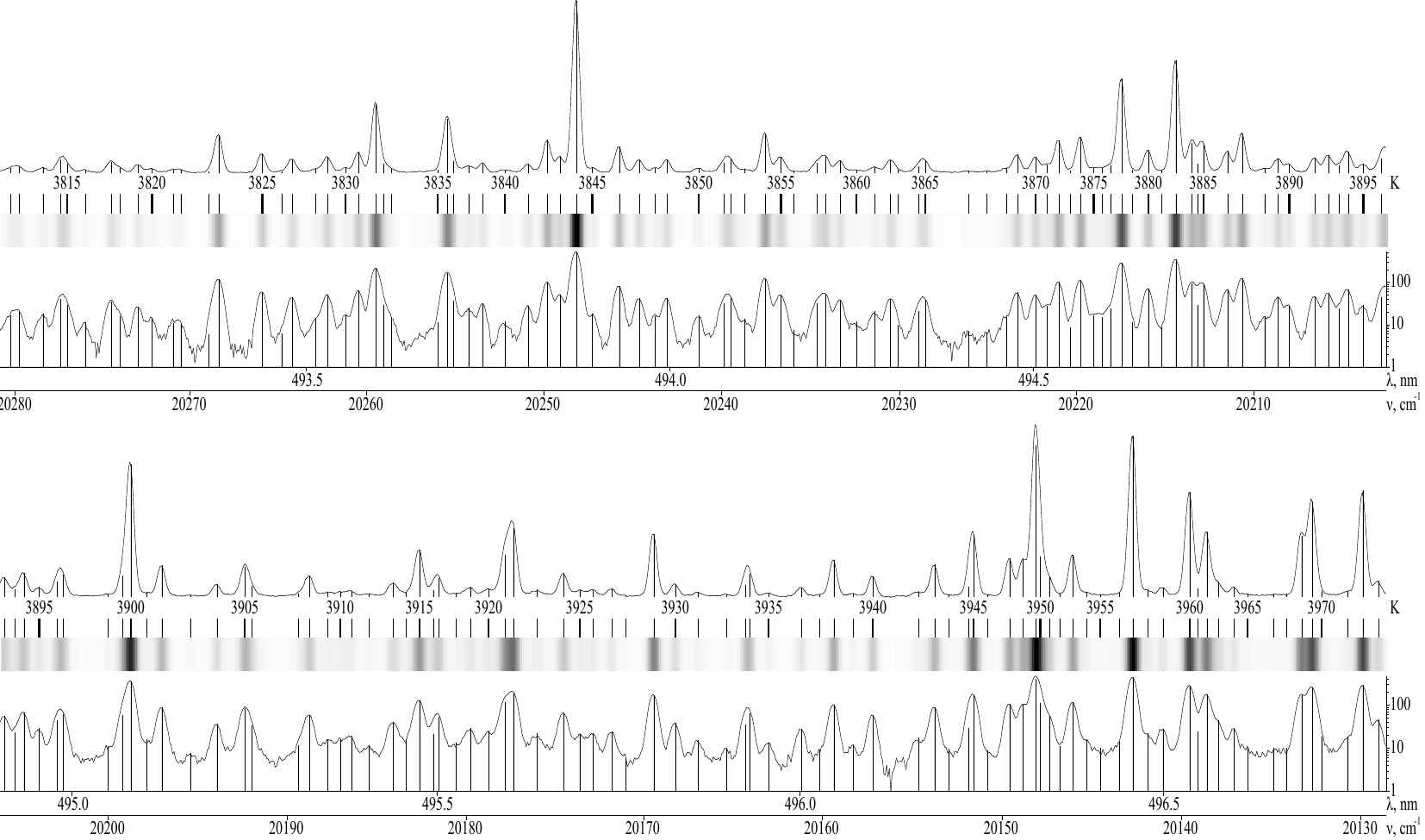}
\end{figure}

\newpage
\begin{figure}[!ht]
\includegraphics[angle=90, totalheight=0.9\textheight]{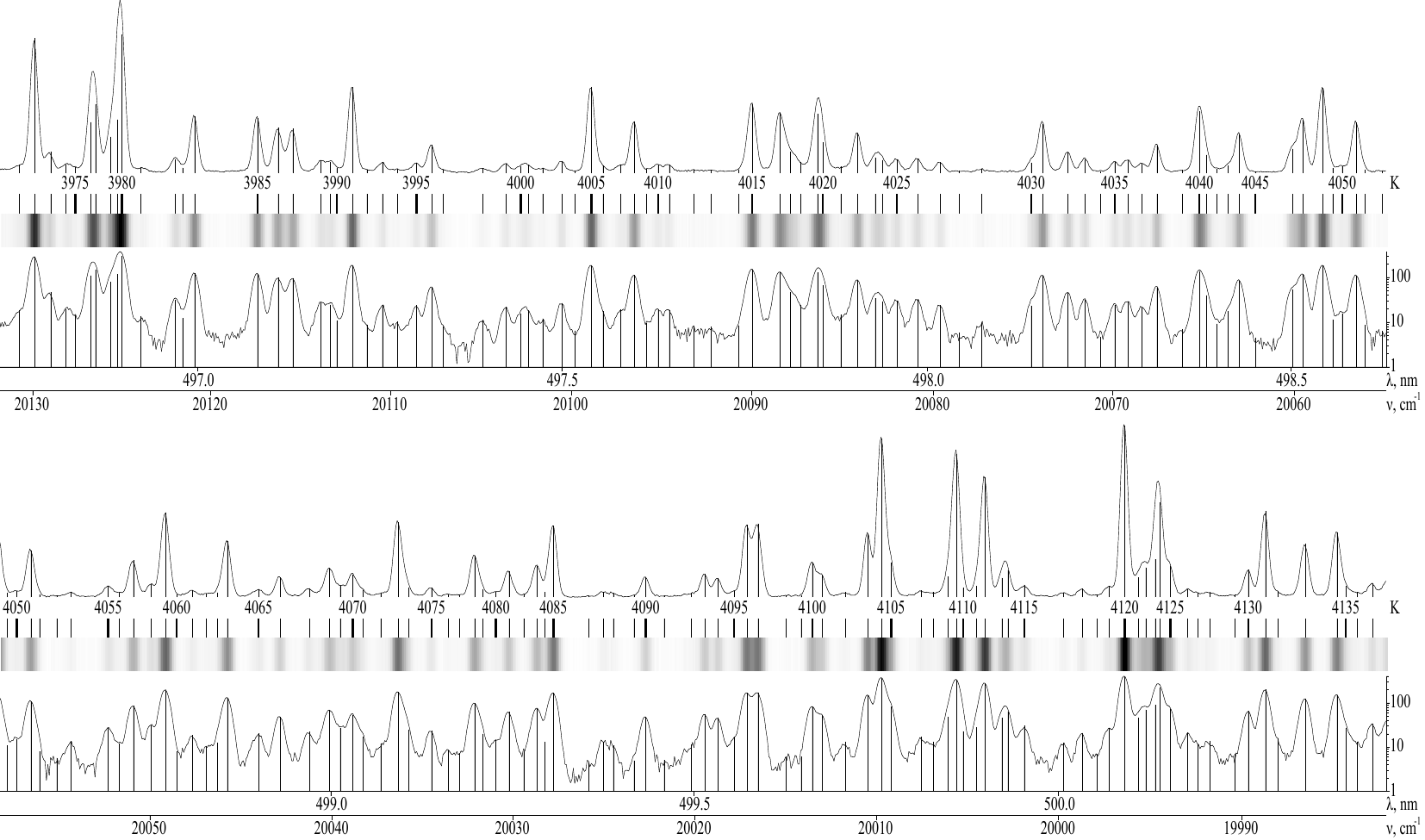}
\end{figure}

\newpage
\begin{figure}[!ht]
\includegraphics[angle=90, totalheight=0.9\textheight]{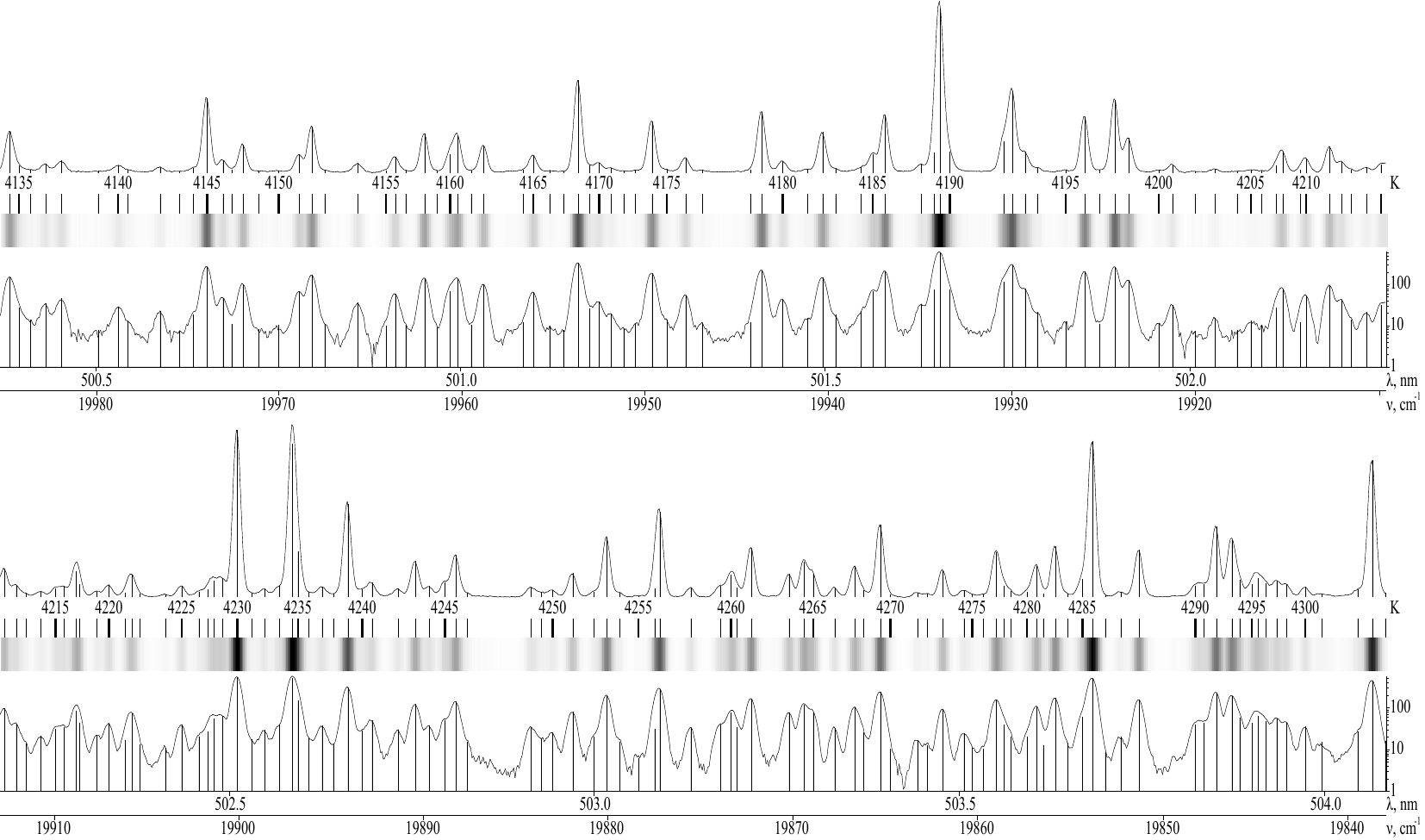}
\end{figure}

\newpage
\begin{figure}[!ht]
\includegraphics[angle=90, totalheight=0.9\textheight]{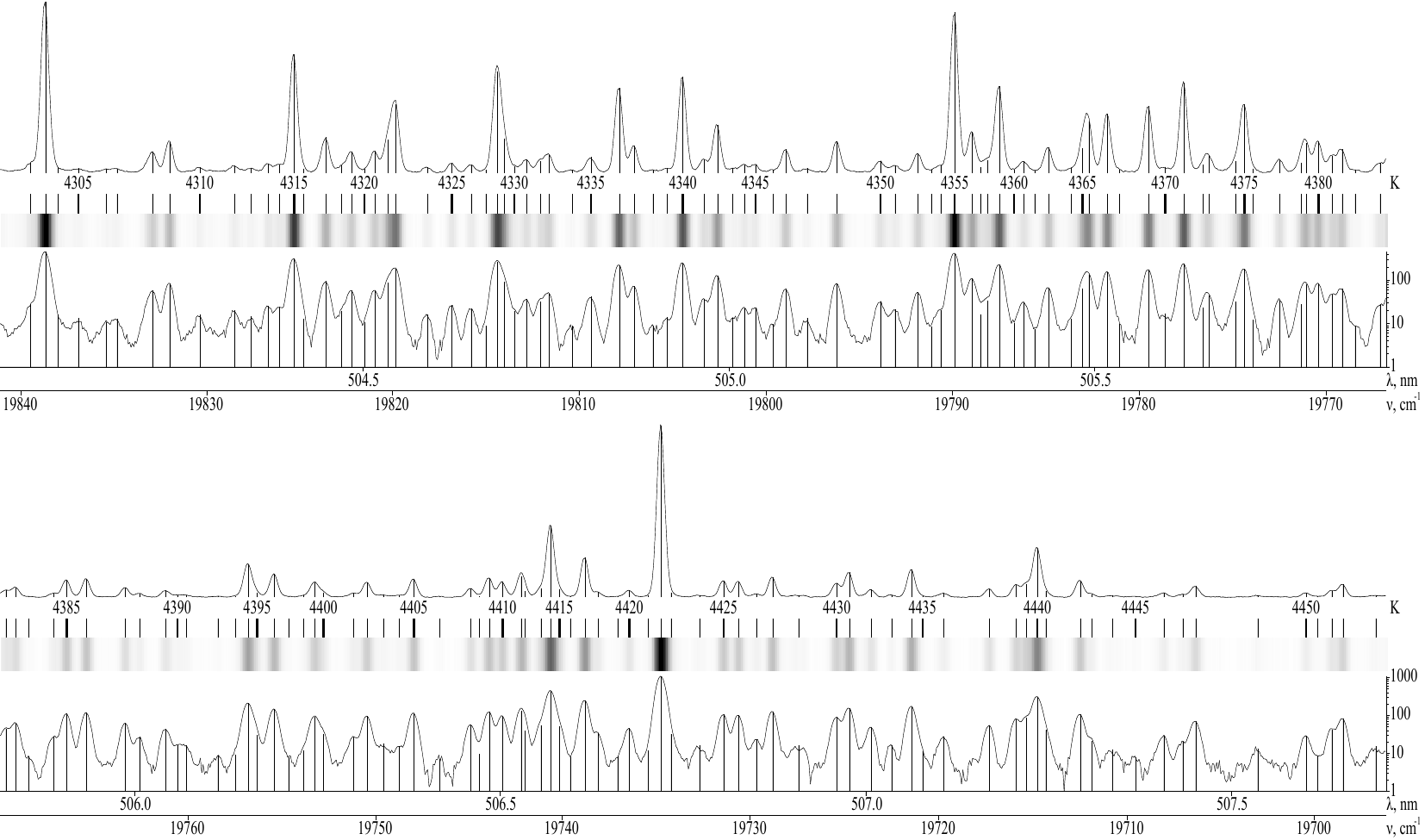}
\end{figure}

\newpage
\begin{figure}[!ht]
\includegraphics[angle=90, totalheight=0.9\textheight]{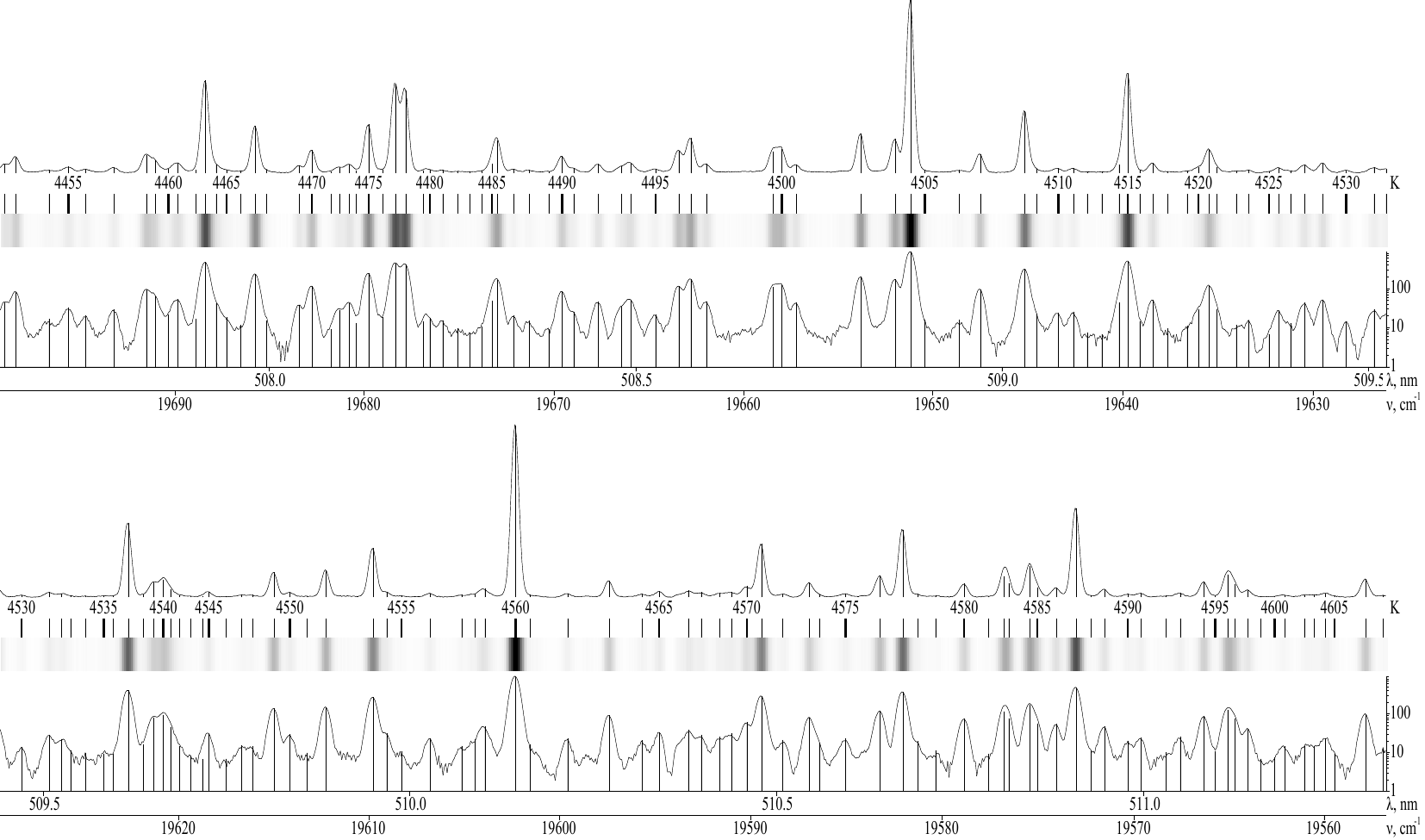}
\end{figure}

\newpage
\begin{figure}[!ht]
\includegraphics[angle=90, totalheight=0.9\textheight]{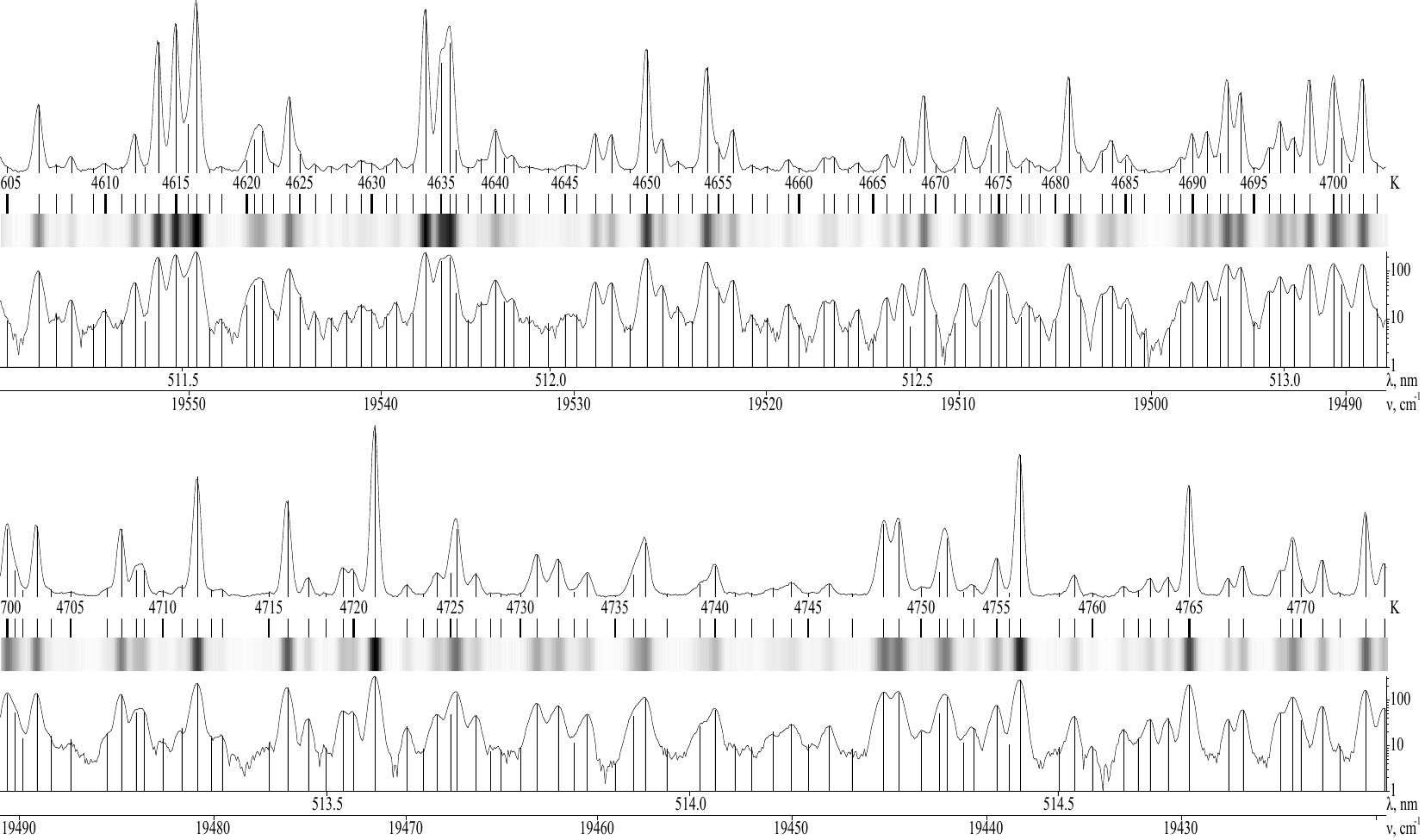}
\end{figure}

\newpage
\begin{figure}[!ht]
\includegraphics[angle=90, totalheight=0.9\textheight]{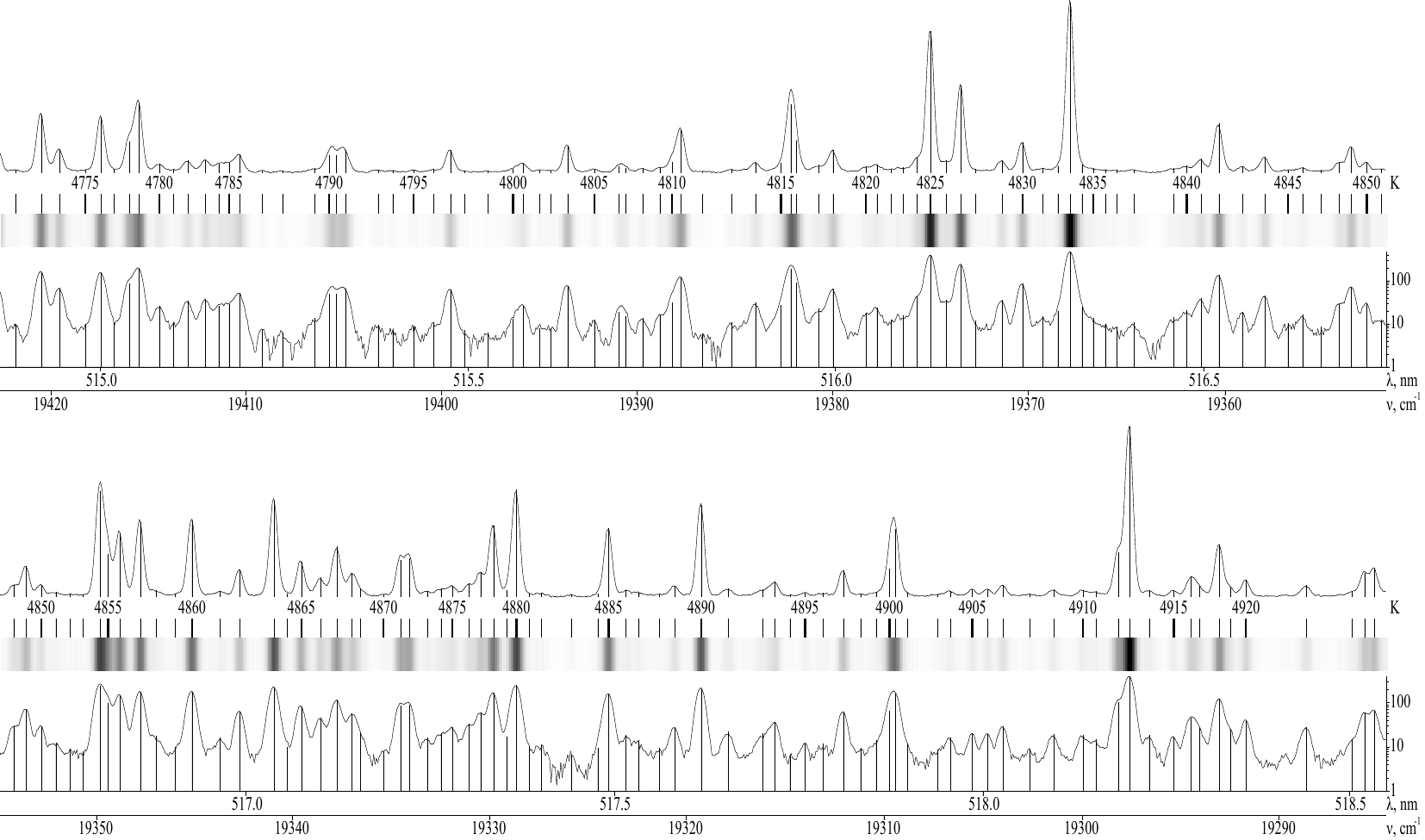}
\end{figure}

\newpage
\begin{figure}[!ht]
\includegraphics[angle=90, totalheight=0.9\textheight]{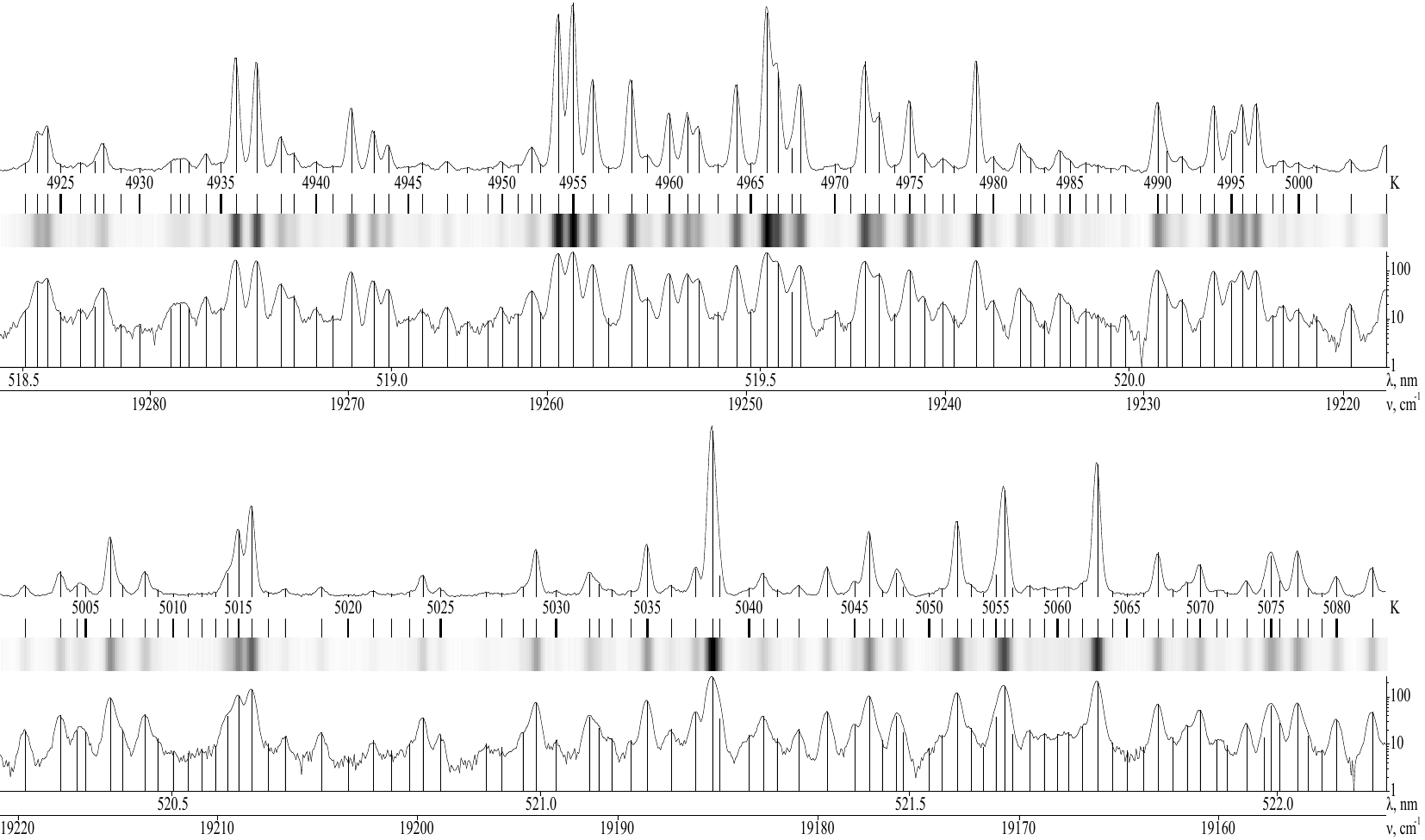}
\end{figure}

\newpage
\begin{figure}[!ht]
\includegraphics[angle=90, totalheight=0.9\textheight]{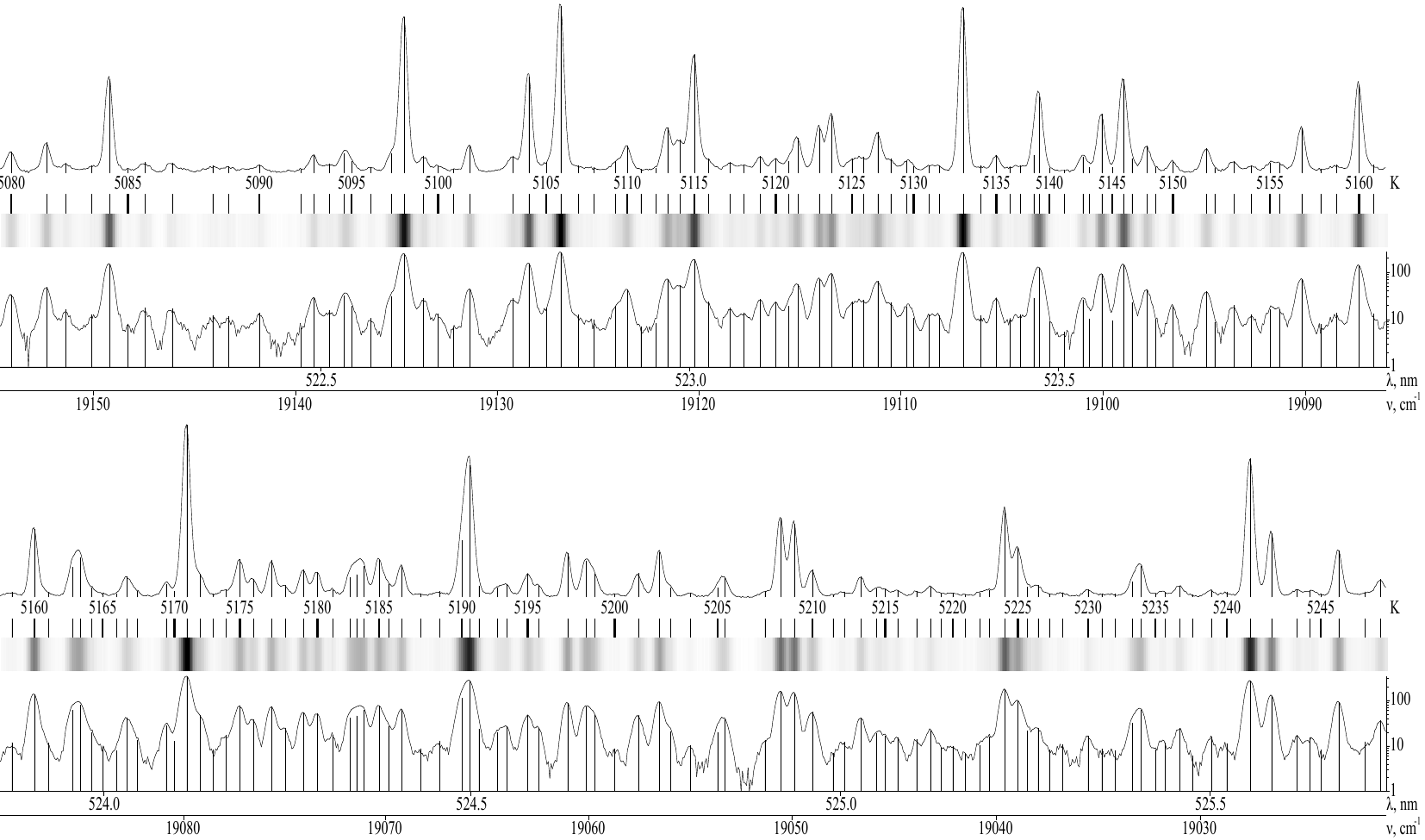}
\end{figure}

\newpage
\begin{figure}[!ht]
\includegraphics[angle=90, totalheight=0.9\textheight]{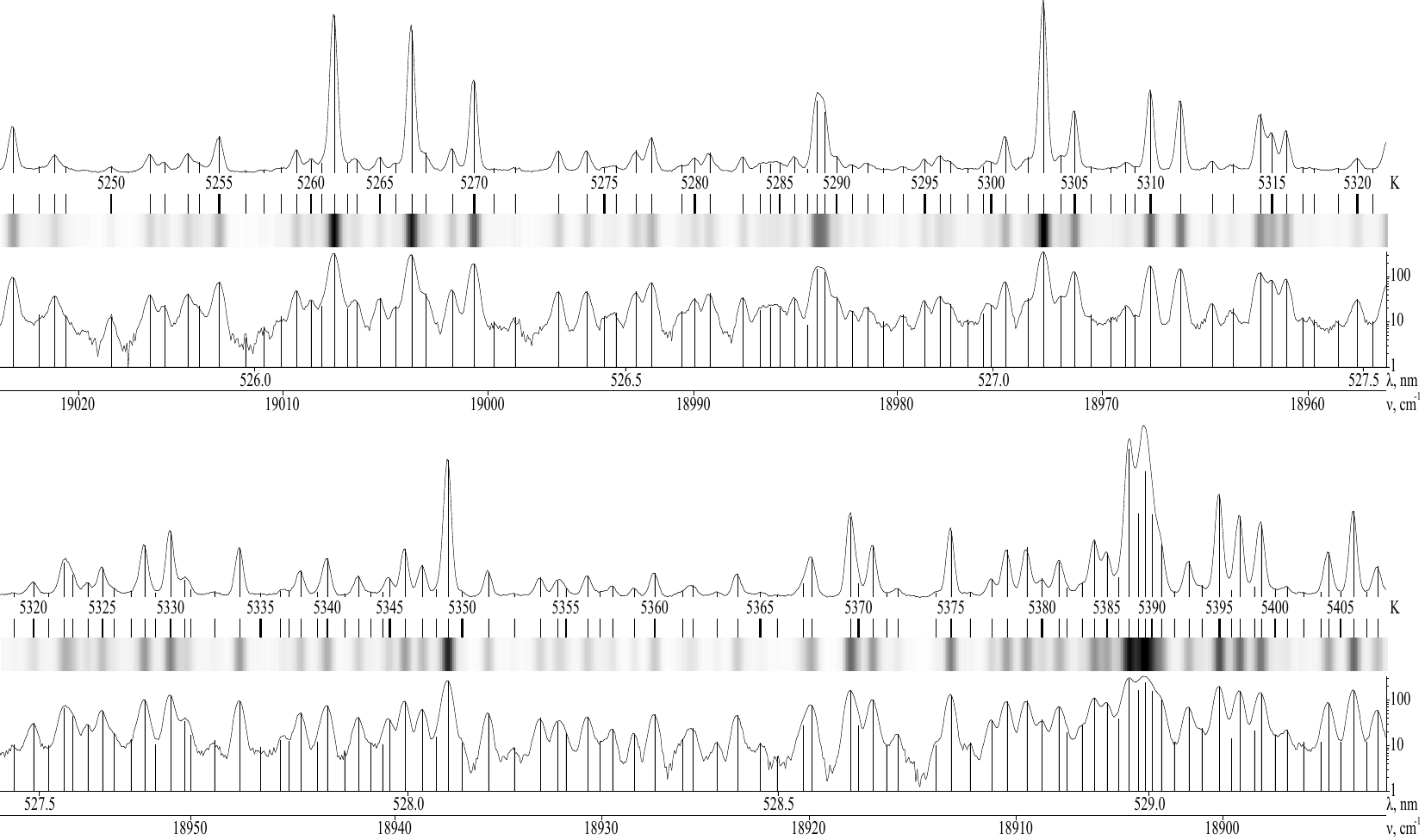}
\end{figure}

\newpage
\begin{figure}[!ht]
\includegraphics[angle=90, totalheight=0.9\textheight]{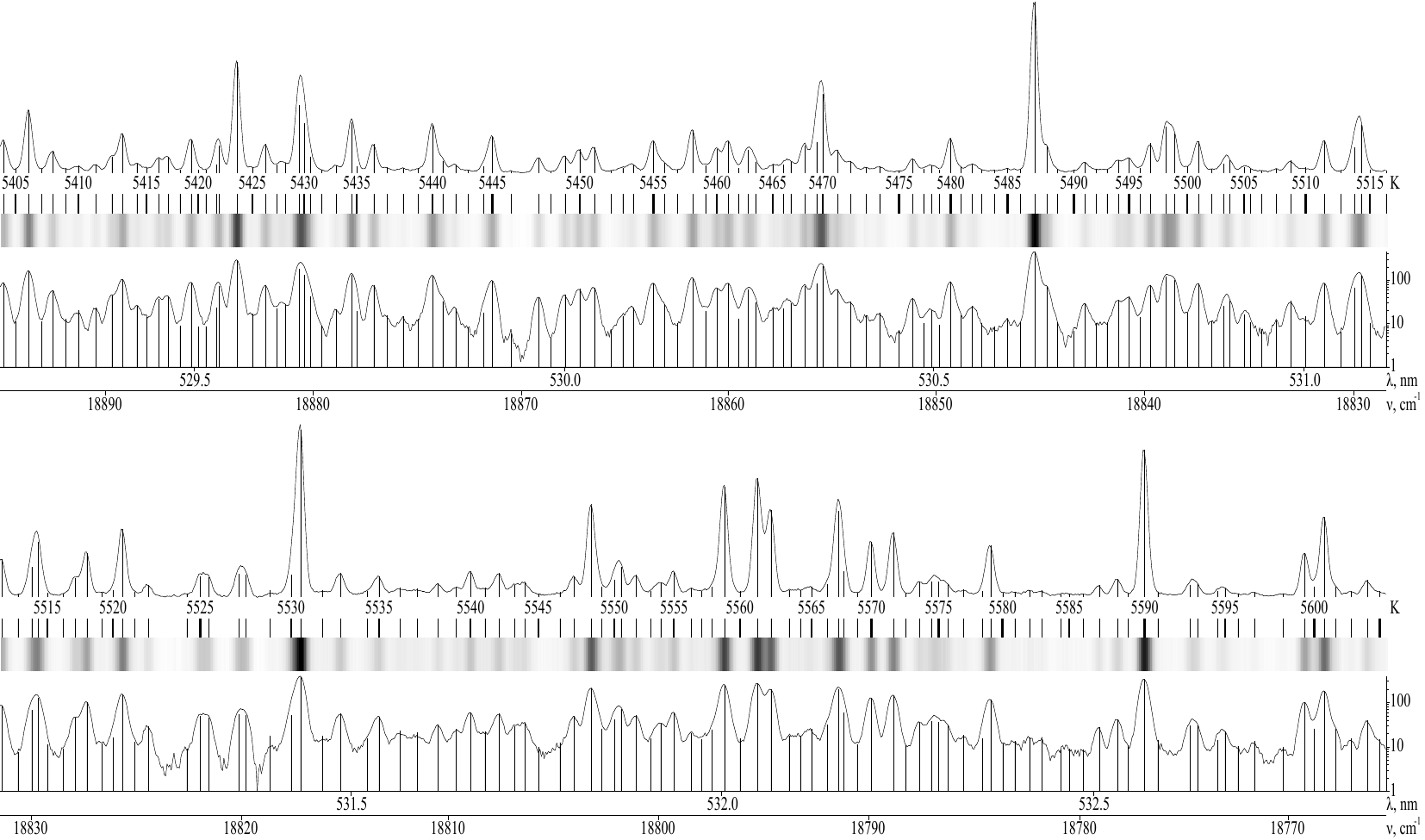}
\end{figure}

\newpage
\begin{figure}[!ht]
\includegraphics[angle=90, totalheight=0.9\textheight]{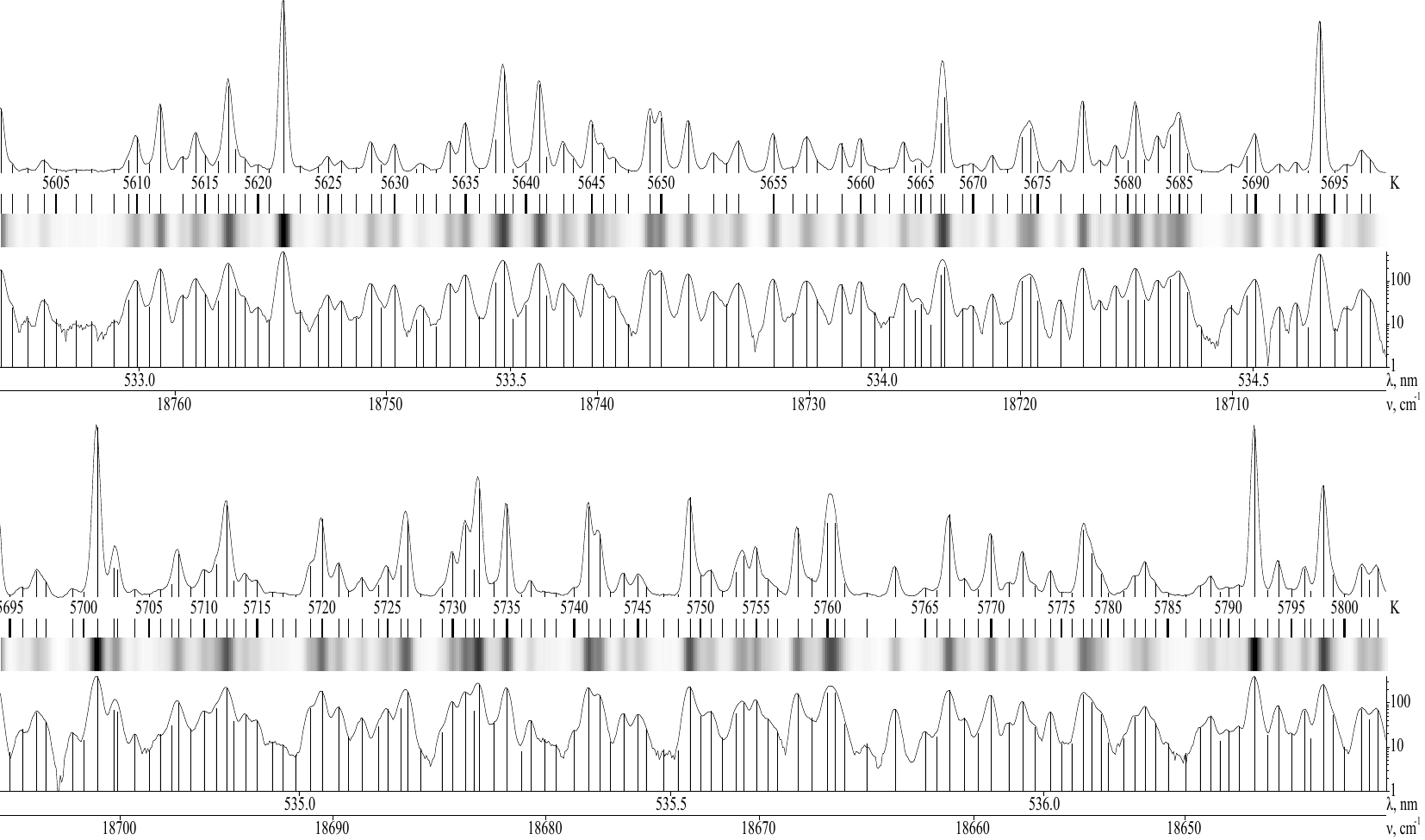}
\end{figure}

\newpage
\begin{figure}[!ht]
\includegraphics[angle=90, totalheight=0.9\textheight]{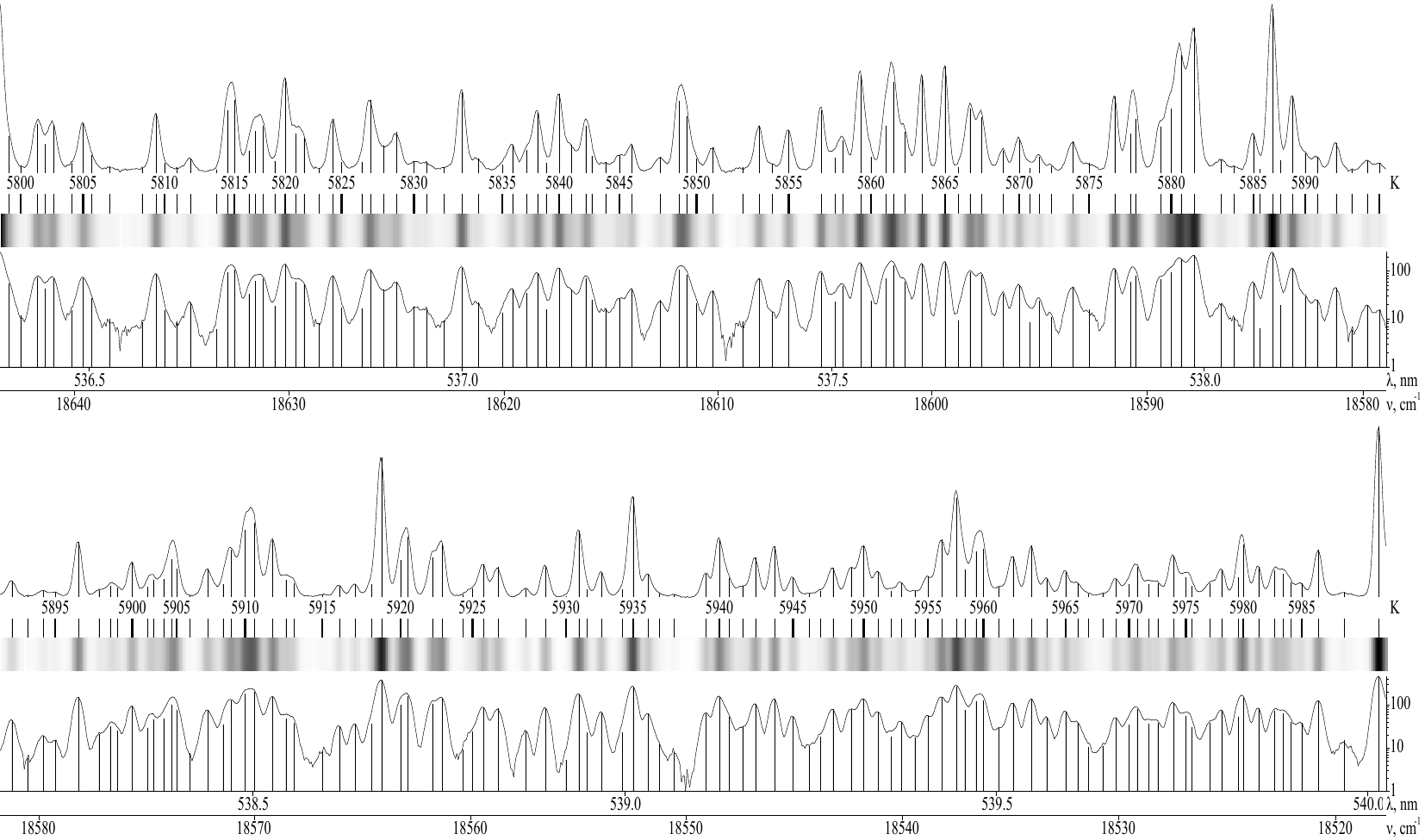}
\end{figure}

\newpage
\begin{figure}[!ht]
\includegraphics[angle=90, totalheight=0.9\textheight]{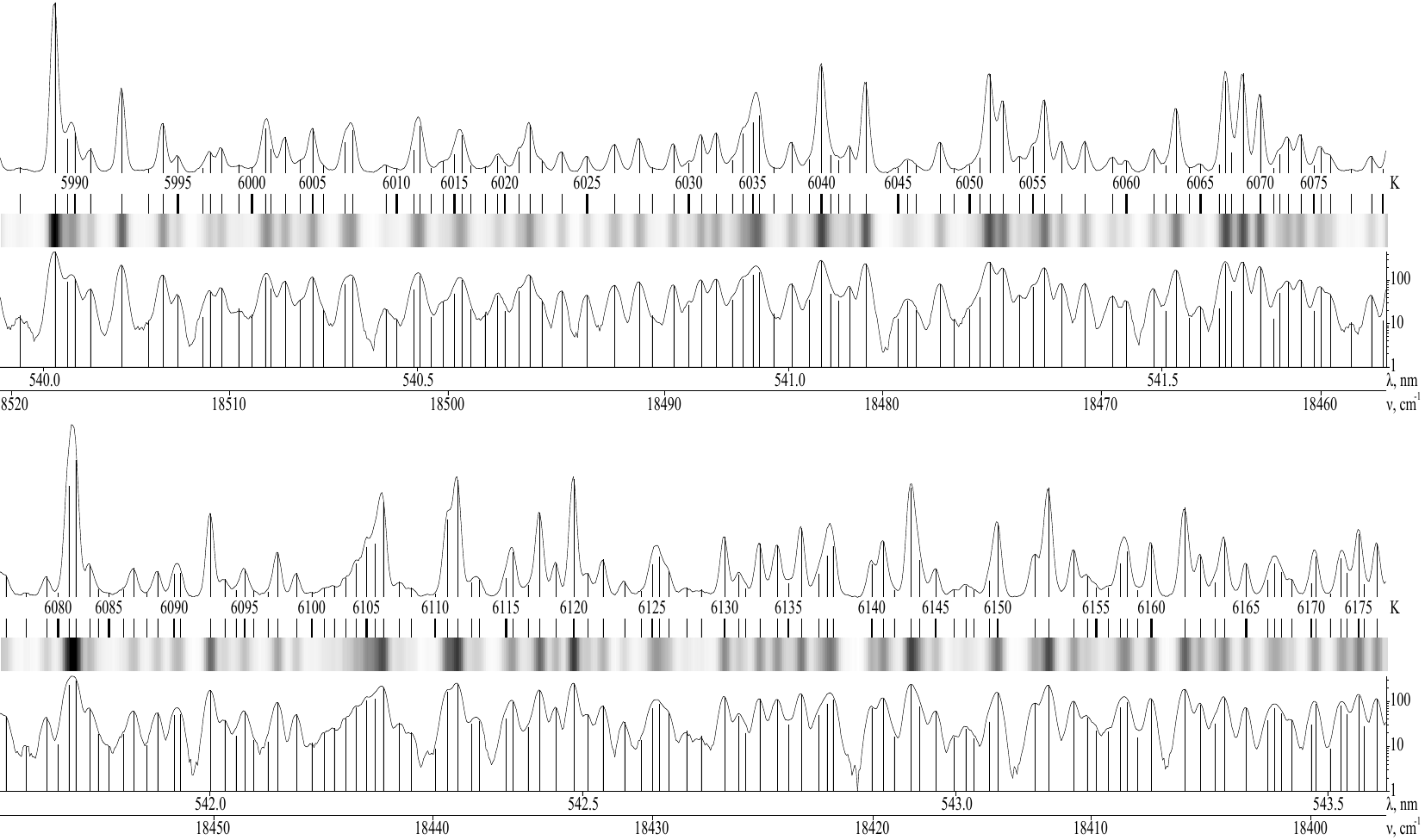}
\end{figure}

\newpage
\begin{figure}[!ht]
\includegraphics[angle=90, totalheight=0.9\textheight]{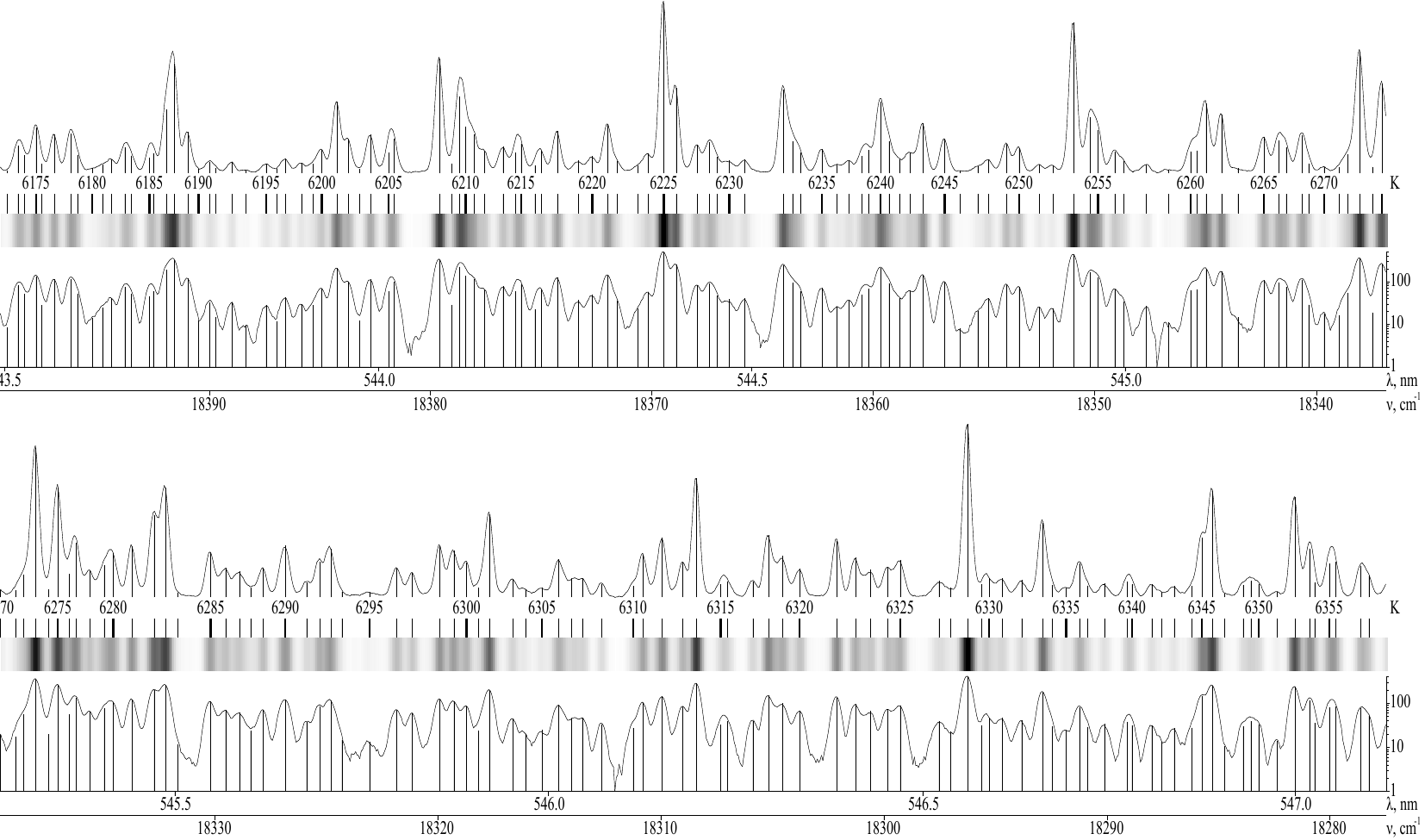}
\end{figure}

\newpage
\begin{figure}[!ht]
\includegraphics[angle=90, totalheight=0.9\textheight]{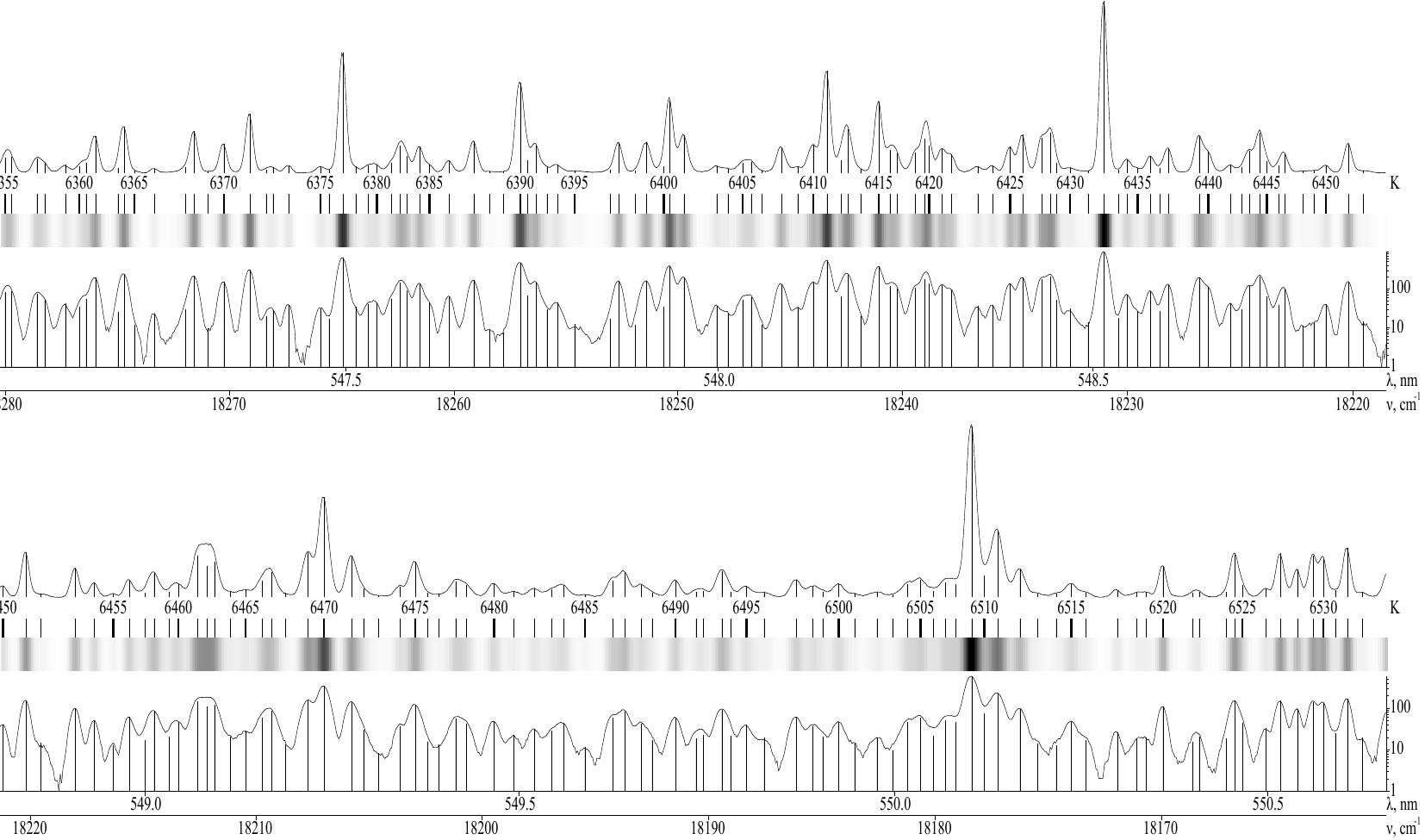}
\end{figure}

\newpage
\begin{figure}[!ht]
\includegraphics[angle=90, totalheight=0.9\textheight]{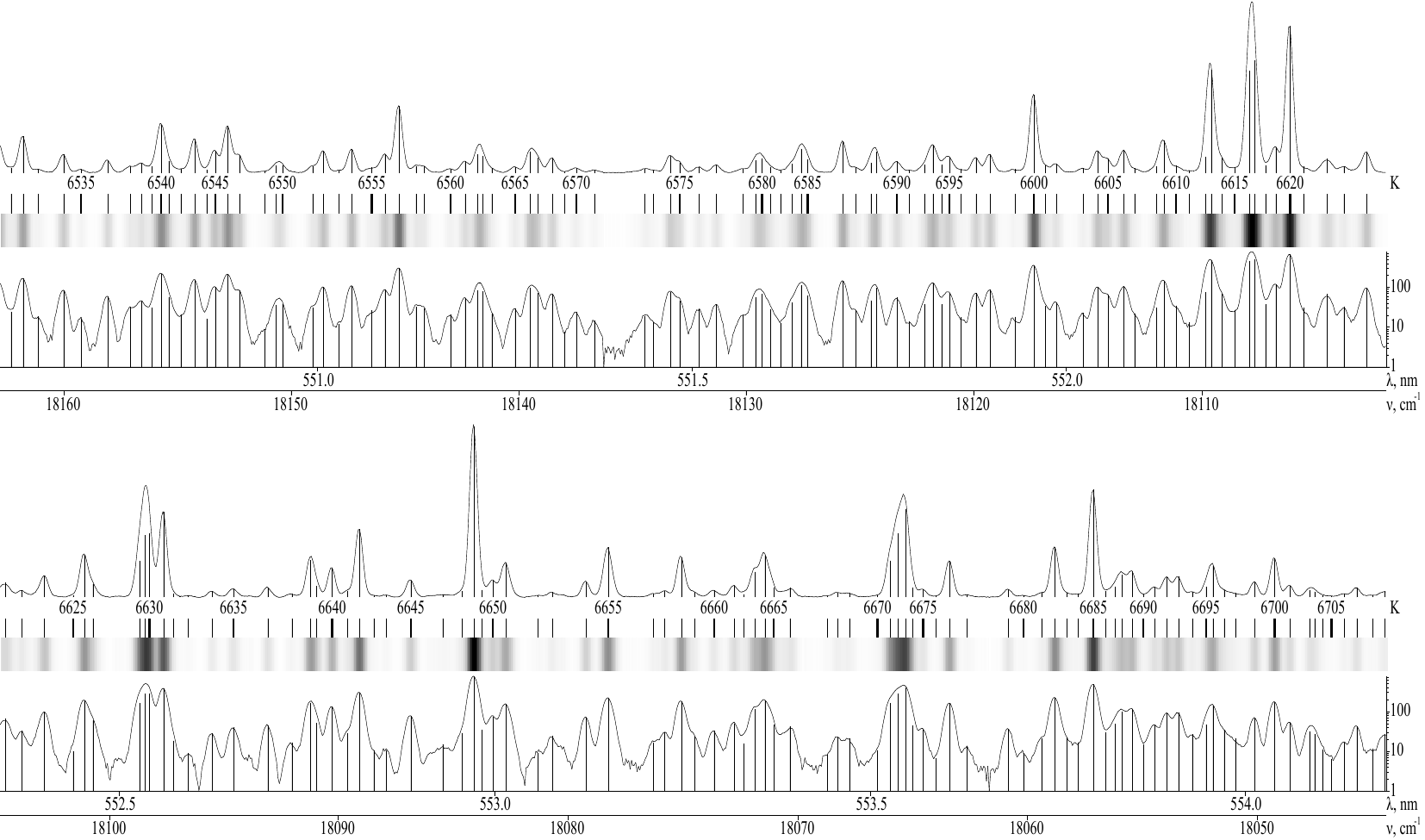}
\end{figure}

\newpage
\begin{figure}[!ht]
\includegraphics[angle=90, totalheight=0.9\textheight]{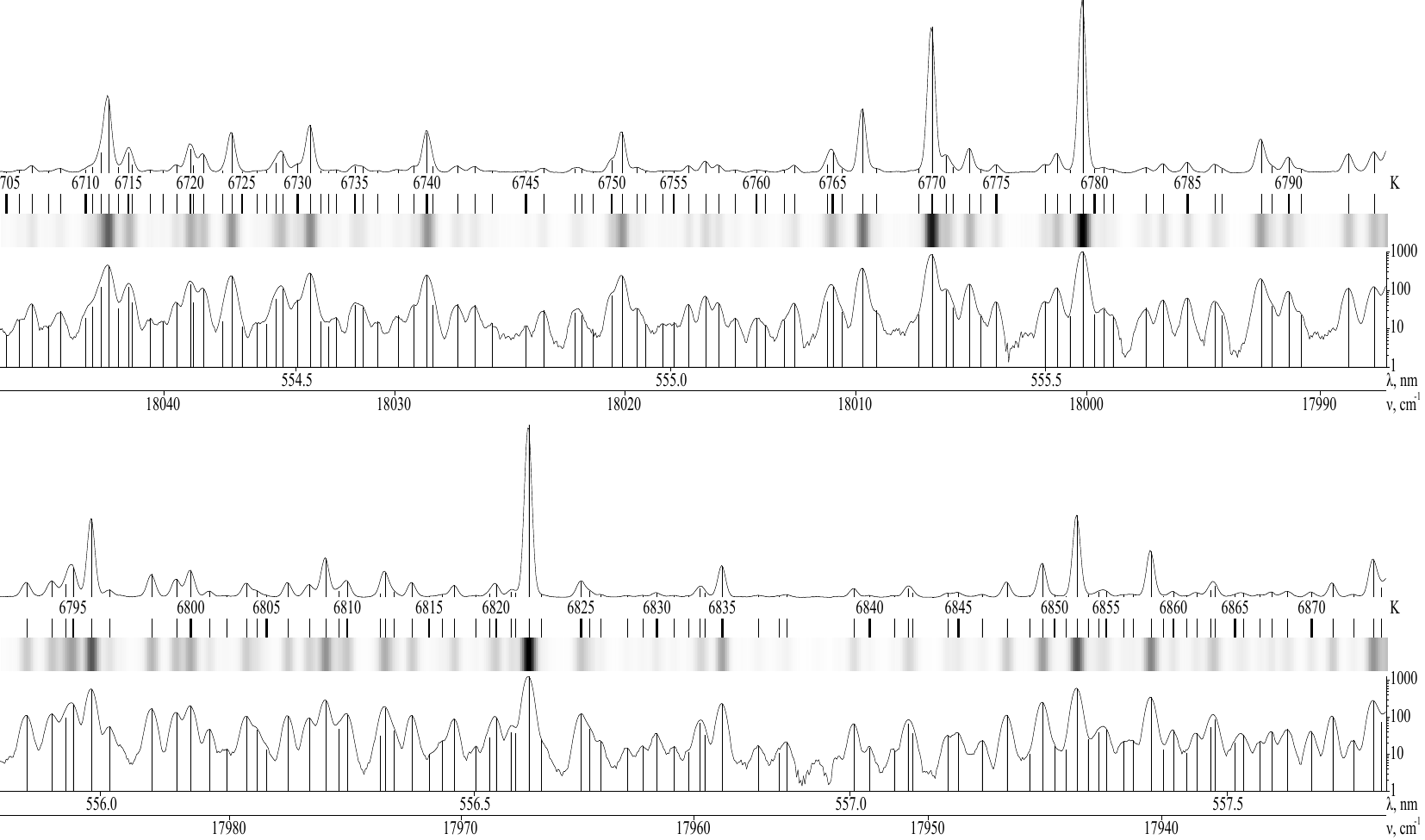}
\end{figure}

\newpage
\begin{figure}[!ht]
\includegraphics[angle=90, totalheight=0.9\textheight]{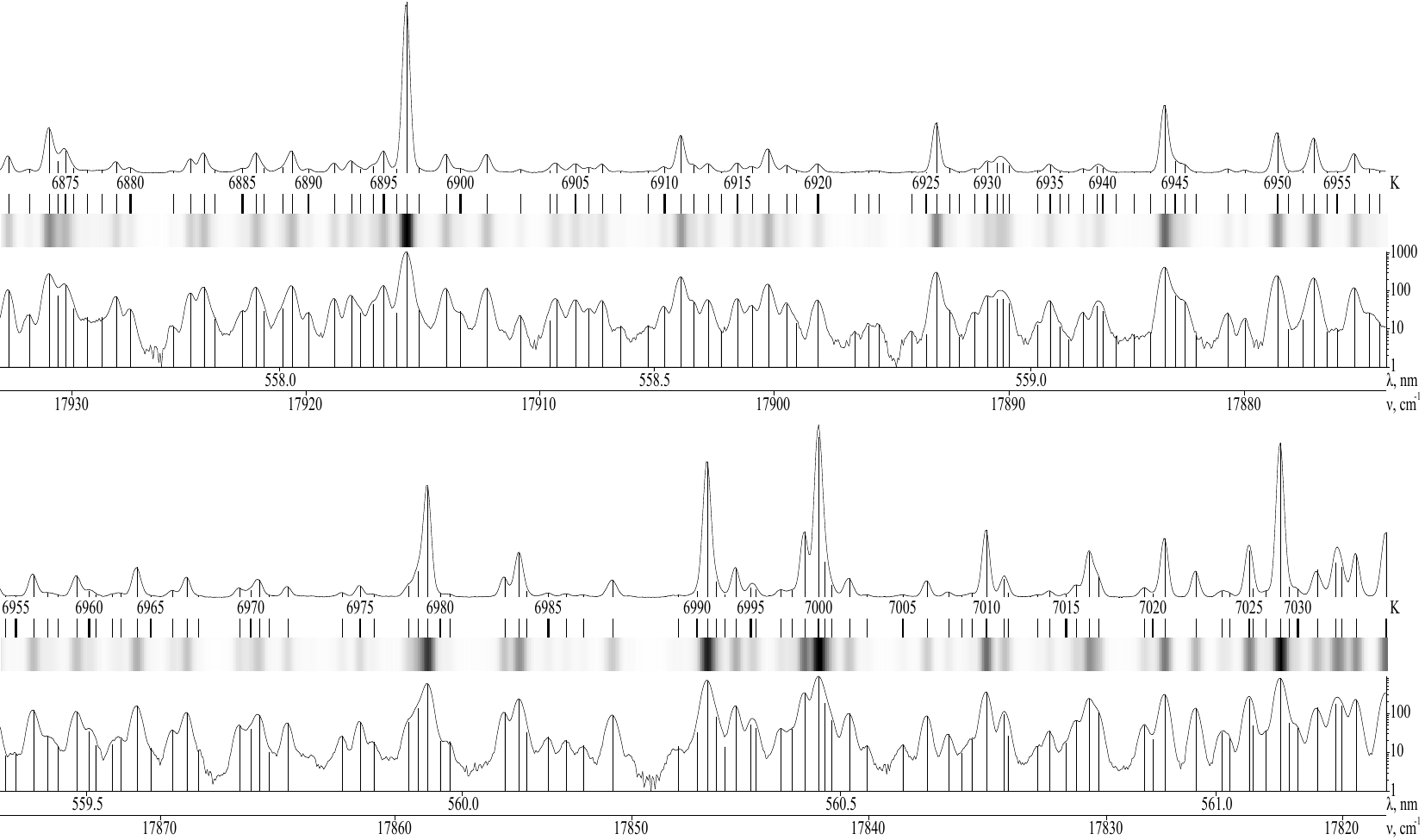}
\end{figure}

\newpage
\begin{figure}[!ht]
\includegraphics[angle=90, totalheight=0.9\textheight]{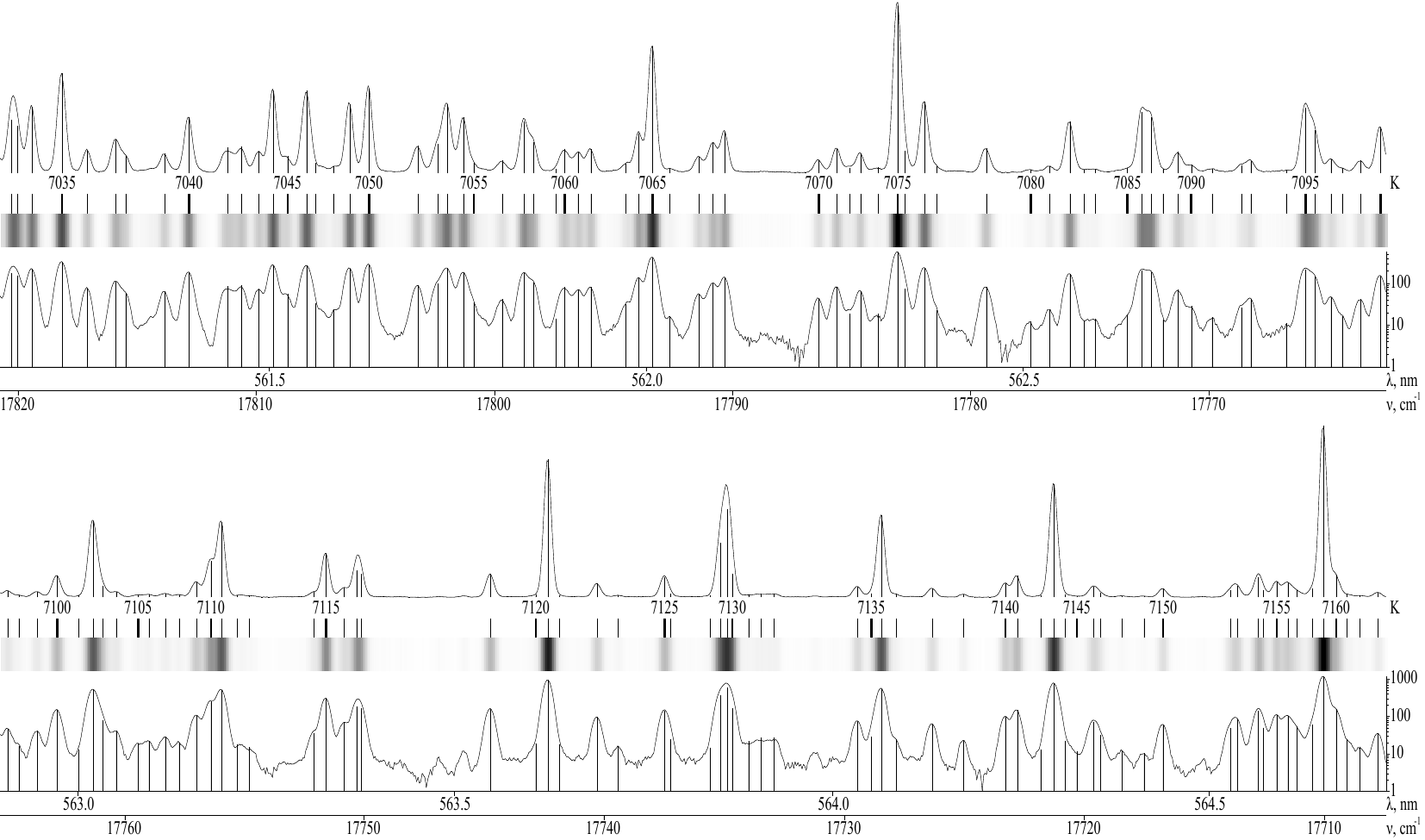}
\end{figure}

\newpage
\begin{figure}[!ht]
\includegraphics[angle=90, totalheight=0.9\textheight]{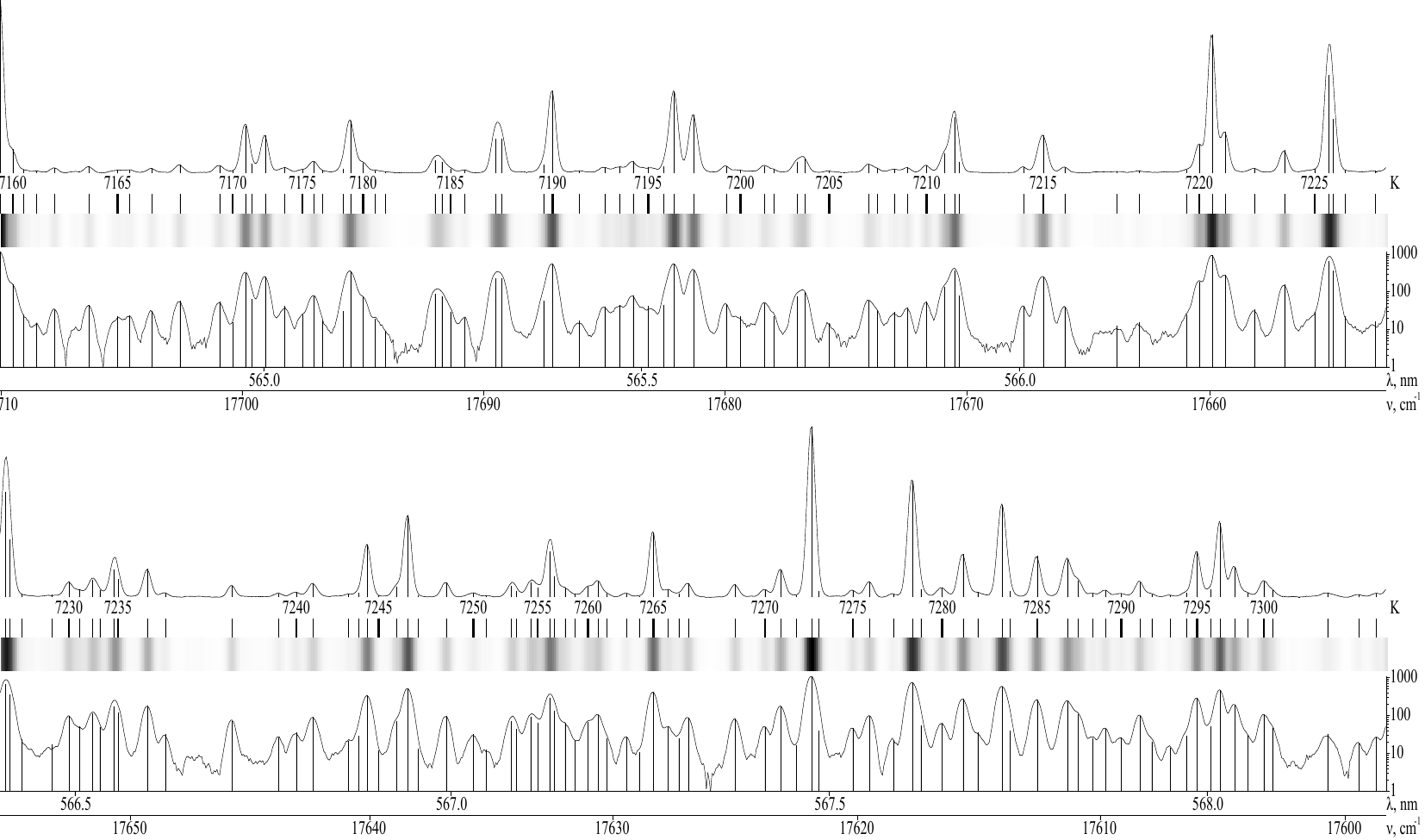}
\end{figure}

\newpage
\begin{figure}[!ht]
\includegraphics[angle=90, totalheight=0.9\textheight]{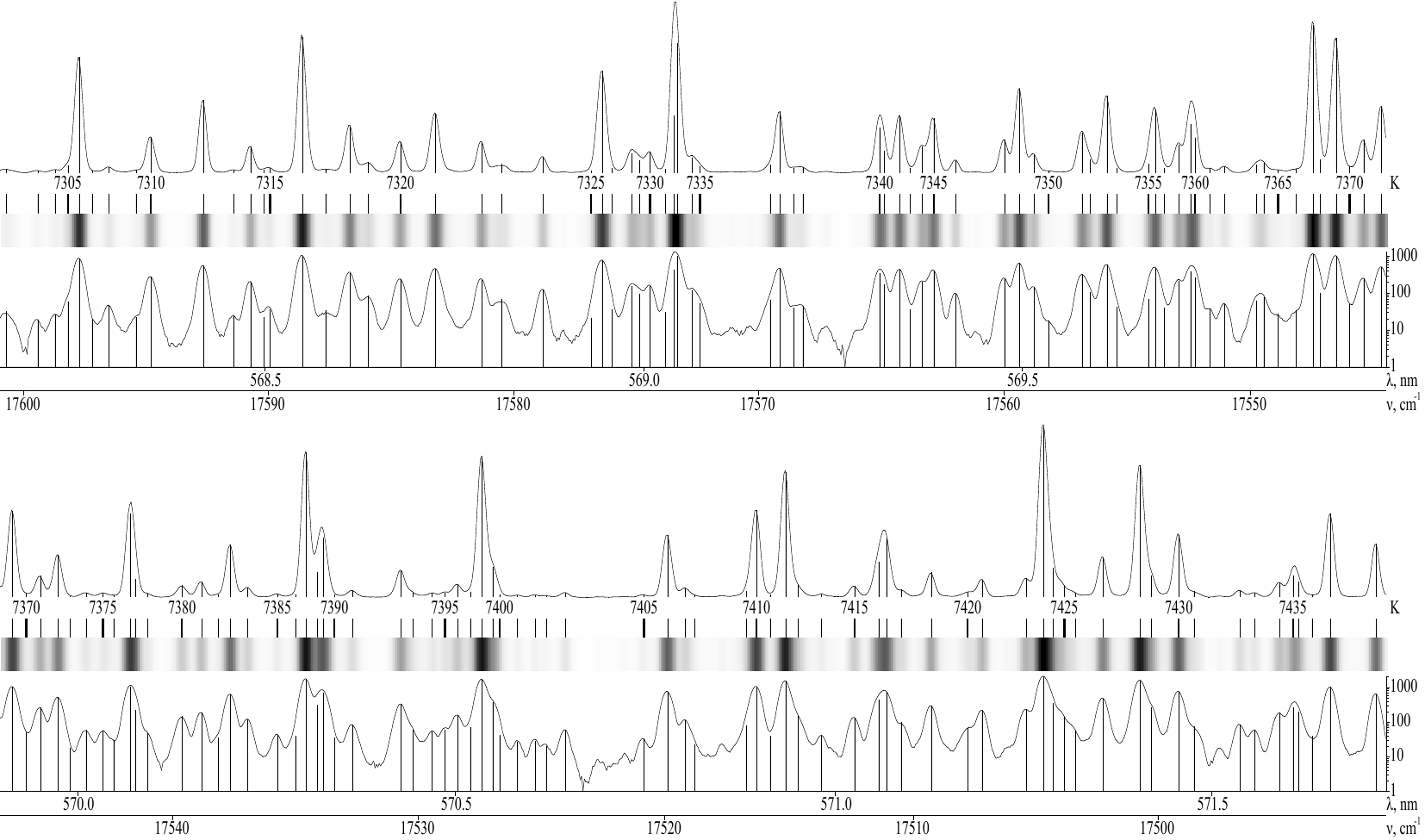}
\end{figure}

\newpage
\begin{figure}[!ht]
\includegraphics[angle=90, totalheight=0.9\textheight]{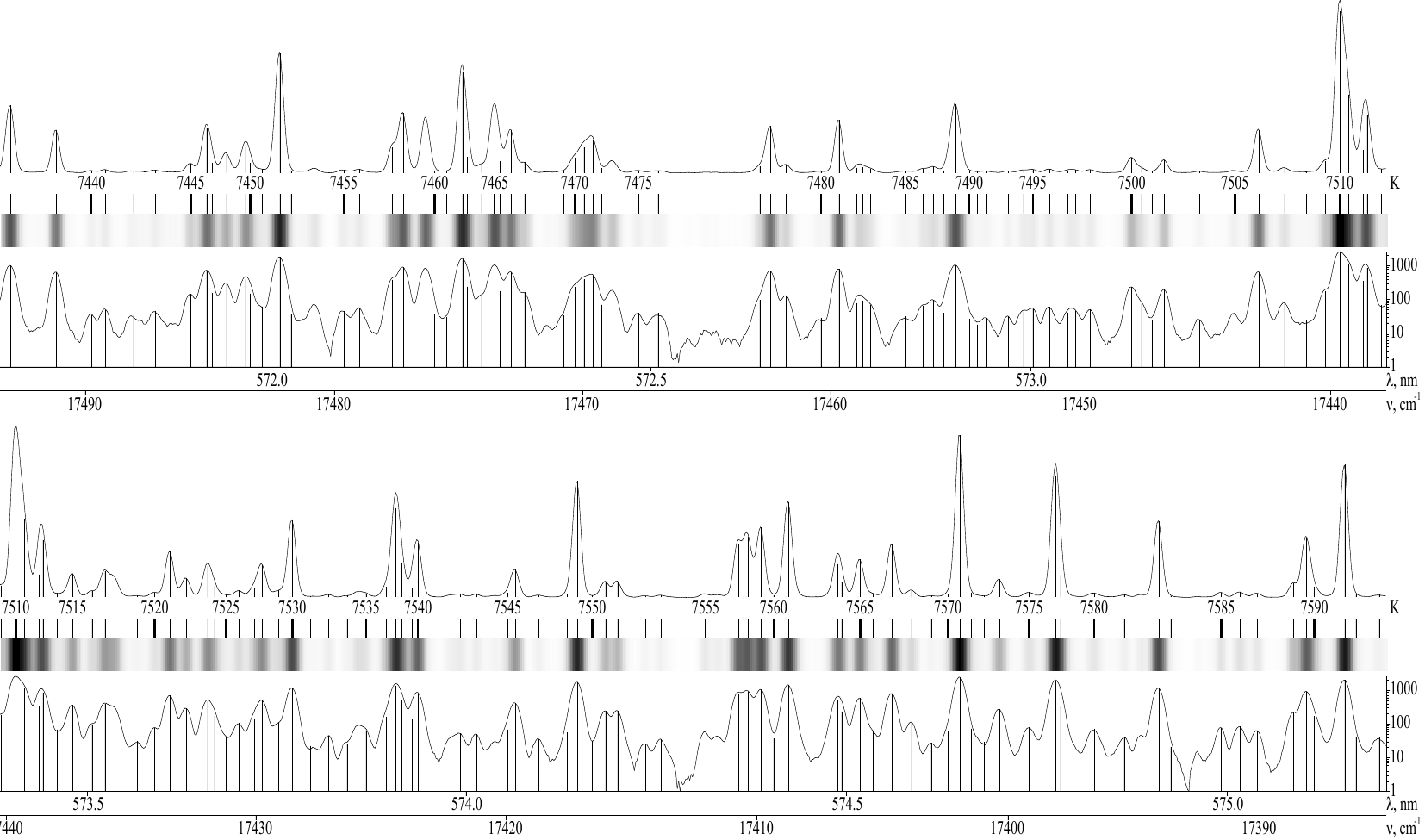}
\end{figure}

\newpage
\begin{figure}[!ht]
\includegraphics[angle=90, totalheight=0.9\textheight]{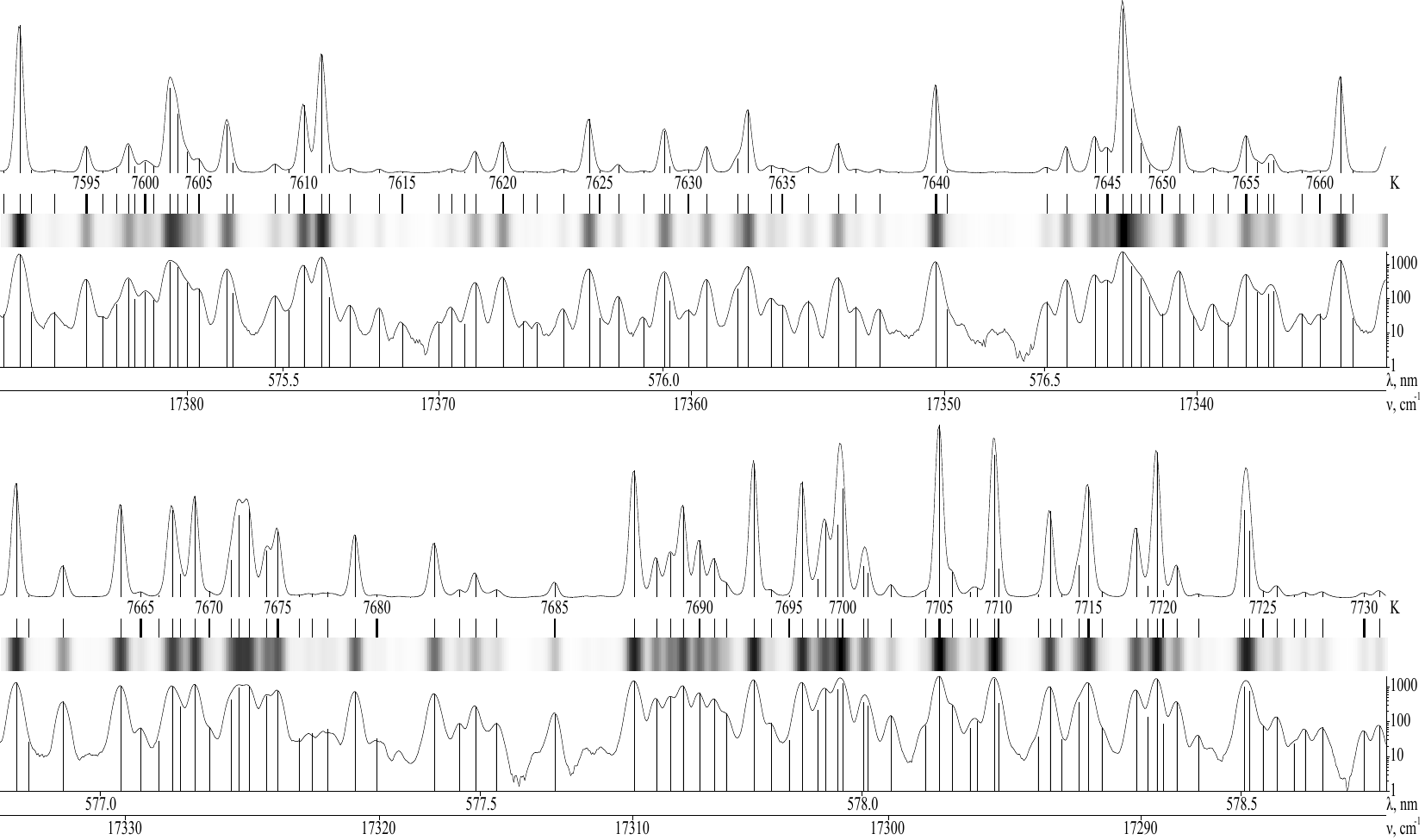}
\end{figure}

\newpage
\begin{figure}[!ht]
\includegraphics[angle=90, totalheight=0.9\textheight]{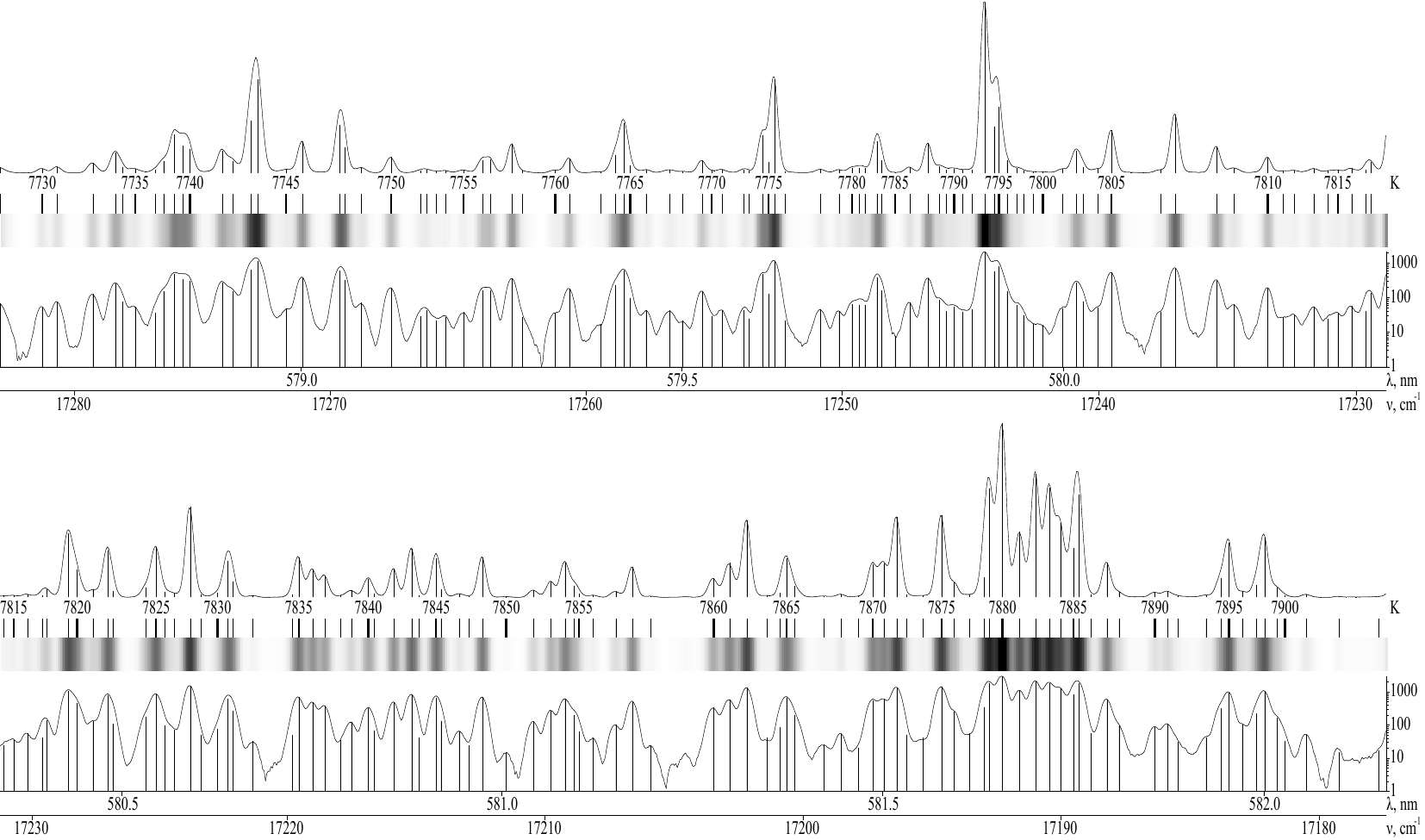}
\end{figure}

\newpage
\begin{figure}[!ht]
\includegraphics[angle=90, totalheight=0.9\textheight]{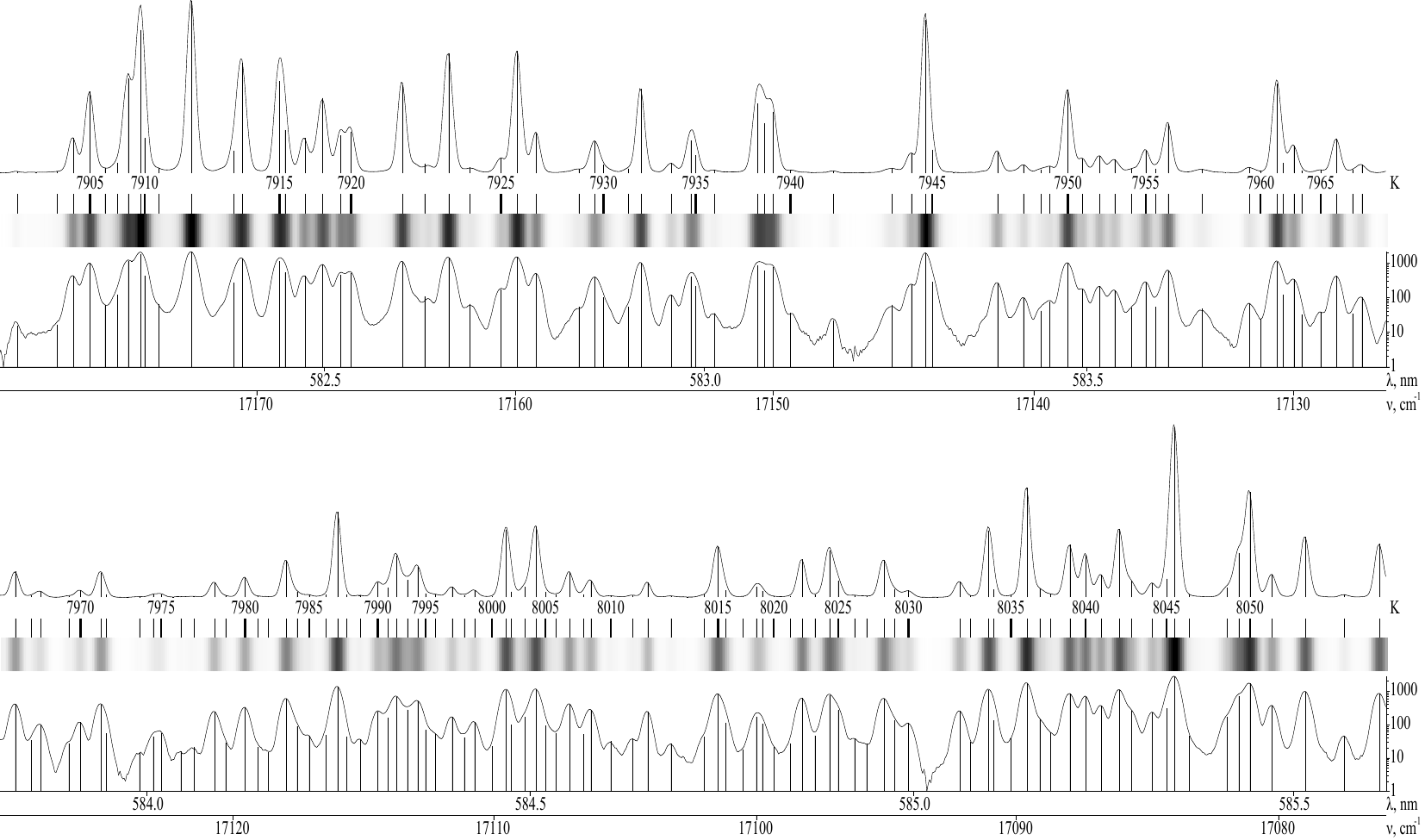}
\end{figure}

\newpage
\begin{figure}[!ht]
\includegraphics[angle=90, totalheight=0.9\textheight]{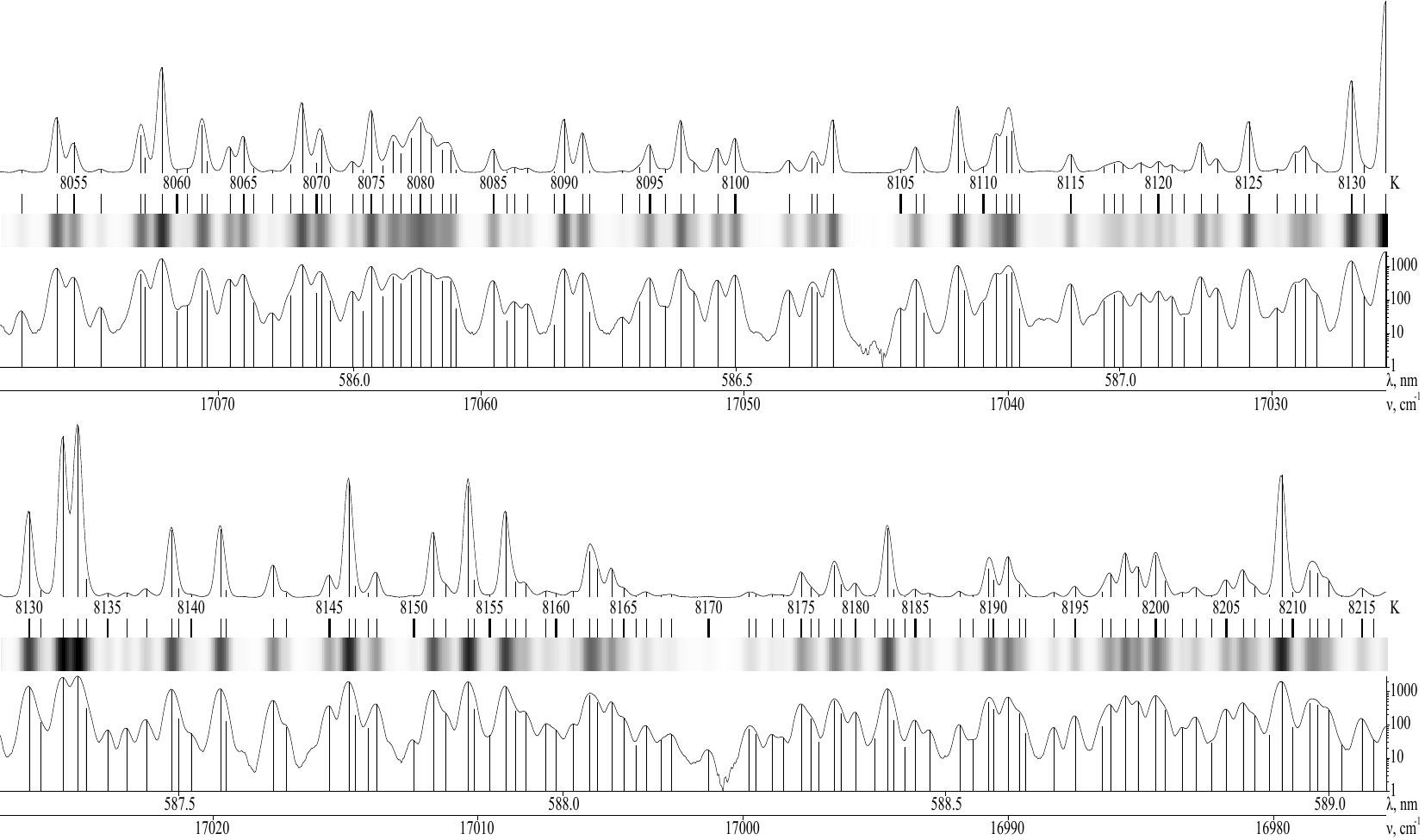}
\end{figure}

\newpage
\begin{figure}[!ht]
\includegraphics[angle=90, totalheight=0.9\textheight]{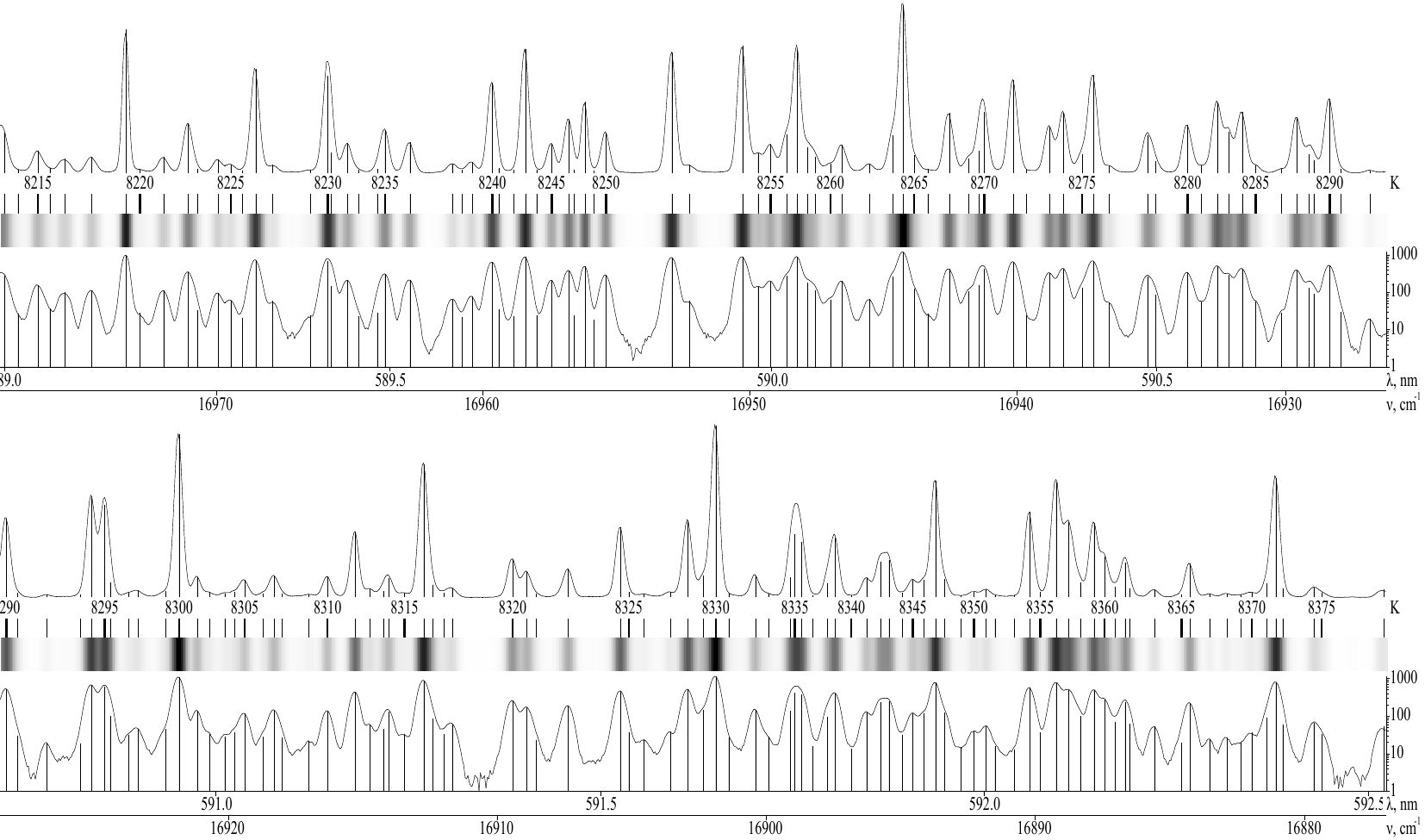}
\end{figure}

\newpage
\begin{figure}[!ht]
\includegraphics[angle=90, totalheight=0.9\textheight]{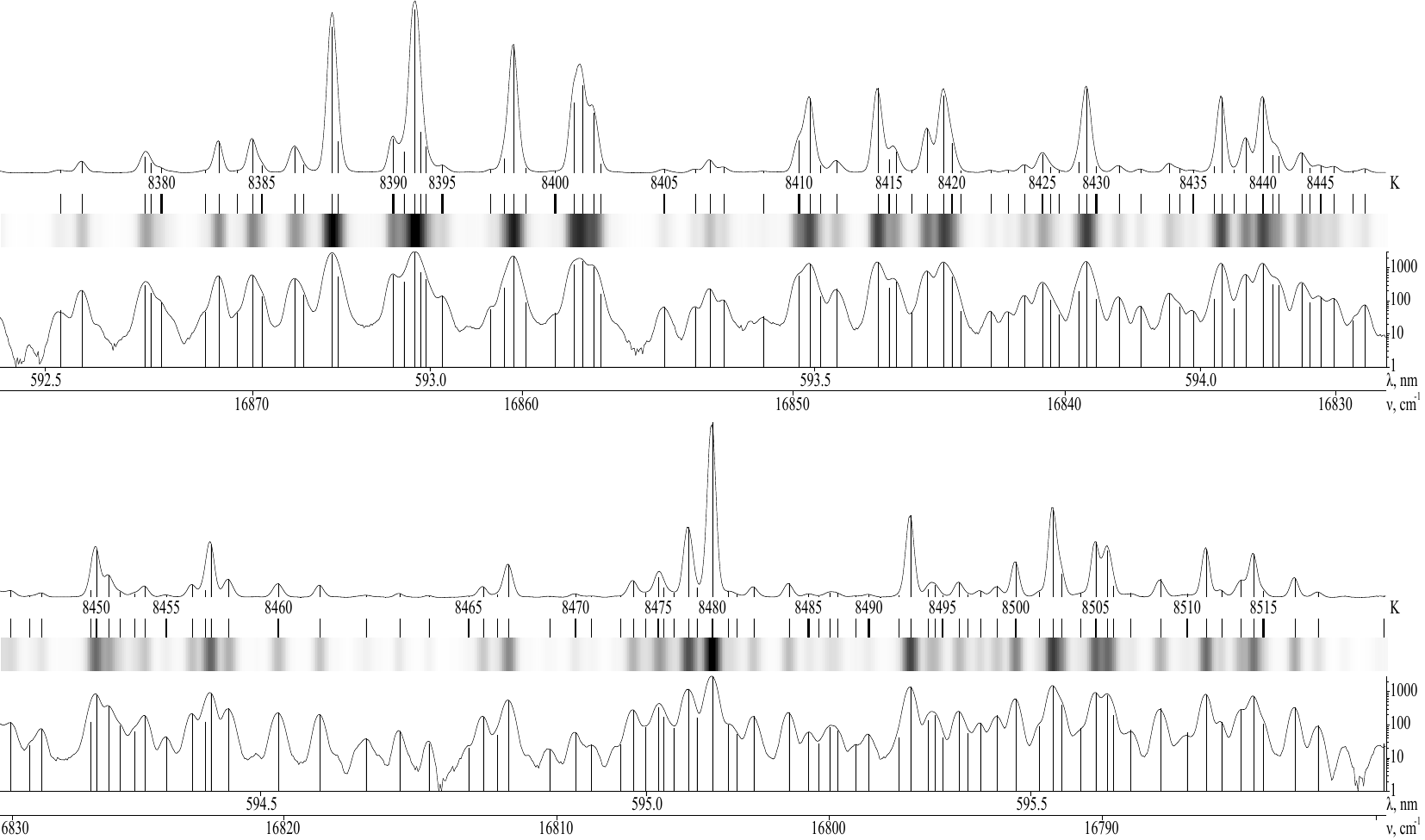}
\end{figure}

\newpage
\begin{figure}[!ht]
\includegraphics[angle=90, totalheight=0.9\textheight]{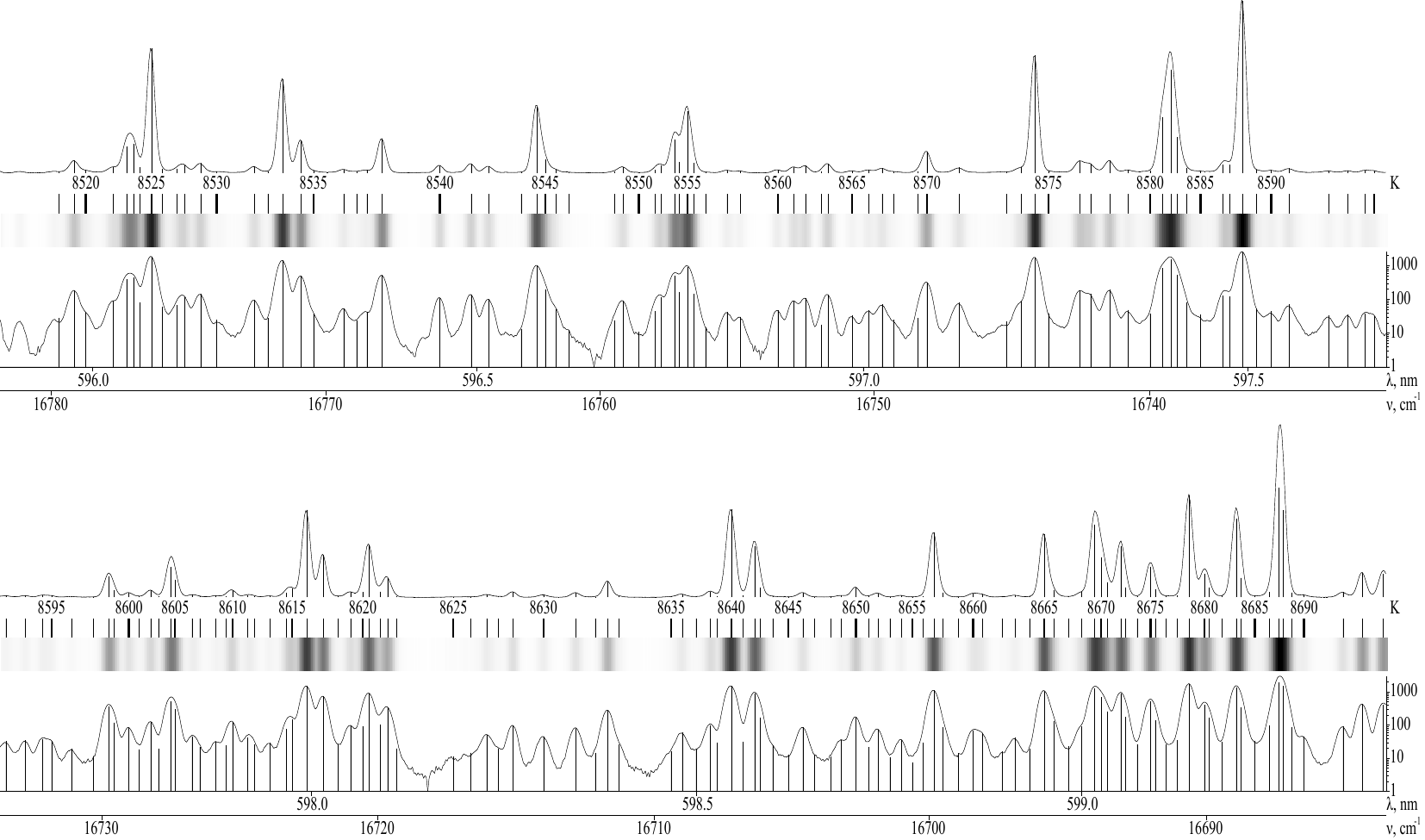}
\end{figure}

\newpage
\begin{figure}[!ht]
\includegraphics[angle=90, totalheight=0.9\textheight]{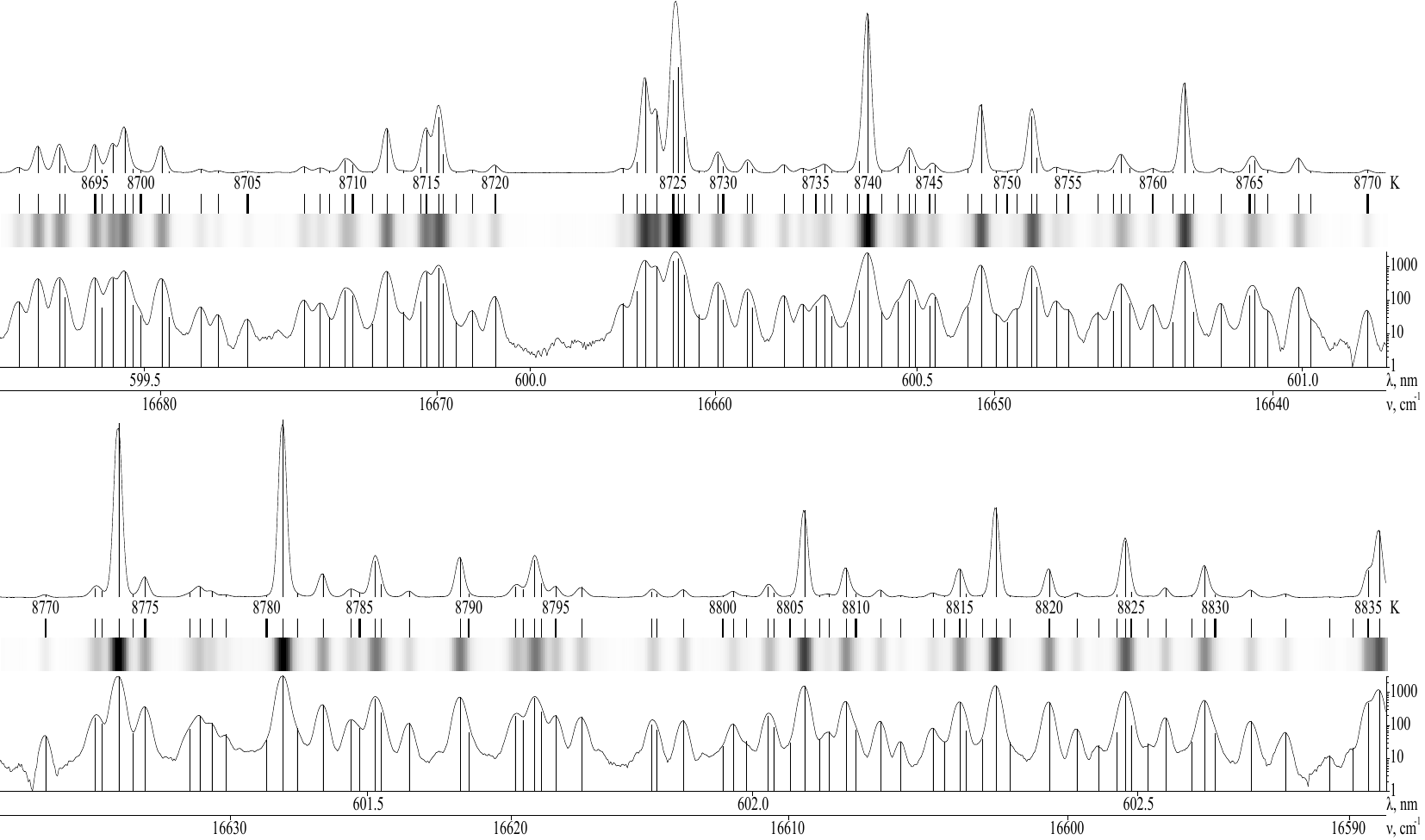}
\end{figure}

\newpage
\begin{figure}[!ht]
\includegraphics[angle=90, totalheight=0.9\textheight]{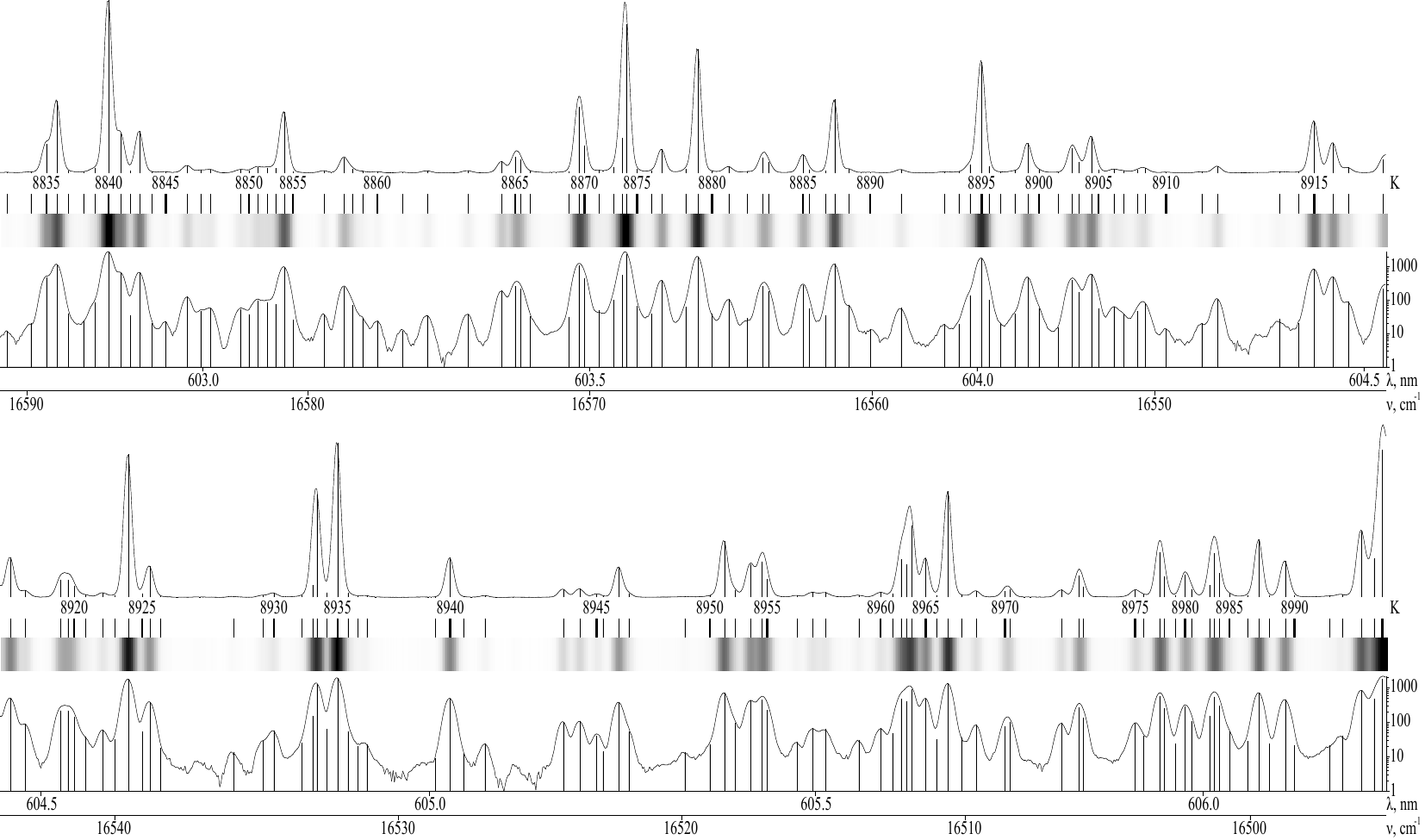}
\end{figure}

\newpage
\begin{figure}[!ht]
\includegraphics[angle=90, totalheight=0.9\textheight]{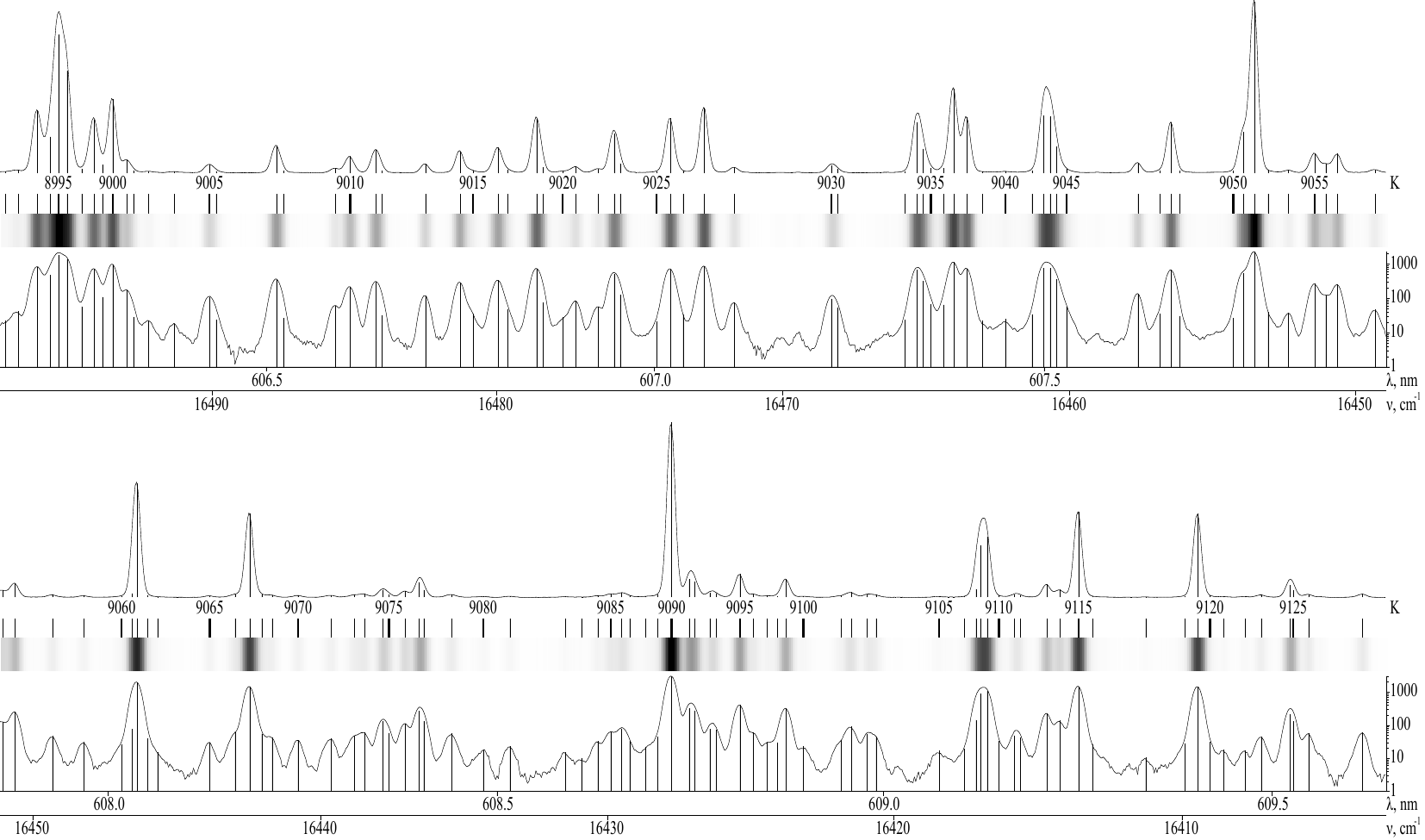}
\end{figure}

\newpage
\begin{figure}[!ht]
\includegraphics[angle=90, totalheight=0.9\textheight]{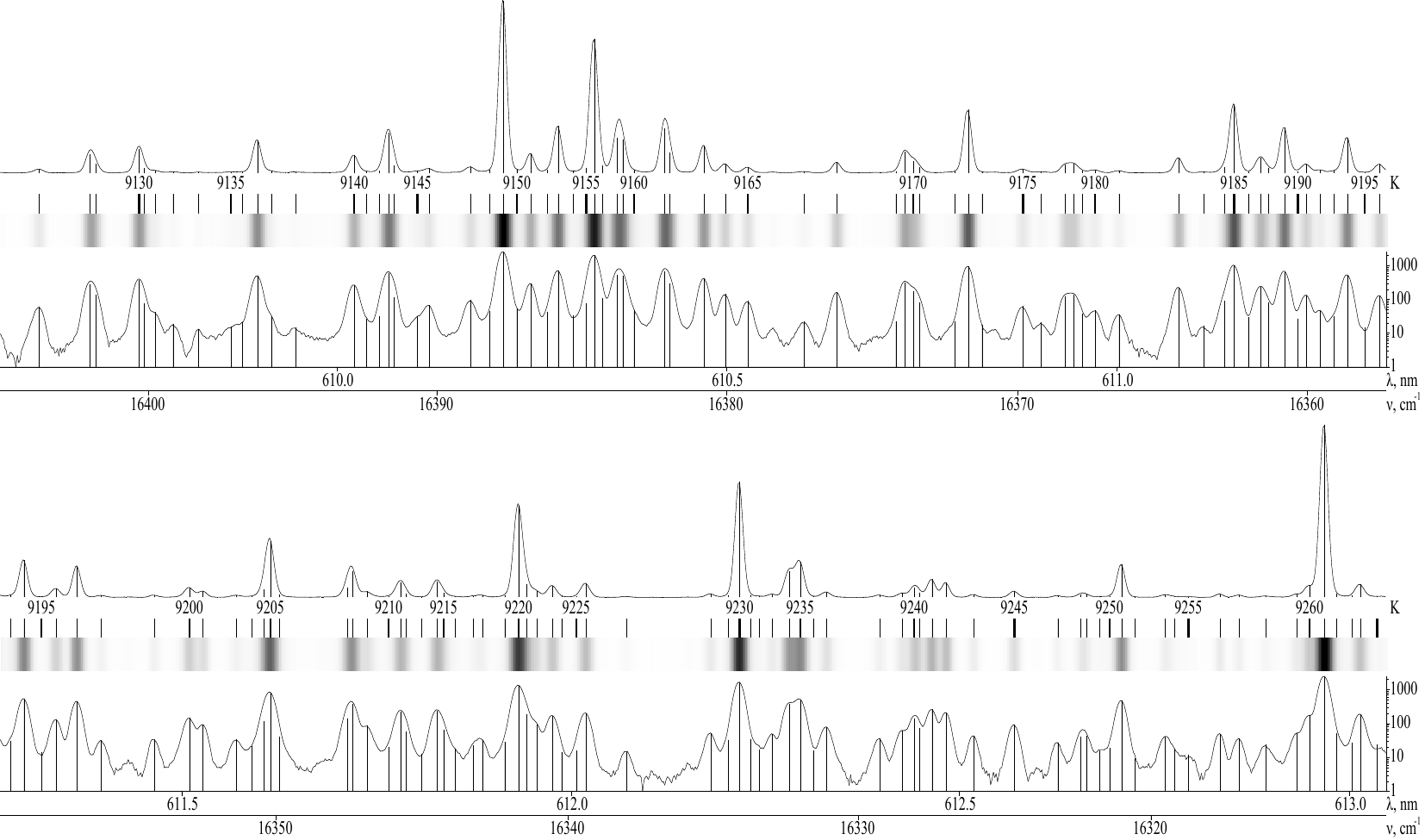}
\end{figure}

\newpage
\begin{figure}[!ht]
\includegraphics[angle=90, totalheight=0.9\textheight]{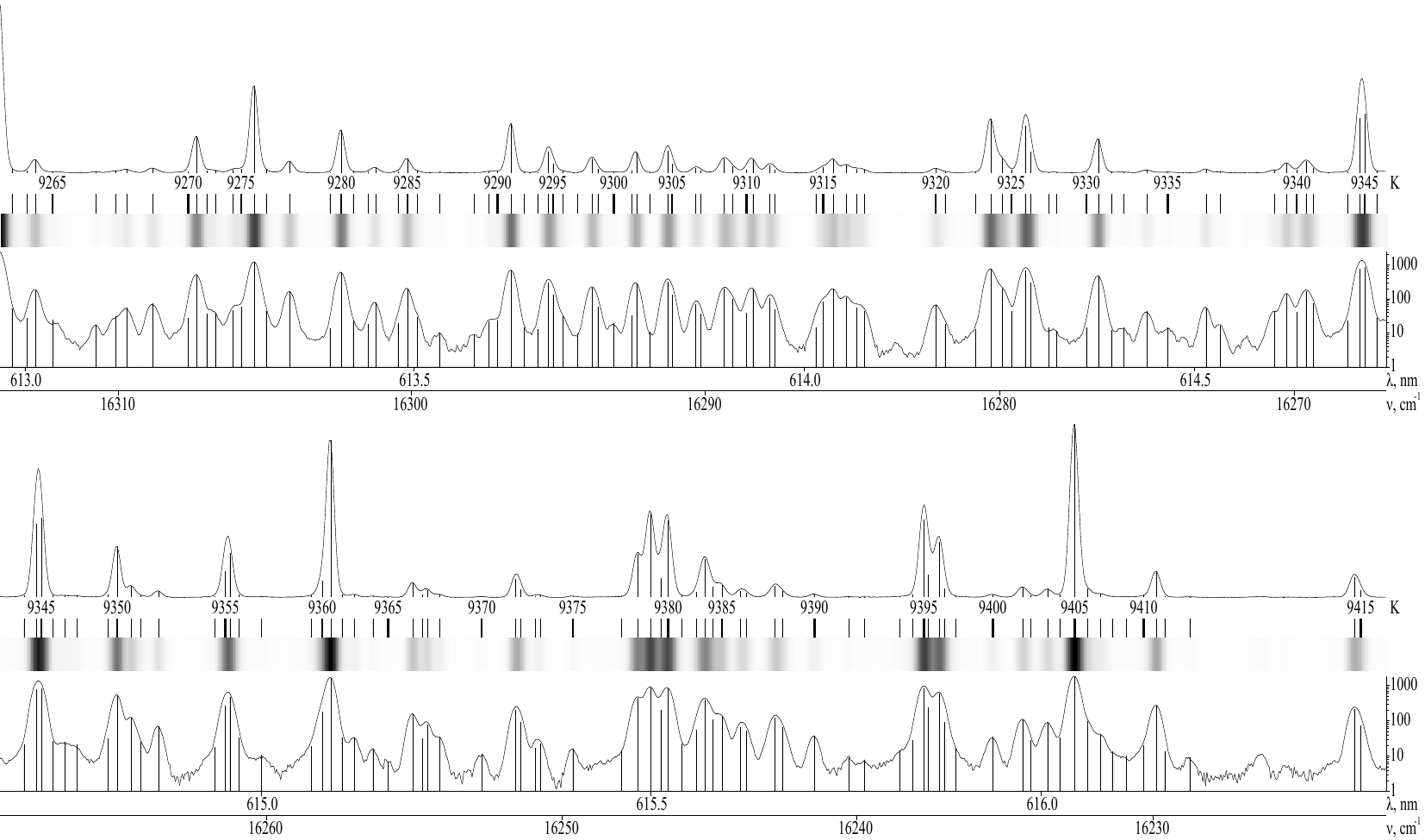}
\end{figure}

\newpage
\begin{figure}[!ht]
\includegraphics[angle=90, totalheight=0.9\textheight]{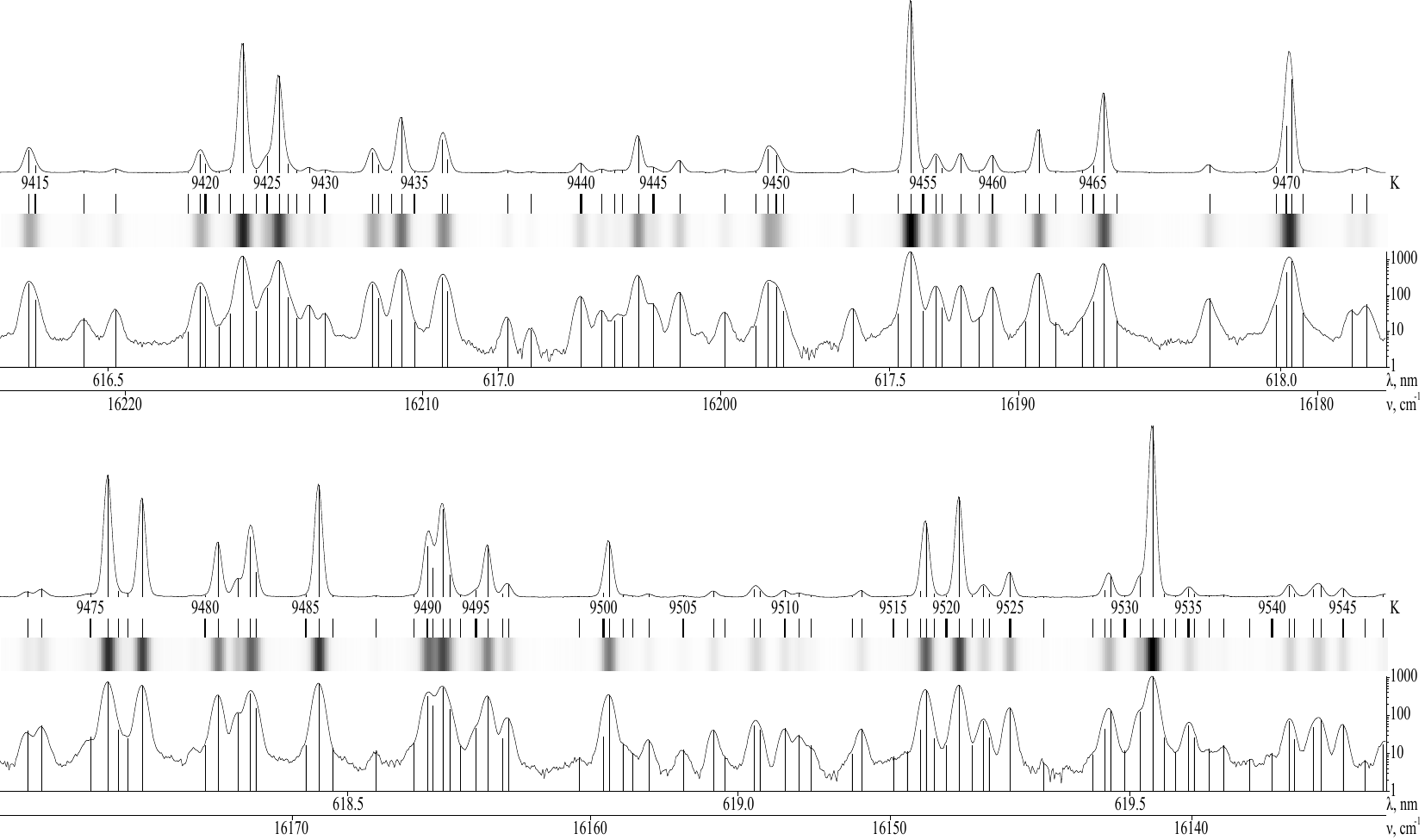}
\end{figure}

\newpage
\begin{figure}[!ht]
\includegraphics[angle=90, totalheight=0.9\textheight]{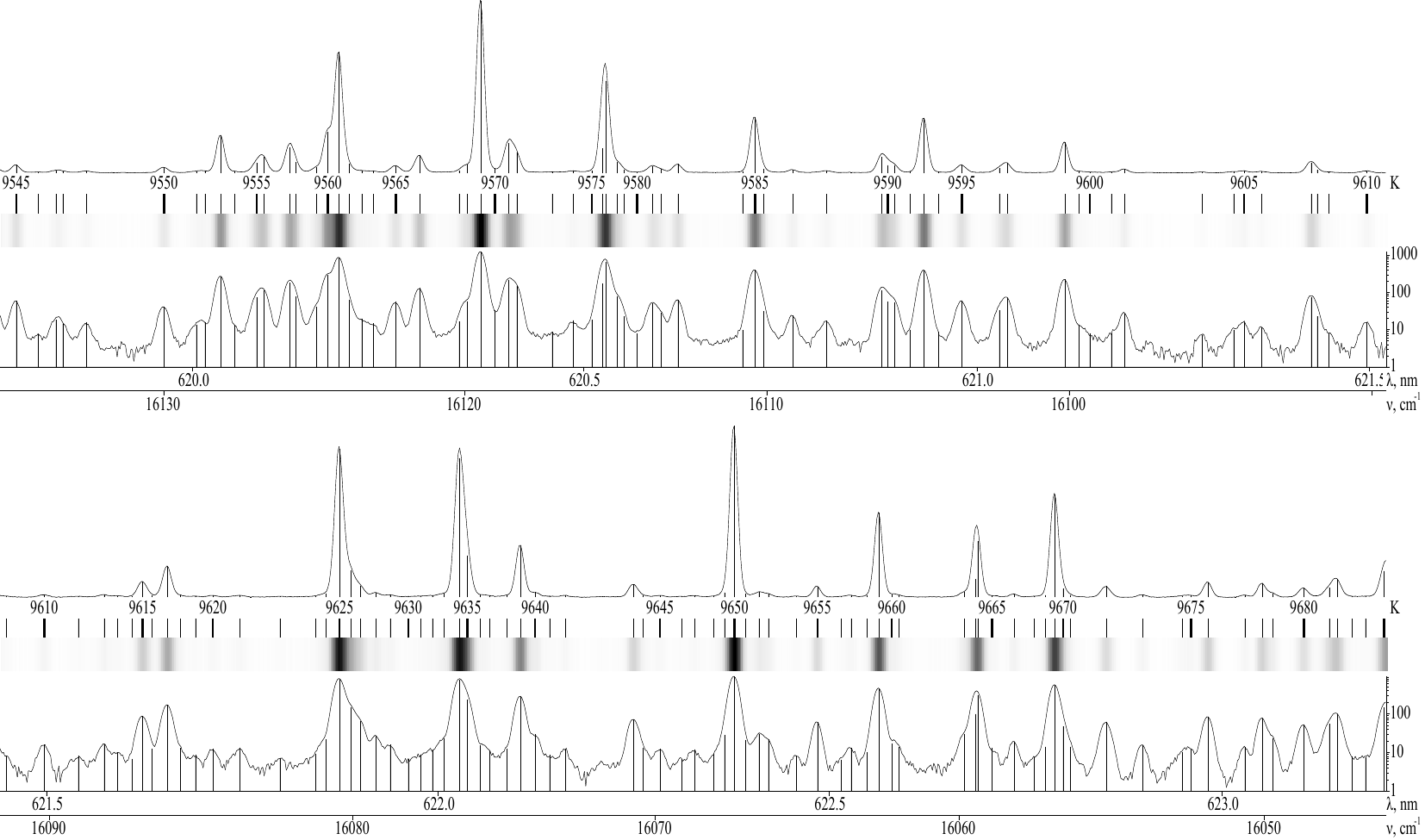}
\end{figure}

\newpage
\begin{figure}[!ht]
\includegraphics[angle=90, totalheight=0.9\textheight]{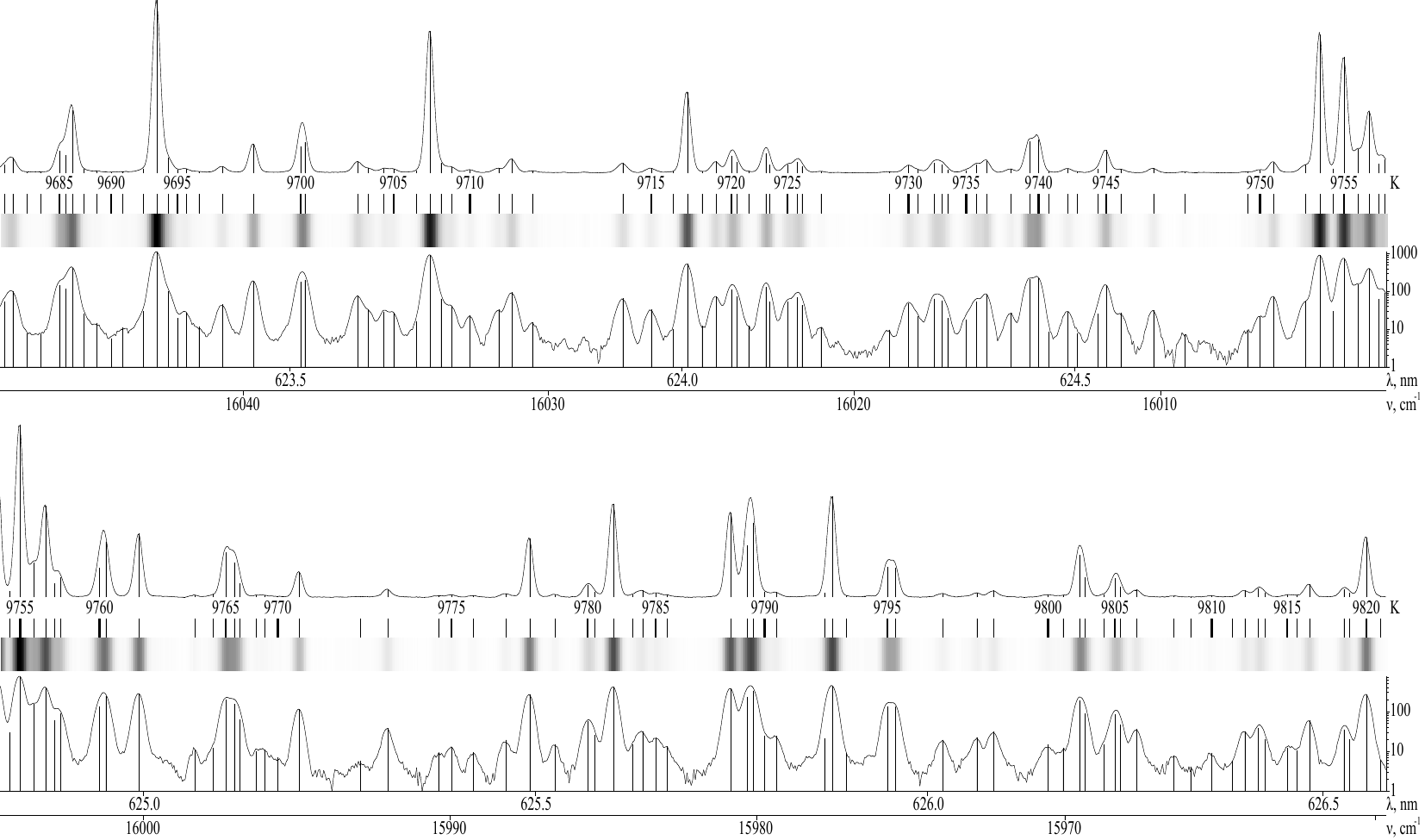}
\end{figure}

\newpage
\begin{figure}[!ht]
\includegraphics[angle=90, totalheight=0.9\textheight]{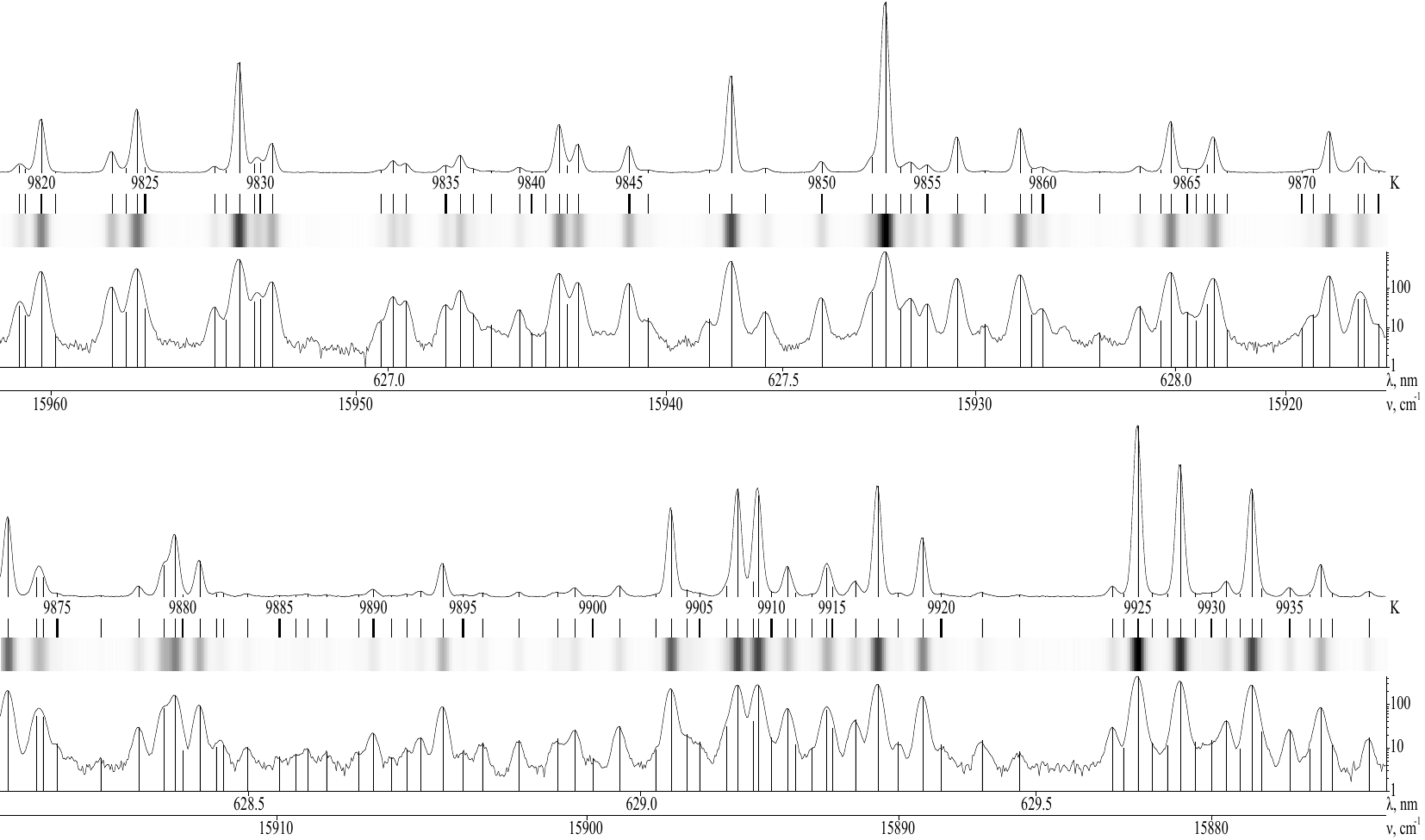}
\end{figure}

\newpage
\begin{figure}[!ht]
\includegraphics[angle=90, totalheight=0.9\textheight]{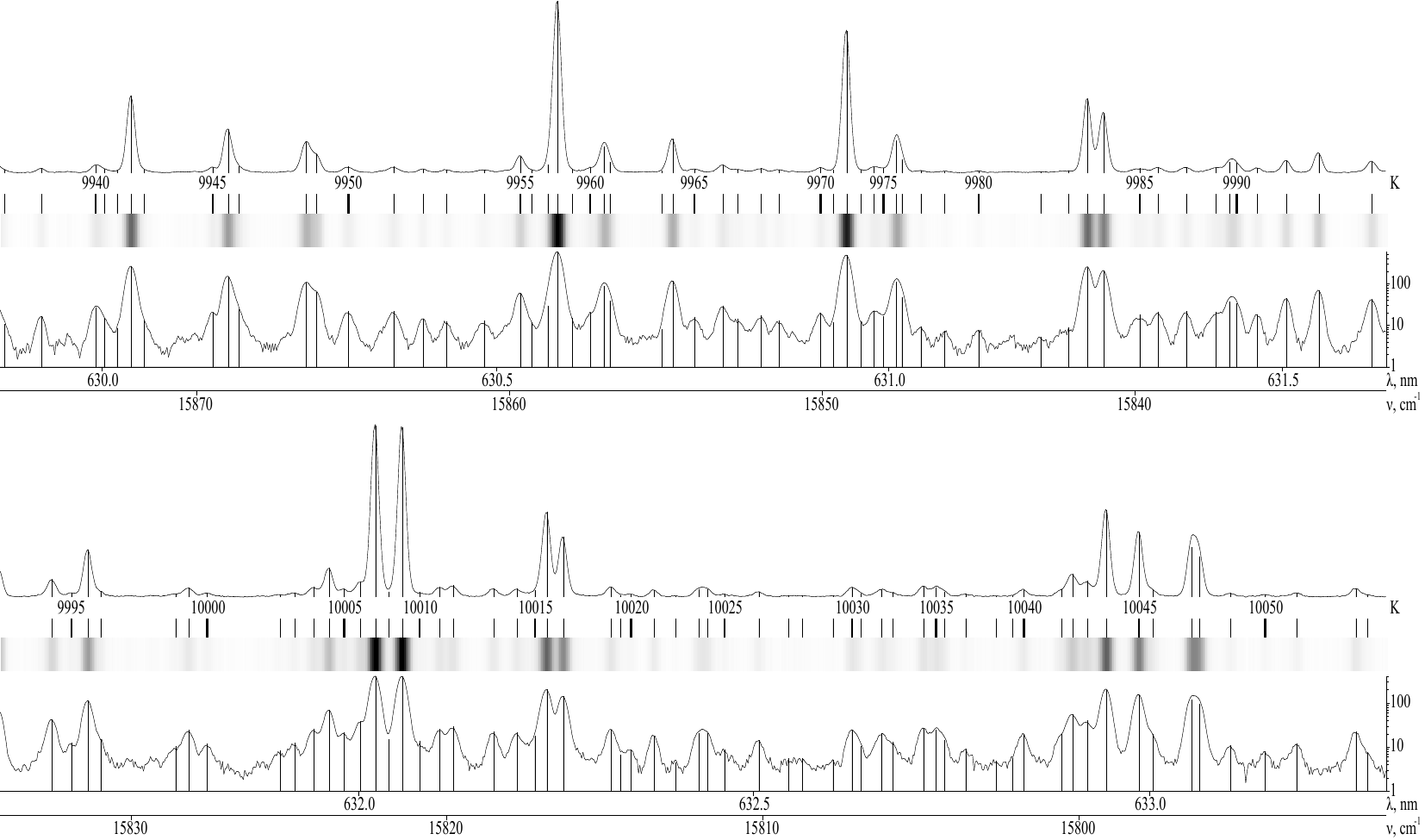}
\end{figure}

\newpage
\begin{figure}[!ht]
\includegraphics[angle=90, totalheight=0.9\textheight]{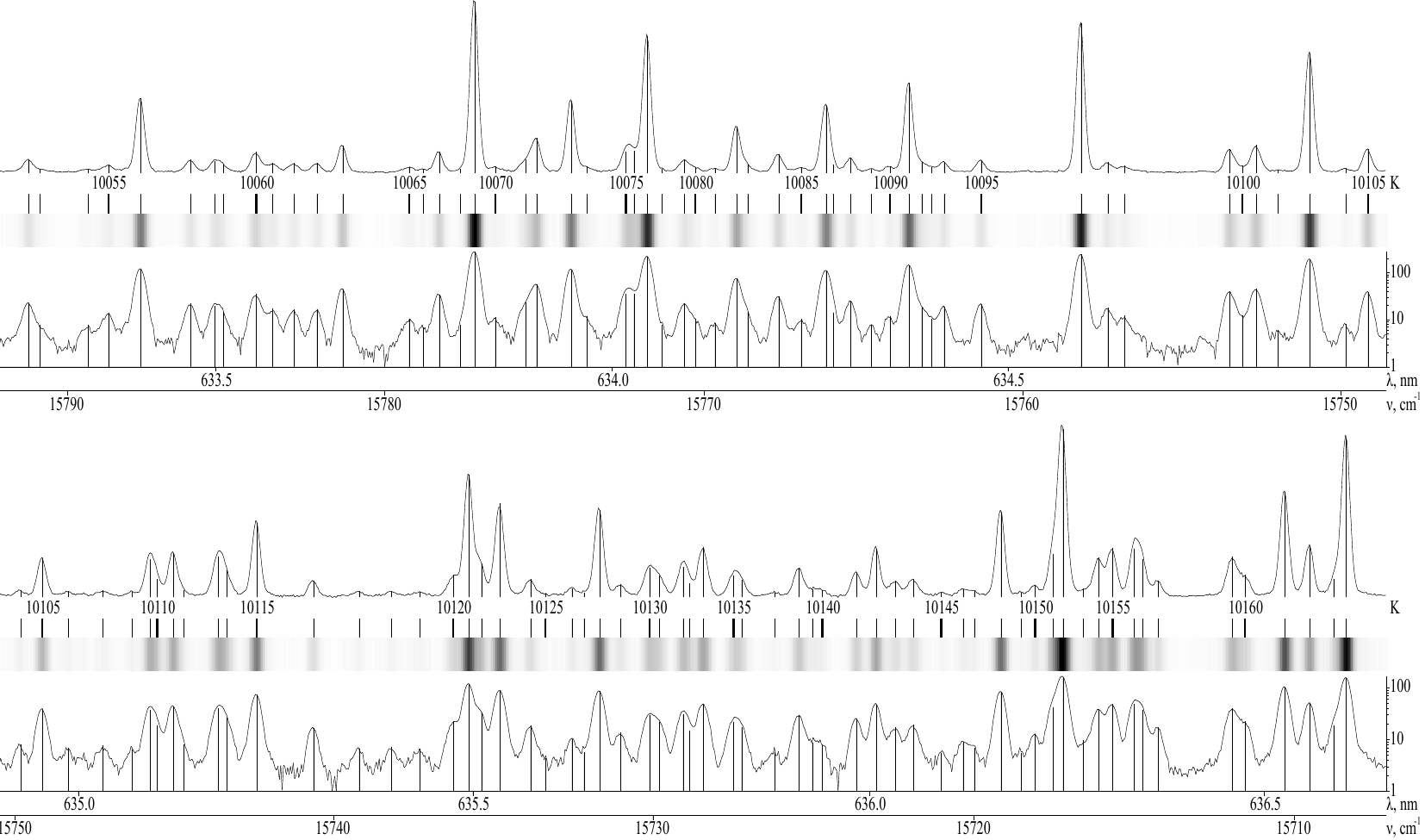}
\end{figure}

\newpage
\begin{figure}[!ht]
\includegraphics[angle=90, totalheight=0.9\textheight]{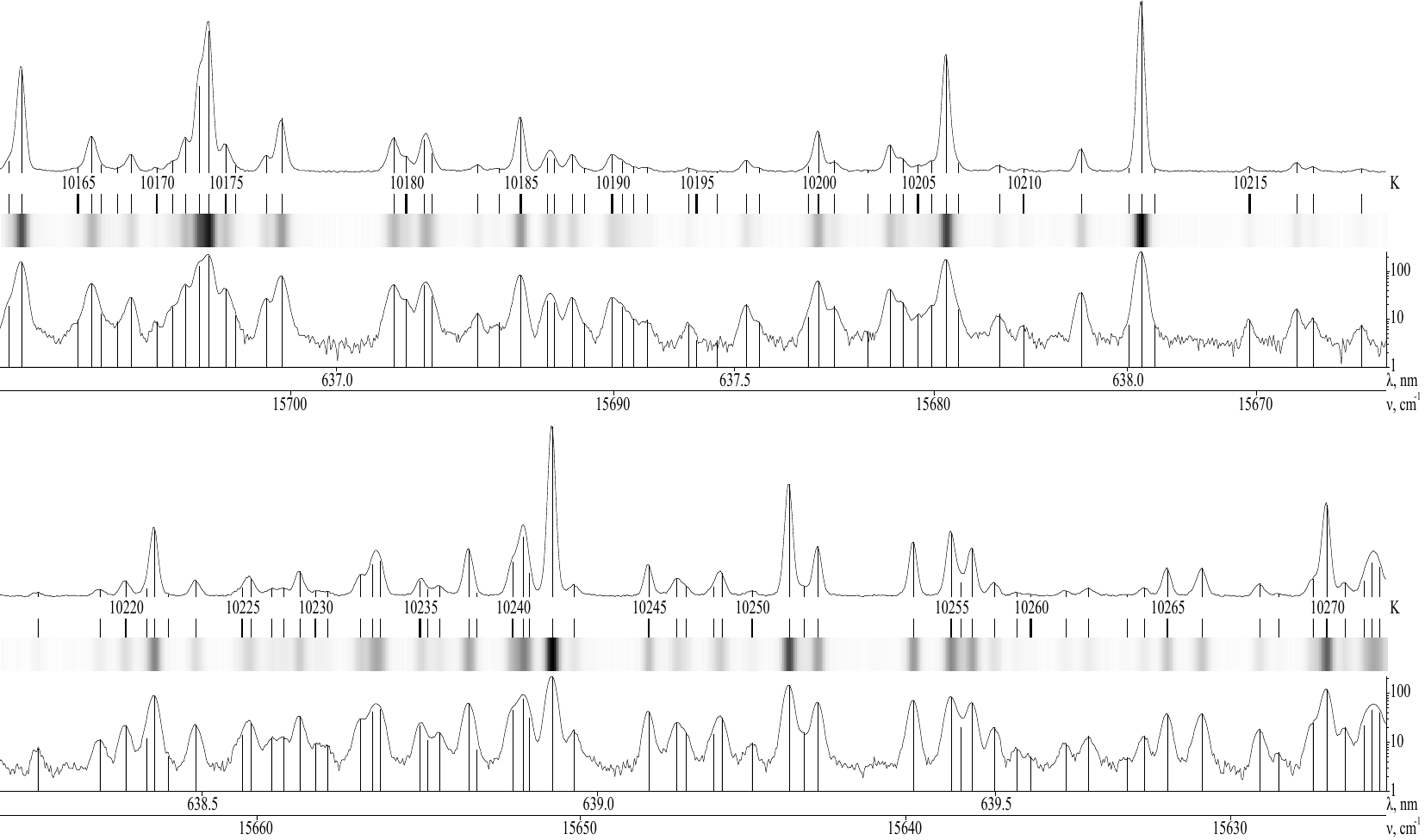}
\end{figure}

\newpage
\begin{figure}[!ht]
\includegraphics[angle=90, totalheight=0.9\textheight]{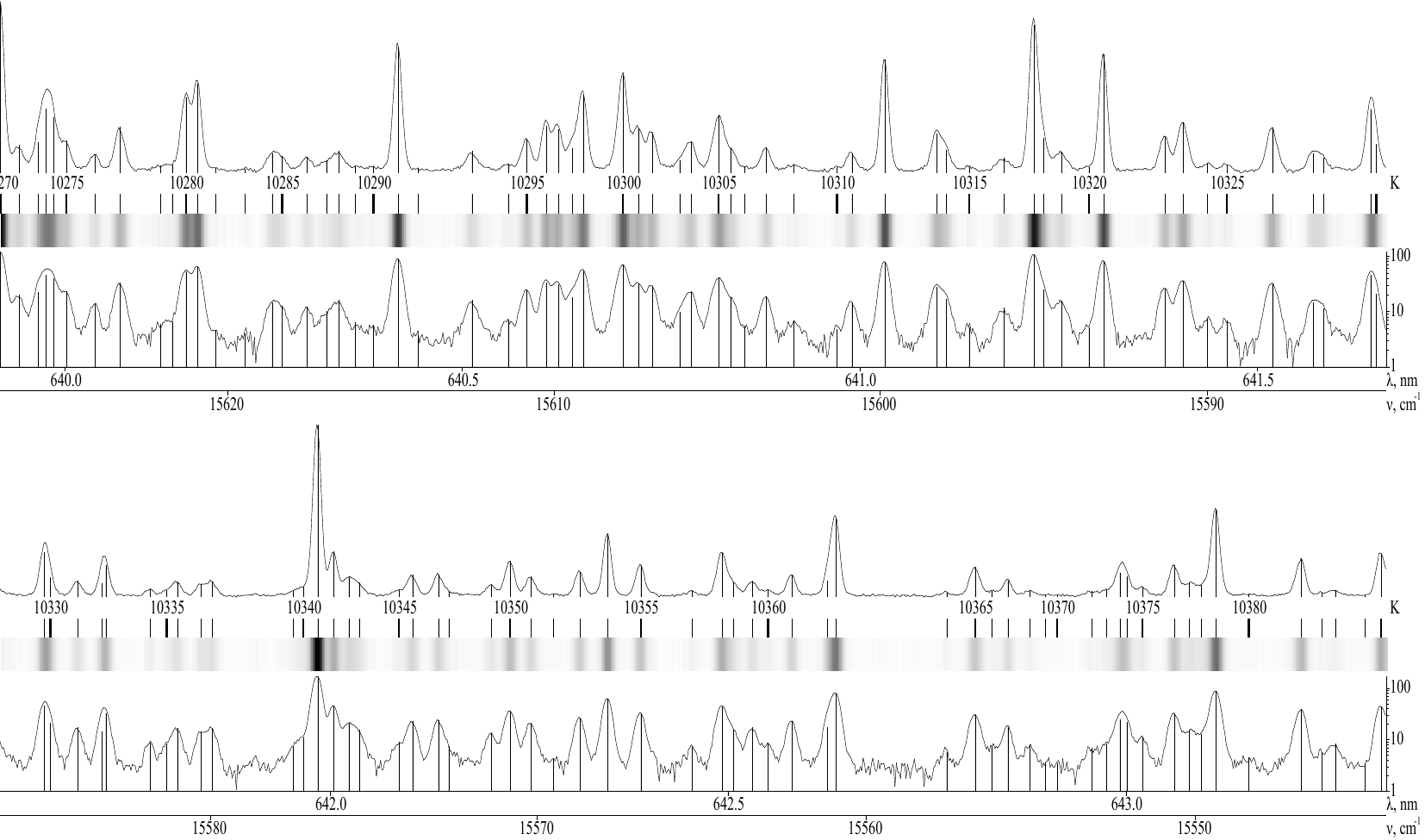}
\end{figure}

\newpage
\begin{figure}[!ht]
\includegraphics[angle=90, totalheight=0.9\textheight]{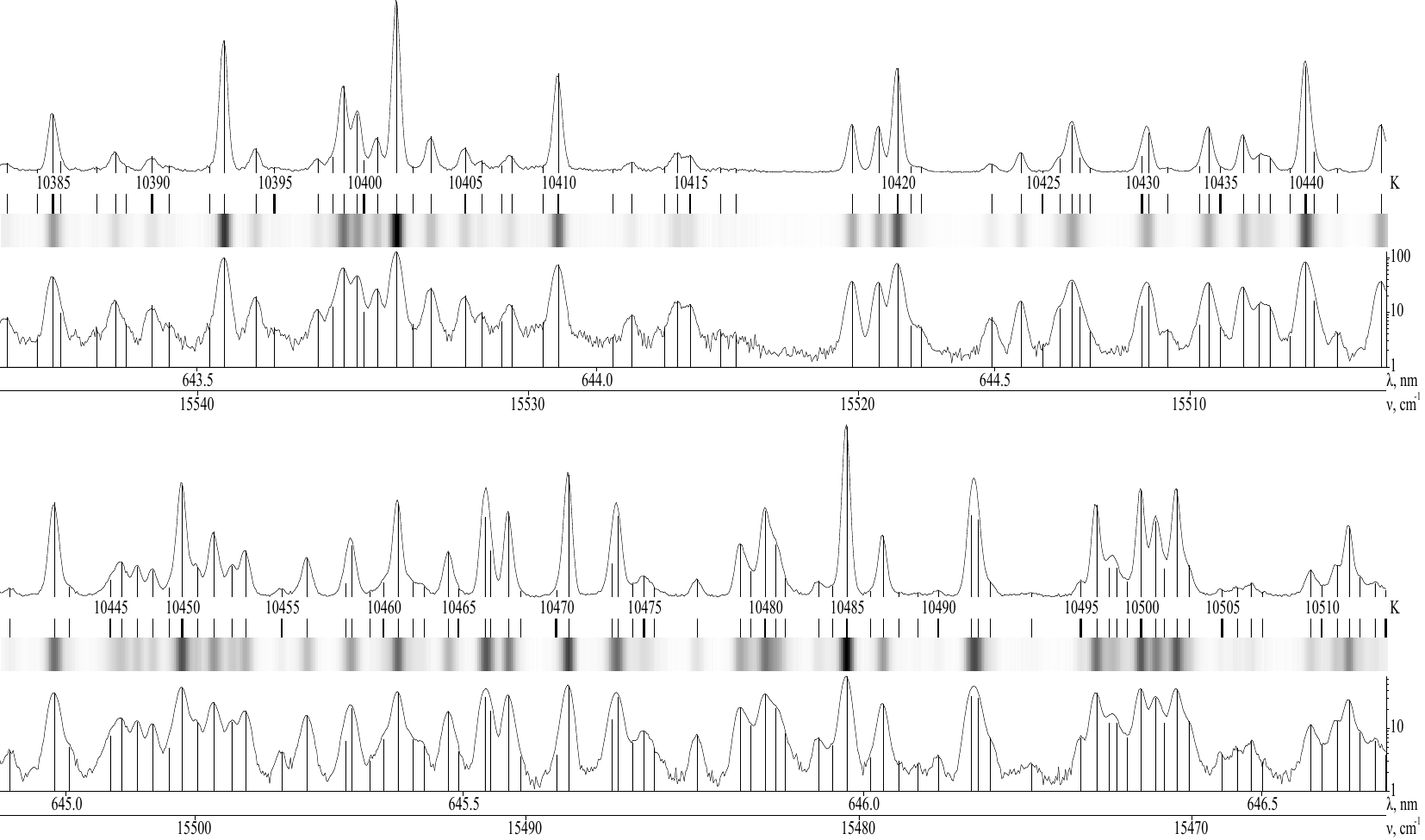}
\end{figure}

\newpage
\begin{figure}[!ht]
\includegraphics[angle=90, totalheight=0.9\textheight]{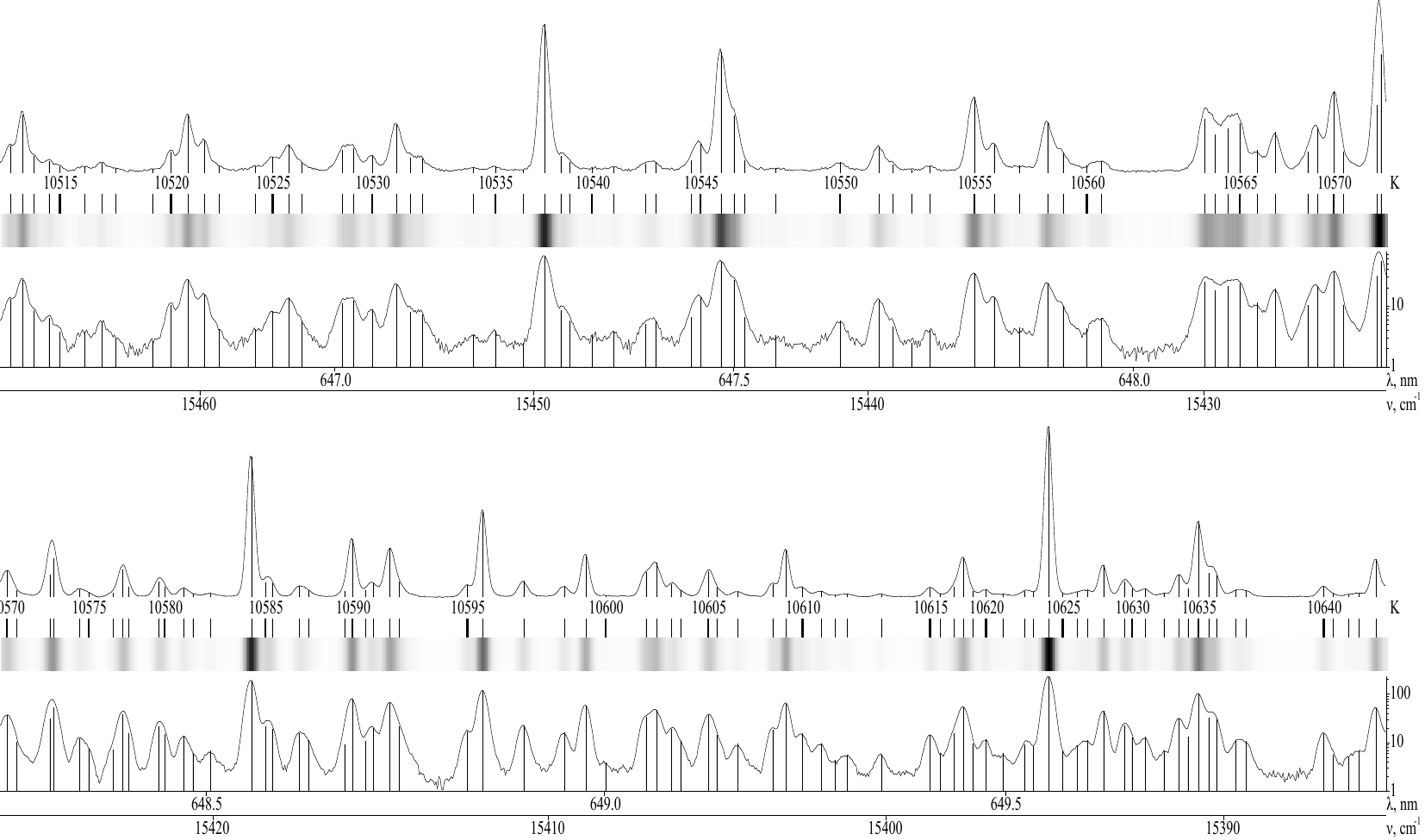}
\end{figure}

\newpage
\begin{figure}[!ht]
\includegraphics[angle=90, totalheight=0.9\textheight]{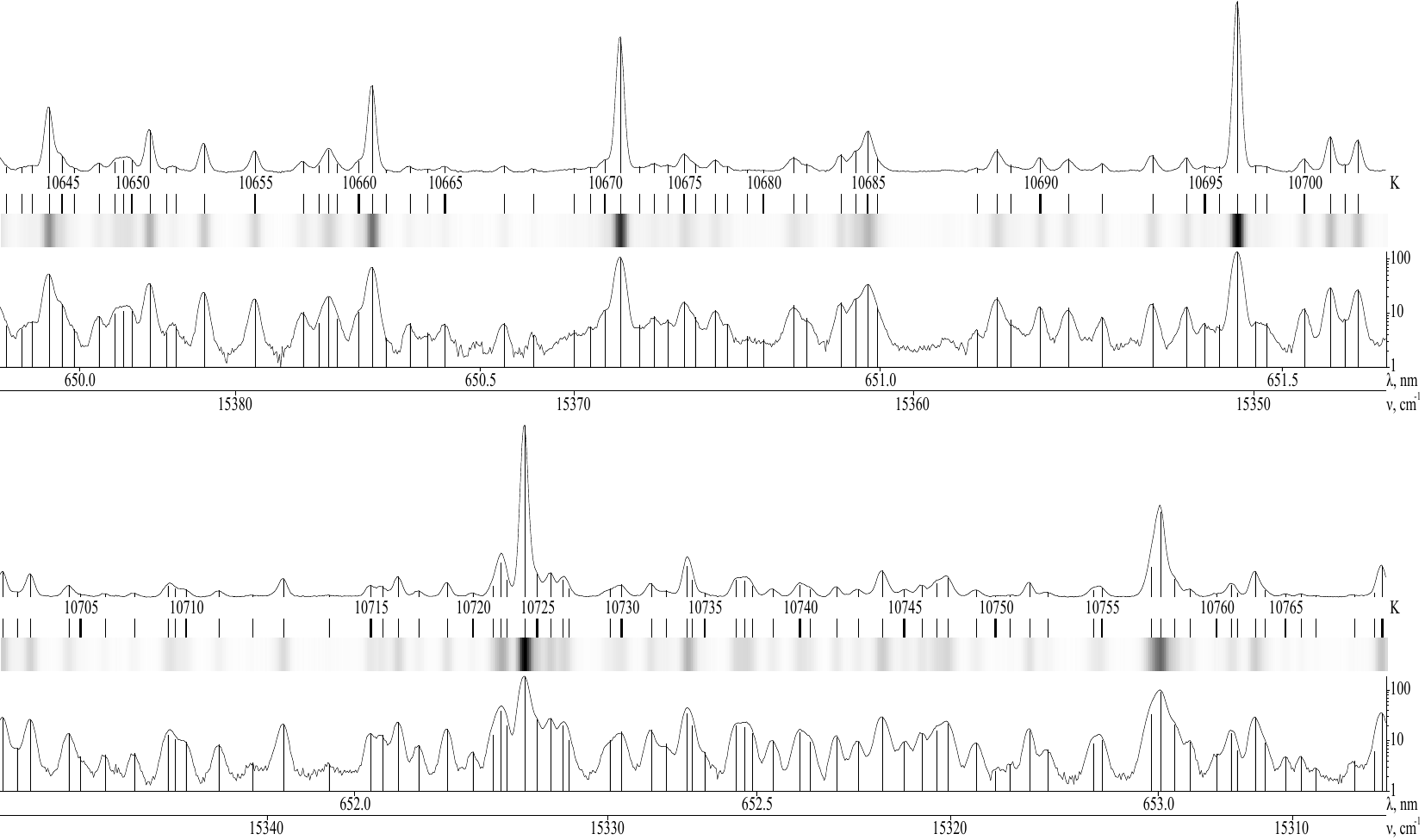}
\end{figure}

\newpage
\begin{figure}[!ht]
\includegraphics[angle=90, totalheight=0.9\textheight]{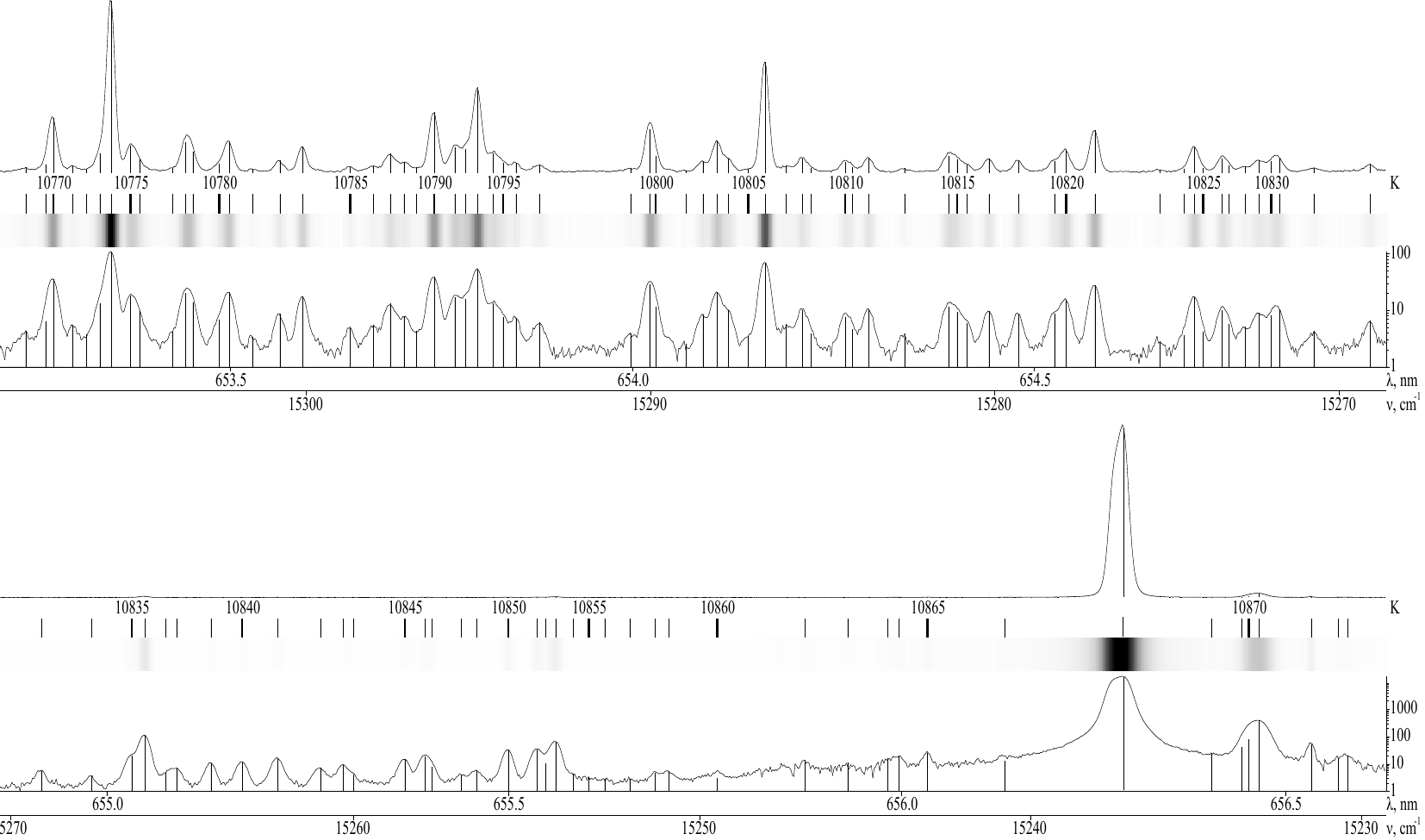}
\end{figure}

\newpage
\begin{figure}[!ht]
\includegraphics[angle=90, totalheight=0.9\textheight]{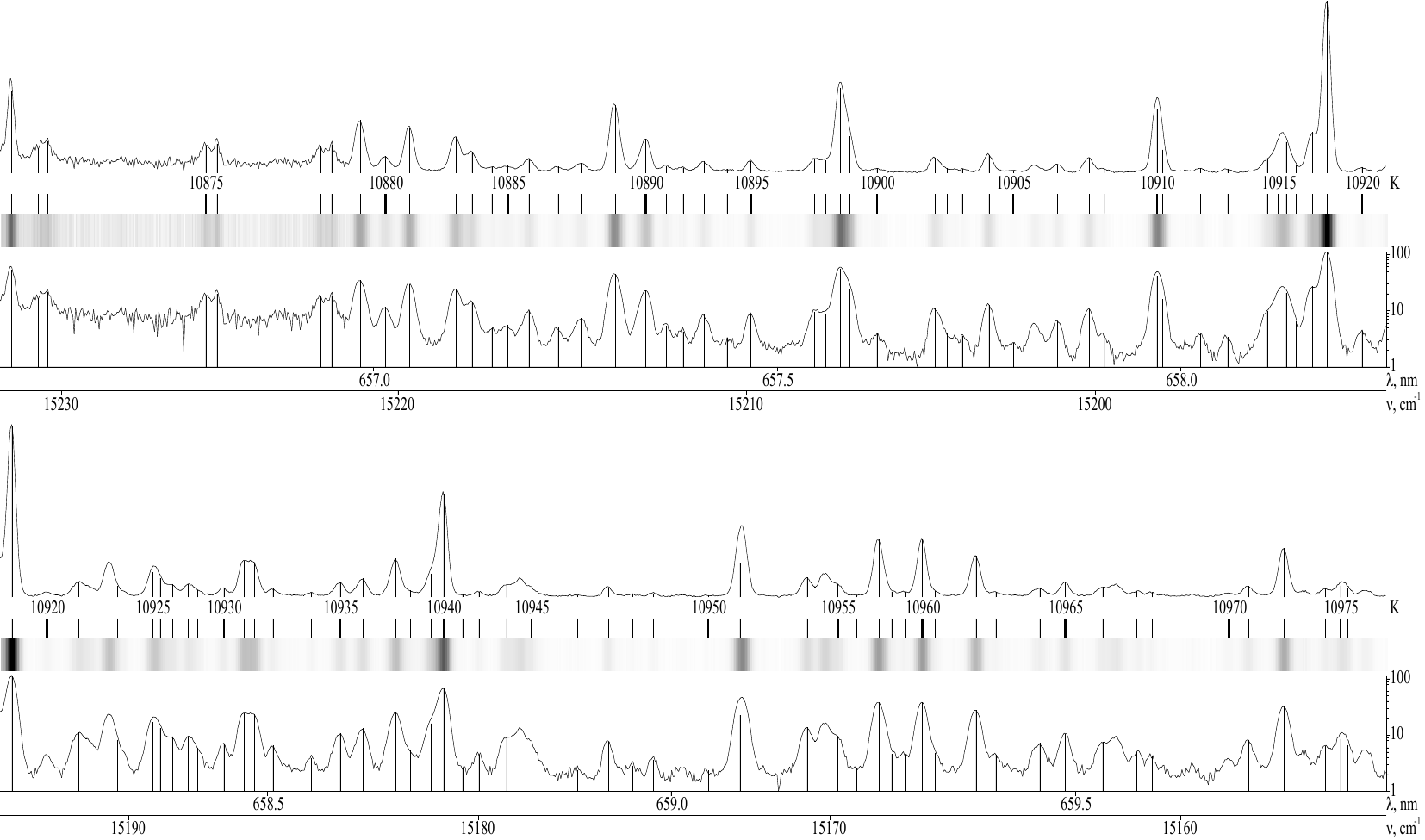}
\end{figure}

\newpage
\begin{figure}[!ht]
\includegraphics[angle=90, totalheight=0.9\textheight]{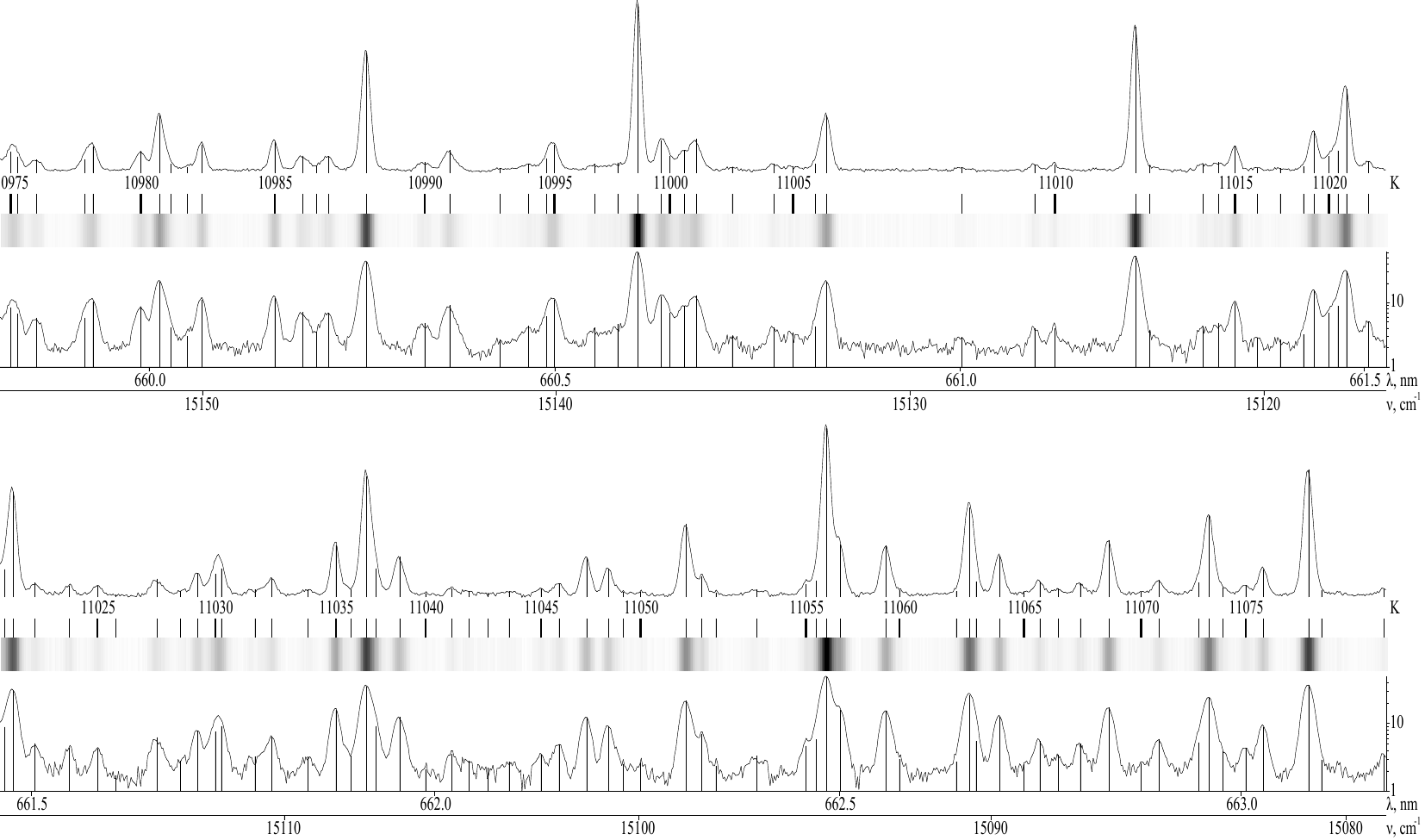}
\end{figure}

\newpage
\begin{figure}[!ht]
\includegraphics[angle=90, totalheight=0.9\textheight]{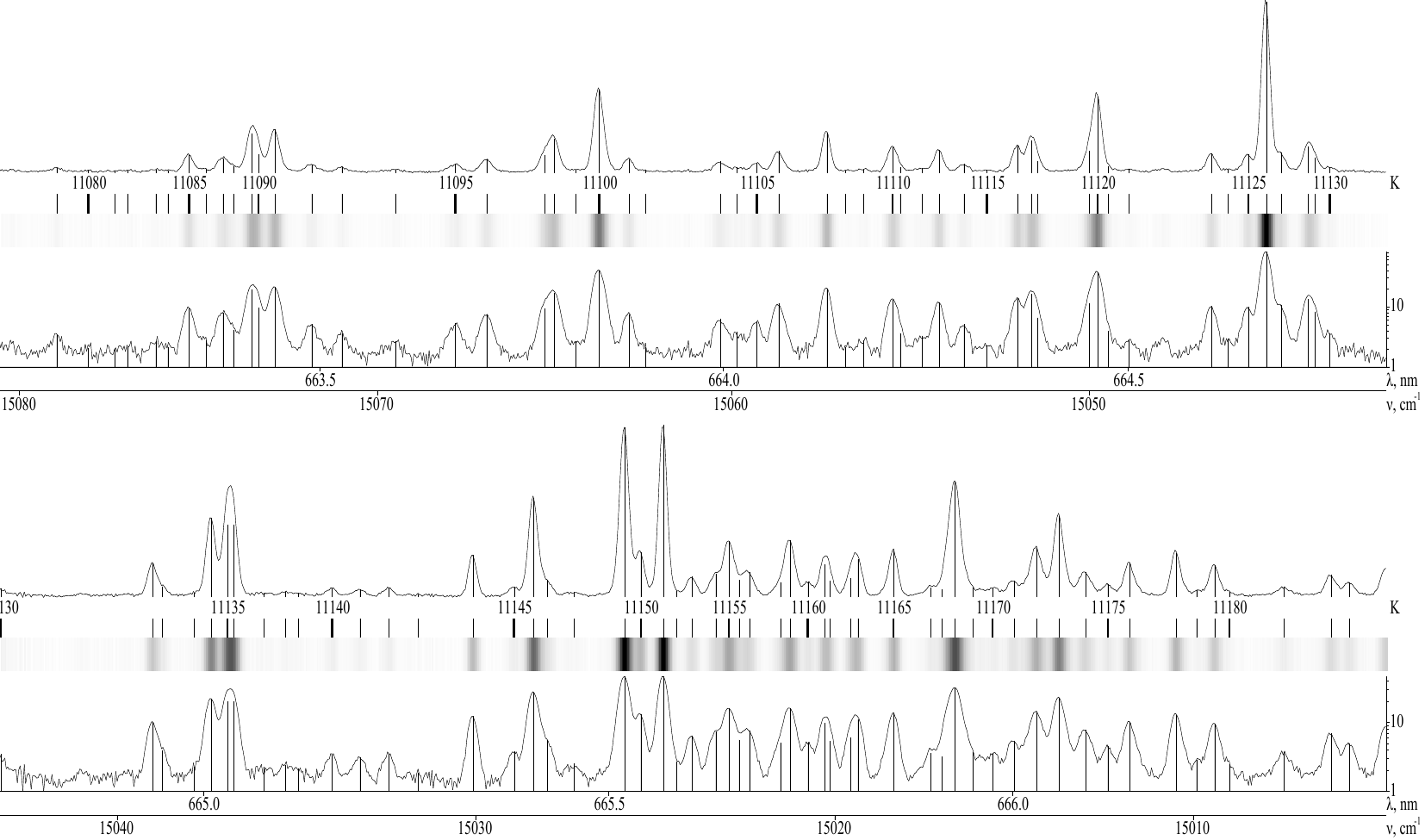}
\end{figure}

\newpage
\begin{figure}[!ht]
\includegraphics[angle=90, totalheight=0.9\textheight]{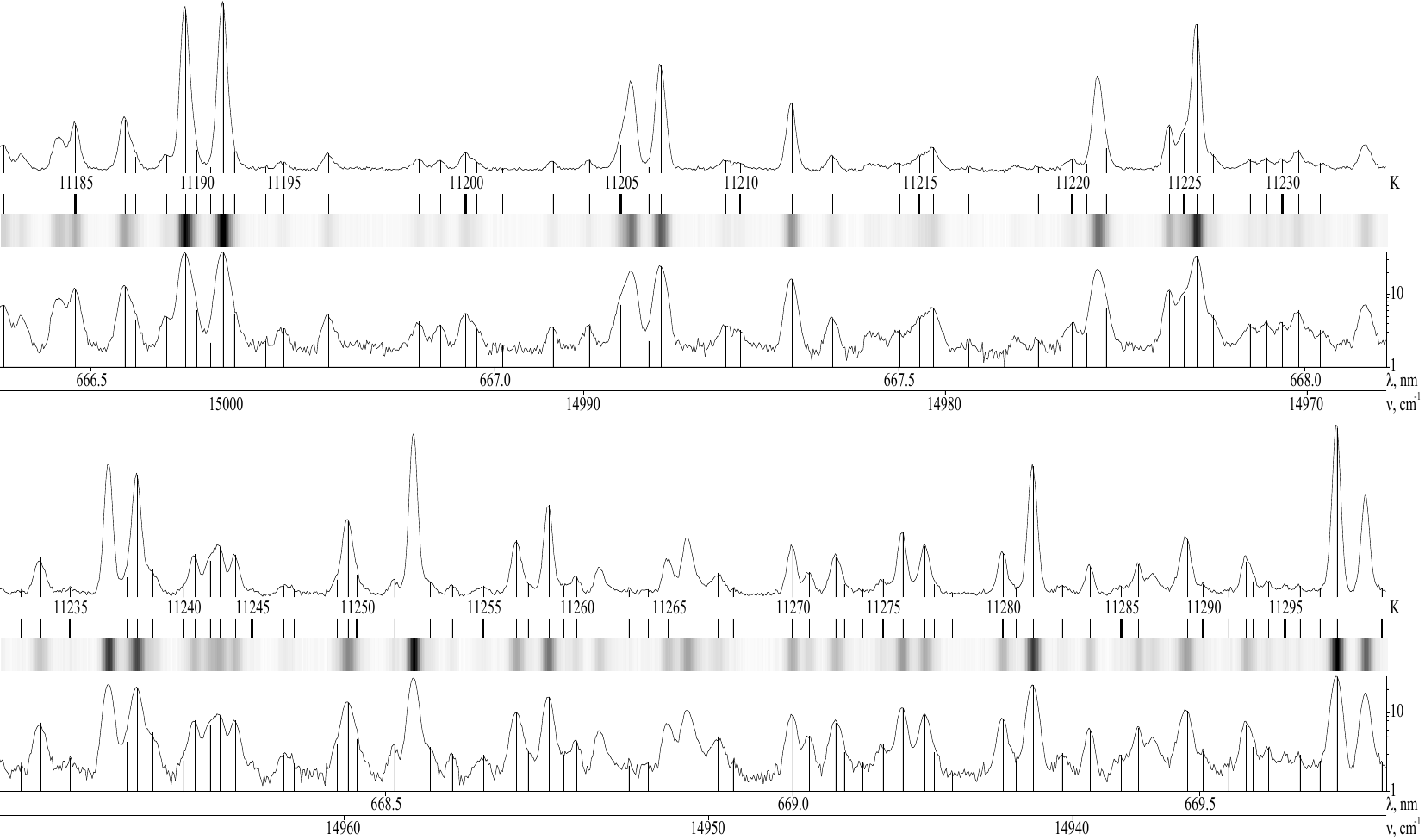}
\end{figure}

\newpage
\begin{figure}[!ht]
\includegraphics[angle=90, totalheight=0.9\textheight]{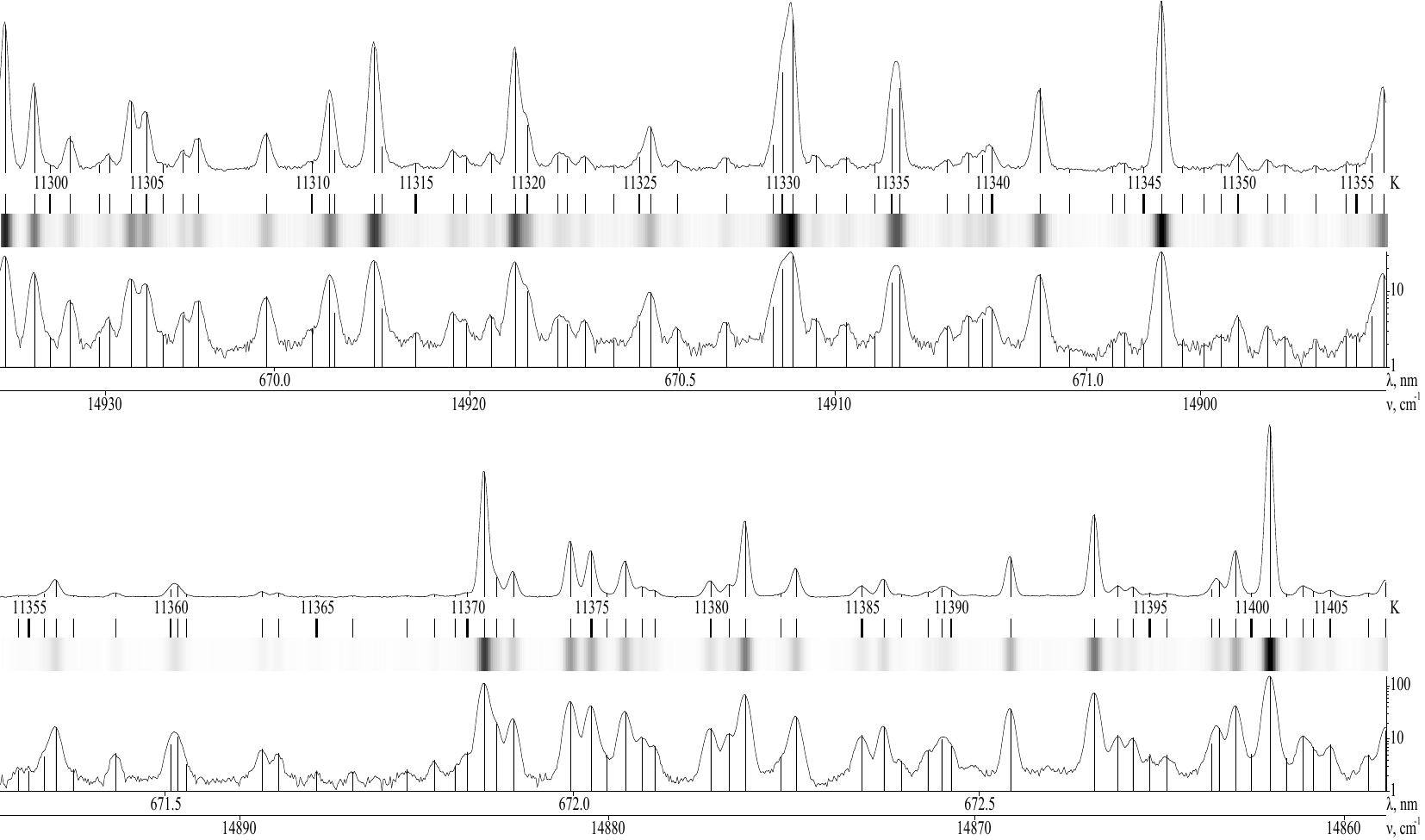}
\end{figure}

\newpage
\begin{figure}[!ht]
\includegraphics[angle=90, totalheight=0.9\textheight]{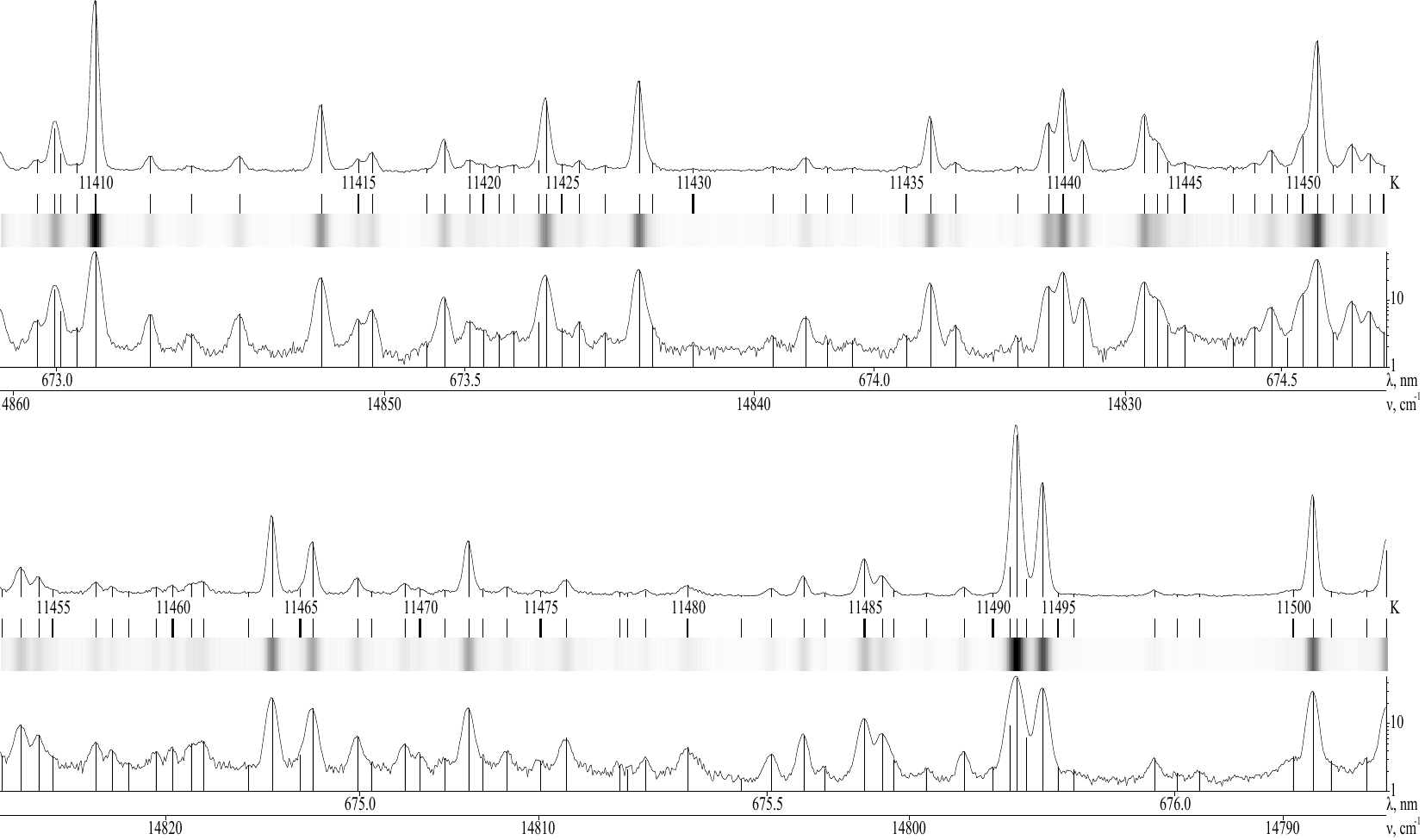}
\end{figure}

\newpage
\begin{figure}[!ht]
\includegraphics[angle=90, totalheight=0.9\textheight]{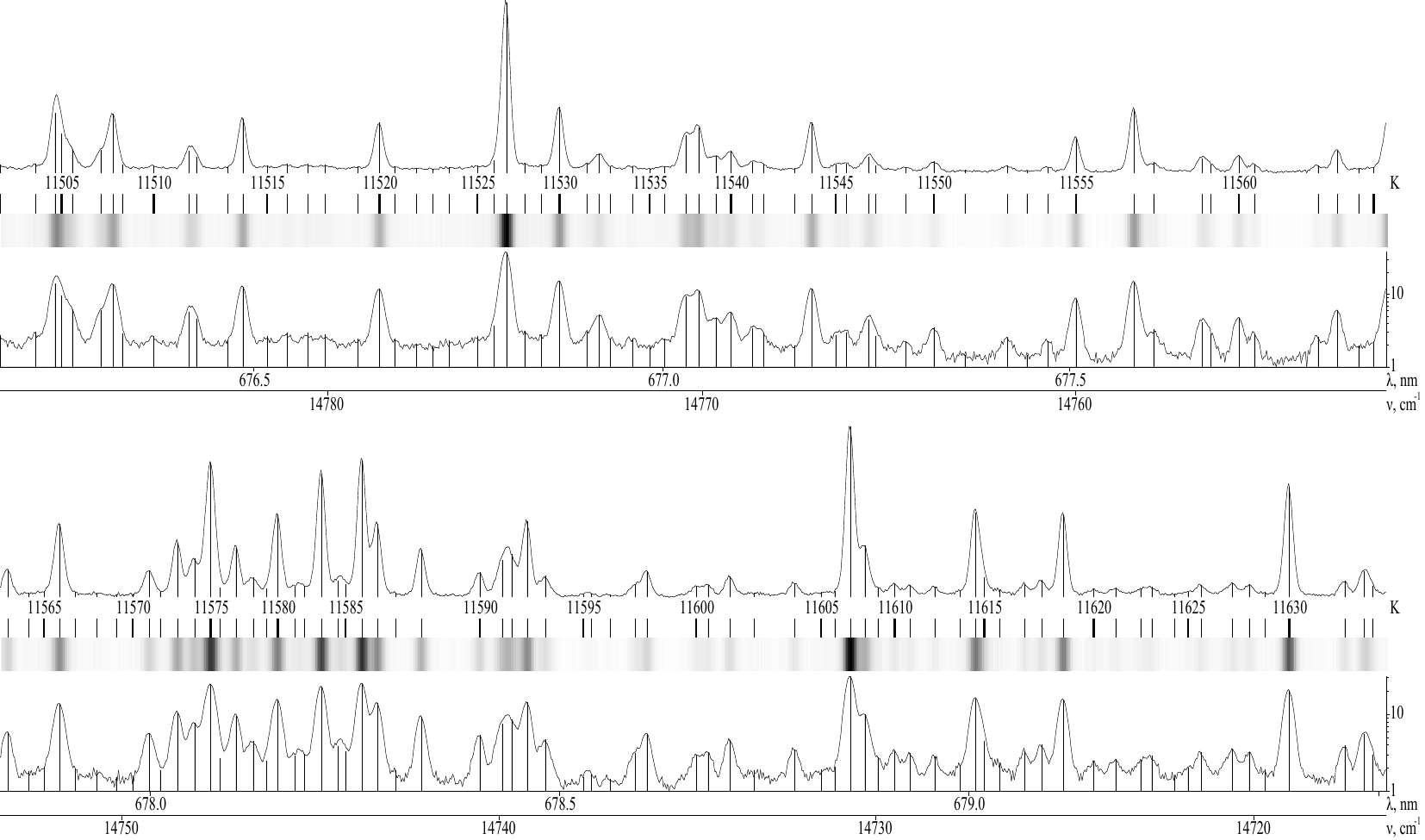}
\end{figure}

\newpage
\begin{figure}[!ht]
\includegraphics[angle=90, totalheight=0.9\textheight]{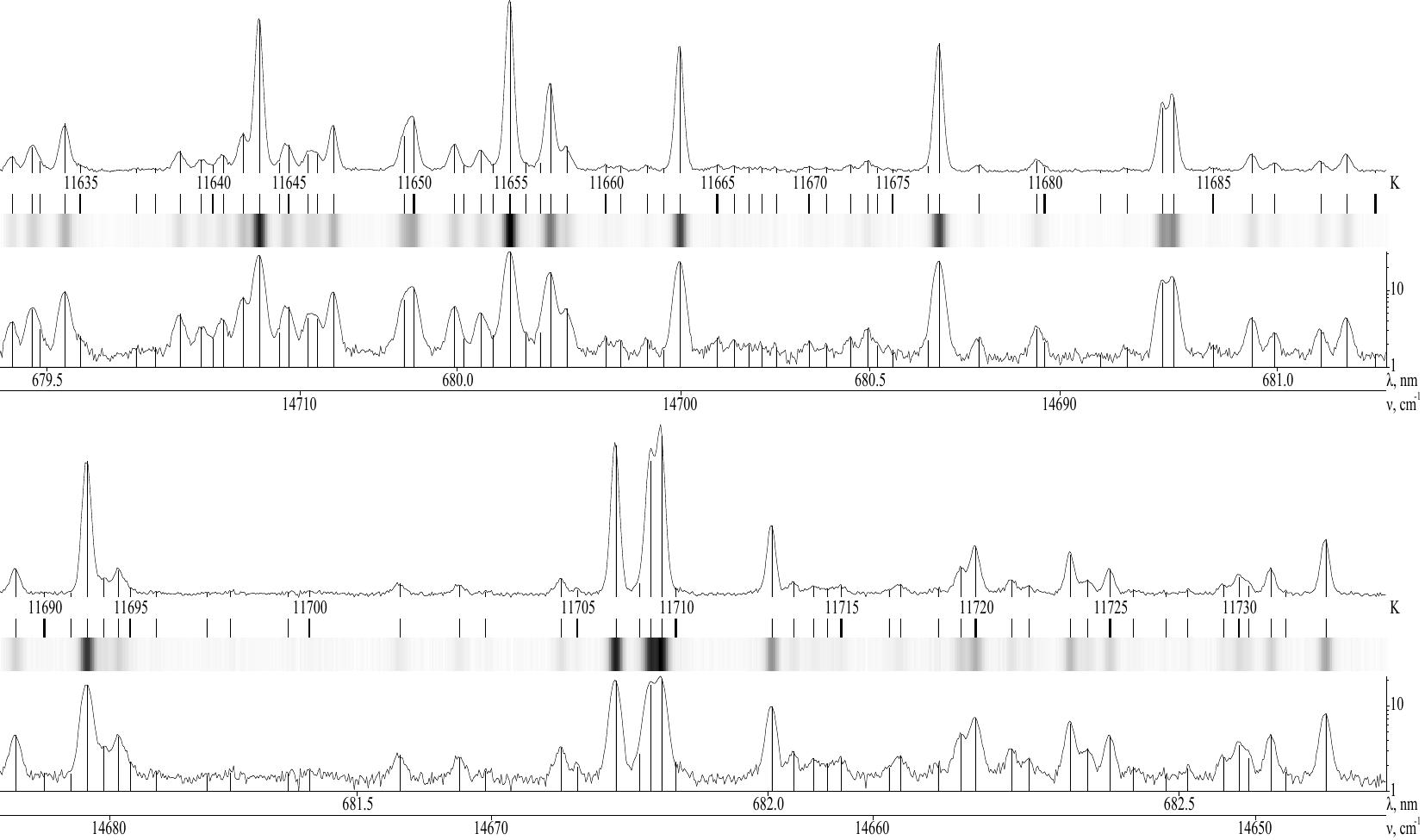}
\end{figure}

\newpage
\begin{figure}[!ht]
\includegraphics[angle=90, totalheight=0.9\textheight]{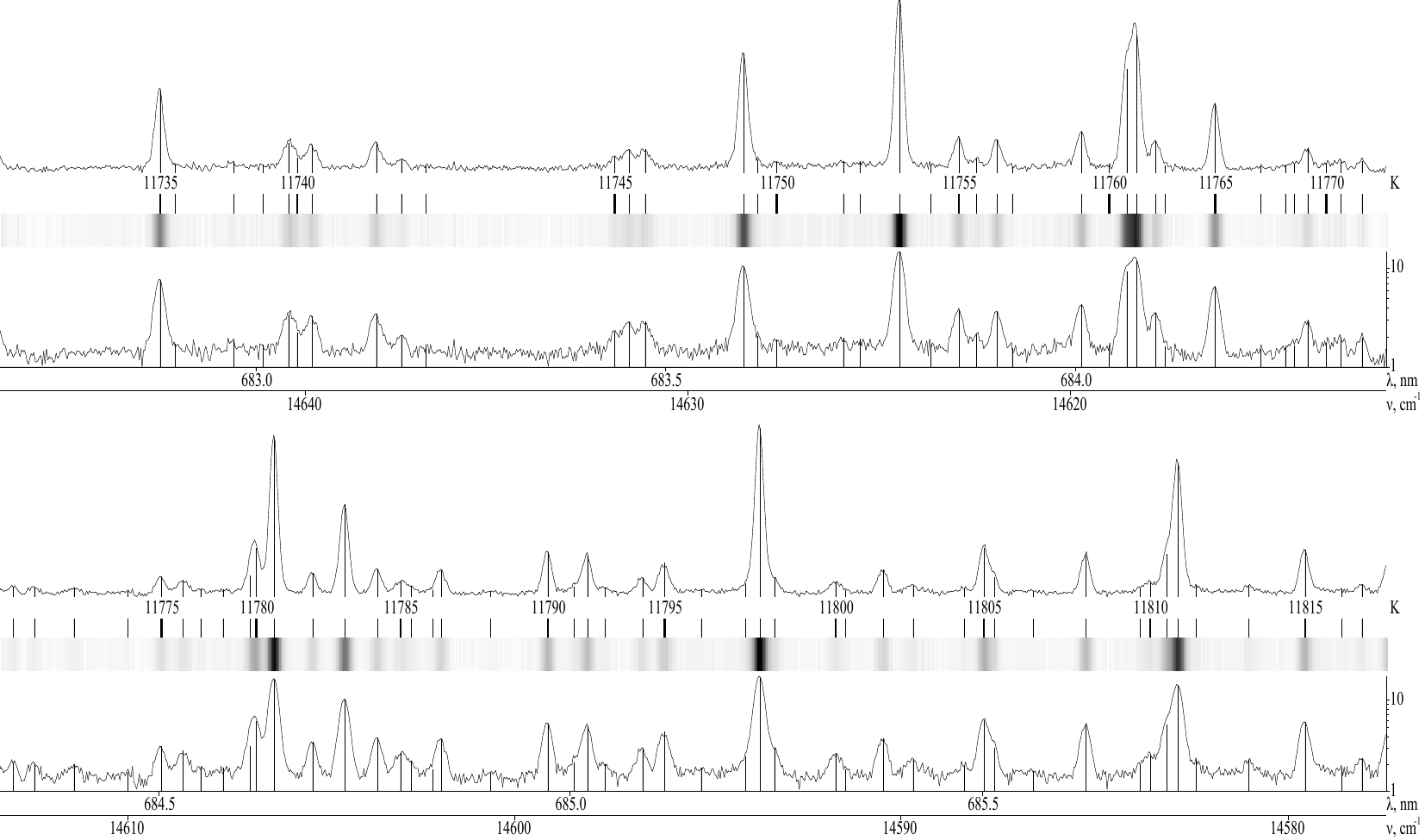}
\end{figure}

\newpage
\begin{figure}[!ht]
\includegraphics[angle=90, totalheight=0.9\textheight]{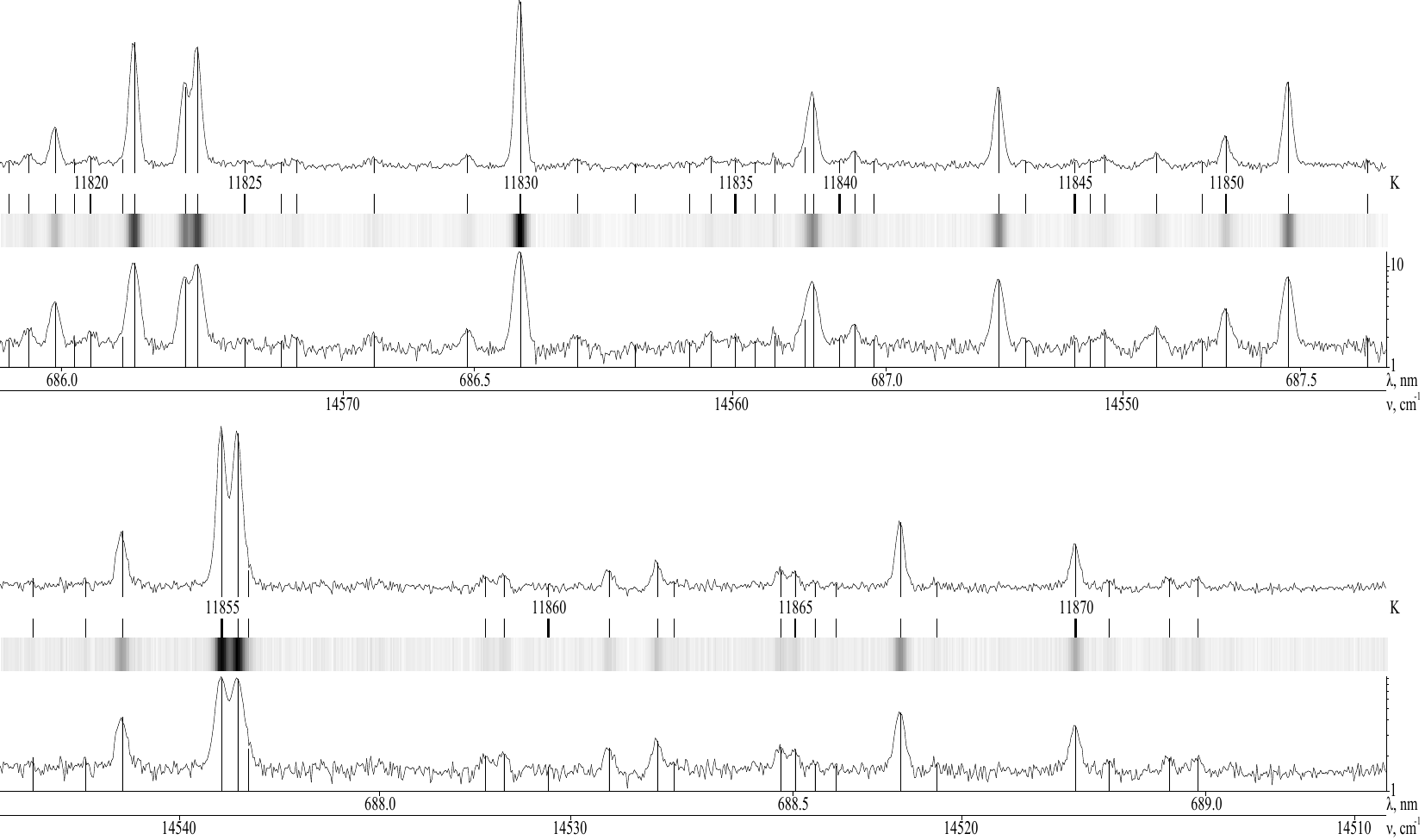}
\end{figure}

\newpage
\begin{figure}[!ht]
\includegraphics[angle=90, totalheight=0.9\textheight]{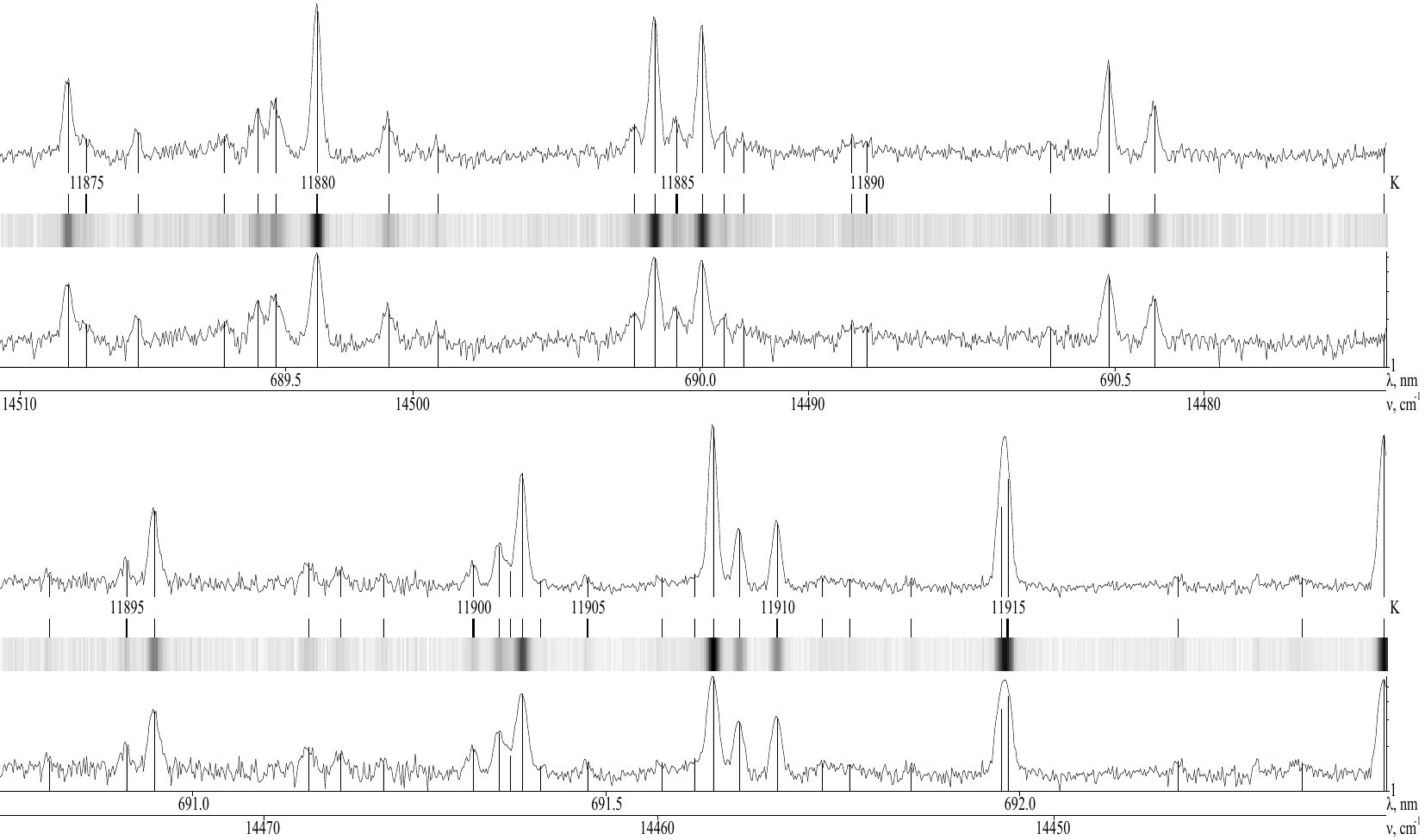}
\end{figure}

\newpage
\begin{figure}[!ht]
\includegraphics[angle=90, totalheight=0.9\textheight]{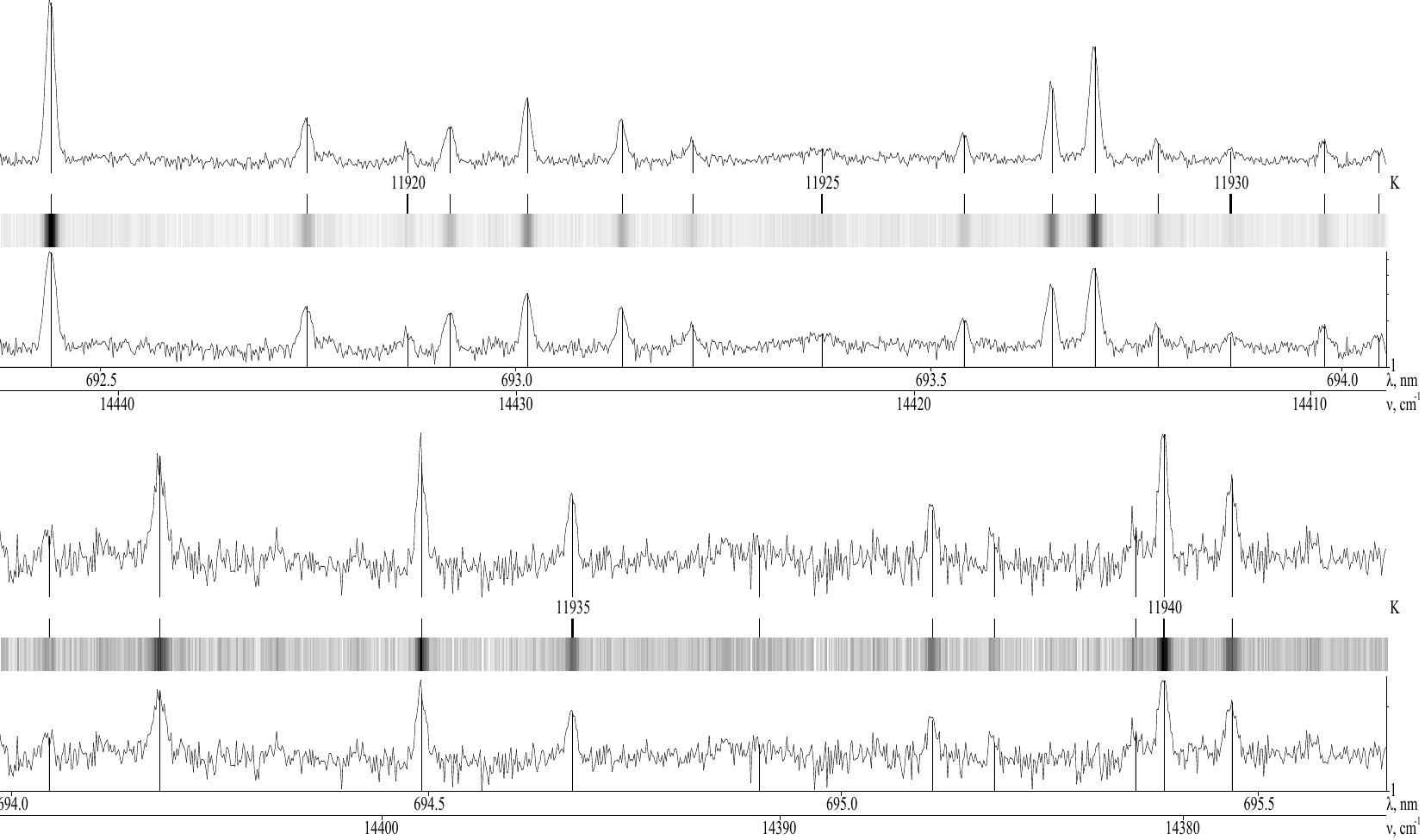}
\end{figure}

\end{document}